\documentclass[aps,prl,showpacs,preprintnumbers,amsmath,amssymb,superscriptaddress,twocolumn,longbibliography]{revtex4-1} 
\usepackage{amsmath}
\usepackage{graphicx}
\usepackage{rotating} 
\usepackage{float}
\usepackage{placeins} 
\usepackage{ulem}
\usepackage{physics}
\usepackage{comment}
\usepackage{xcolor}
\usepackage{amsfonts}
\usepackage{amssymb}
\usepackage{booktabs}   
\usepackage{multirow}   
\usepackage{siunitx}    
\usepackage{lipsum}
\usepackage{tabularx}

\definecolor{gold}{rgb}{0.83, 0.69, 0.22}

\newcommand\COMMENTED[1] {}

\begin{document}

\title{Translating Spin-Adapted RPA to Spin-Adapted TDDFT}

\author{Xiaoyu Zhang}
\email{zhangxiaoyu@stu.pku.edu.cn}

\affiliation{College of Chemistry and Molecular Engineering, Peking University, Beijing 100871, China}

\author{Tai Wang}
\affiliation{College of Chemistry and Molecular Engineering, Peking University, Beijing 100871, China}

\date{\today}

\begin{abstract}
Linear-response TDDFT based on open-shell references may produce spin-mixed excited states, particularly when the response space is not spin complete. Spin-adapted RPA was developed using tensor equation-of-motion and applying the Wigner–Eckart theorem with tensor decoupling. Casting the RPA Fock matrix and kernels as energy derivatives gives a TDDFT extension. To restore spin-component degeneracy and eliminate multi-counting of spin correlation in pure exchange-correlation functionals, we use hybrids combining HF exchange with spin-unpolarized pure XC parts. Given that the mapping from spin-adapted RPA to spin-adapted TDDFT is \textit{ad hoc}, we introduce a first-principles framework for evaluating the expectation value $\langle \hat{S}^2 \rangle$ of excited states in order to rigorously prove their spin purification. In addition, we derive explicit working equations for the computation of the single-particle density matrix of excited states, thereby enabling a natural orbital analysis of their electronic structure. Benchmarks, diradicals/triradicals, Cr$_2$ dissociation, and phenol O–H conical intersections demonstrate the method.
\end{abstract}

\maketitle
\section{Introduction}
Linear-response TDDFT is prevalent due to its balance between efficiency and accuracy.\cite{tddft} TDDFT formulated on a closed-shell reference state is typically free from spin contamination. However, when TDDFT is applied using an open-shell reference state, approximate response states may not be eigenfunctions of $\hat{S}^2$. \cite{IPATOV200960} Care is required, however, when interpreting spin contamination within Kohn–Sham DFT. Since $\hat{S}^2$
 contains a two-electron contribution, its expectation value cannot be determined from the one-particle density alone; the commonly reported ground-state $\langle \hat{S}^2 \rangle$ is that of the auxiliary Kohn–Sham determinant rather than the unknown interacting state yielding the same density. Moreover, different $\alpha$ and $\beta$ orbitals may be required to describe physical spin polarization in open-shell systems, and the reported spin contamination might not be accurate. \cite{https://doi.org/10.1002/qua.560560414} This should be distinguished from spin incompleteness, in which the response space lacks the spin-complementary configurations required to construct an $\hat{S}^2$ eigenfunction. \cite{doi:https://doi.org/10.1002/9781119356059.ch4}
Spin-flip TDDFT formulated on an open-shell reference state has been demonstrated to be particularly valuable, first for the diradicals \cite{10.1063/1.1545679}, and then for the description of conical intersections, charge-transfer excited states, nonadiabatic derivative couplings, and bond dissociation processes.\cite{doi:10.1021/acs.jctc.5c01272,10.1039/c9cp06507e} Thus, eliminating spin contamination in TDDFT, which entails performing spin adaptation, is of interest to us.

For spin-adapted TDDFT, several early works involved spin-step operators. \cite{10.1063/1.436028,10.1063/1.436297,FS9841900085,ODDERSHEDE198433,10.1063/1.2566733} However, evaluating the equation-of-motion (EOM) with spin-step operators in the matrix representation of second quantization is very complicated, which hinders the derivation of more detailed equations and implementations. We denote the total spin of the targeted (final) state as $S_{\mathrm{f}}$, and the total spin of the reference (initial) state as $S_{\mathrm{i}}$.
Subsequently, Sears, Sherrill, and Krylov \cite{10.1063/1.1568735} obtained spin-pure states corresponding to $S_{\mathrm{f}} - S_{\mathrm{i}} = 0,-1$ with $S_{\mathrm{i}} = 1$ within the configuration-interaction-singles (CIS) framework. Building on this advancement, Casanova and Head-Gordon \cite{10.1063/1.2965131} extended the spin purification procedure to access states with $S_{\mathrm{f}} - S_{\mathrm{i}} = 1, 0, -1$ for $S_{\mathrm{i}} = 1$ in CIS.
Later, Li and Liu \cite{10.1063/1.3463799} constructed spin-pure excited states characterized by $S_{\mathrm{f}} - S_{\mathrm{i}} = 1, 0,-1$ for arbitrary initial spin $S_{\mathrm{i}}$ within the random-phase approximation (RPA) and the time-dependent density functional theory (TDDFT). For $S_{\mathrm{f}}=S_{\mathrm{i}}-1$, they need additional truncation because of the spin-adapted nature of OO-OO blocks.  However, their own study identified limitations related to both the accuracy and the treatment of spin degeneracy.\cite{10.1063/1.3660688} Zhang and Herbert \cite{10.1063/1.4937571} constructed spin-pure excited states characterized by $S_{\mathrm{f}} - S_{\mathrm{i}} = -1$ at the RPA level for arbitrary initial spin $S_{\mathrm{i}}$ within the excitation manifold (the deexcitation manifold does not contribute to results in this case), and they translated it to TDA by introducing DFT Fock matrix elements. This method includes an empirical parameter $p_J$ and an empirical function $p(N_O)$, which need to be optimized for each functional. In the practical implementation, $p(N_O)$ is set to zero, and $p_J$ is fixed as $1-c_{\mathrm{HF}}$. If noncollinear or collinear kernels are applied, their spin-adapted TDDFT approach sometimes acquires unphysically negative roots (see table I in ref. \citenum{10.1063/5.0275059}).

To address this issue, the X-TDA ($S_{\mathrm{f}}=S_{\mathrm{i}}$) \cite{10.1063/1.3660688} and X-SF-TDA ($S_{\mathrm{f}}=S_{\mathrm{i}}-1$) \cite{Zhao18022026} approaches are introduced, in which a spin-adapted correction term is appended to the conventional TDA. X-TDA has no modifications to spin-flip-up TDA for $S_{\mathrm{f}}=S_{\mathrm{i}}+1$. X-TDA contains no additional fitted parameters, while X-SF-TDA introduces a global scaling function $g_X(c_X)$ containing the empirical parameter $g_{\mathrm{LDA}}$. Chibueze and Visscher \cite{10.1063/5.0275059} proposed a partially spin-adapted TDDFT called Q-SF-TDA. However, Q-SF-TDA is only spin-adapted in OO-OO blocks, where TDDFT has been spin-adapted.
Lee \textit{et al.} \cite{10.1063/1.5044202} introduced a mixed-reference scheme that yields electronic states with an enhanced degree of spin purification for transitions characterized by $S_{\mathrm{f}} - S_{\mathrm{i}} = 0,-1$ with $S_{\mathrm{i}} = 1$, although full spin purification is not attained. Nevertheless, MRSF-TDDFT continues to rely on collinear exchange–correlation kernels, following the formulation of Shao \textit{et al.} \cite{10.1063/1.1545679}. The use of a collinear kernel is fundamentally problematic because it does not possess global spin-rotational invariance and, consequently, does not maintain the degeneracy between the $|10\rangle$ and $|11\rangle$ states when $|11\rangle$ is selected as the reference state and a conventional spin-flip TDDFT framework is applied. \cite{doi:10.1021/acs.jctc.5c00714} It is noteworthy that the numerical instabilities of noncollinear kernels beyond the noncollinear LDA \cite{10.1063/1.4714499} have been solved by the proposal of the multicollinear approach \cite{PhysRevResearch.5.013036,doi:10.1021/acs.jctc.5c01305,RN69,doi:10.1021/acs.jctc.5c00714}. In addition, MRSF-TDDFT employs an empirical parameter $c$ for the pairing strength, $c_{pq}$. In the practical implementation, $c$ is fixed as $c_{\mathrm{HF}}$. In short, spin contamination in TDDFT remains a puzzle. We make a comparison among previously proposed methods in table \ref{tab:comparison}.

\begin{table}[H]
\caption{Comparison of representative TDDFT-based approaches
for spin purification.}
\label{tab:comparison}
\begin{ruledtabular}
\scriptsize
\setlength{\tabcolsep}{0.8pt}
\renewcommand{\arraystretch}{1.08}
\begin{tabular*}{\columnwidth}
{@{\extracolsep{\fill}}lccccc}
Method
& \shortstack[c]{spin\\purity}
& $S_{\mathrm{i}}$
& $\Delta S$
& \shortstack[c]{empirical\\parameters}
& \shortstack[c]{additional\\fitting} \\
\hline

X-TDA \cite{10.1063/1.3660688}
& Yes & Any & $0,+1$ & No & No \\

X-SF-TDA \cite{Zhao18022026}
& Yes & Any & $-1$ & Yes & Yes \\

Q-SF-TDA \cite{10.1063/5.0275059}
& Partial & Any & $-1$ & No & No \\

SA-SF-DFT \cite{10.1063/1.4937571}
& Yes & Any & $-1$ & Yes & No \\

MRSF-TDDFT \cite{10.1063/1.5044202}
& Partial & $1$ & $0,-1$ & Yes & No \\

This work
& Yes & Any & $+1,0,-1$ & No & No \\

\end{tabular*}
\end{ruledtabular}
\end{table}

In this work, we develop a rigorously spin-adapted formalism that enables the systematic construction of spin-pure excited states characterized by changes in the total spin quantum number $S_{\mathrm{f}} - S_{\mathrm{i}} = 1, 0, -1$ for arbitrary initial spin $S_{\mathrm{i}}$. The central strategy is to enforce spin adaptation at the level of the random-phase approximation (RPA), and subsequently to map the resulting Fock operator and response kernel onto their time-dependent density functional theory (TDDFT) counterparts. Thanks to ref. \citenum{10.1063/1.3463799} and ref. \citenum{10.1063/1.4937571}, we can directly use some of their resulting equations for spin-adapted RPA and focus on the translation. We recover the correct spin degeneracies associated with the individual components of the multiplet and eliminate the necessity of introducing empirical parameters. Furthermore, the approach preserves consistency between the density functional theory (DFT) Fock operators and the corresponding TDDFT kernels on the same functional. We denote this work as SA-TDDFT.

\section{Theory}
\subsection{Spin-Adapted TDDFT}
Our theory is built on spin adapted RPA, which was developed in Ref. \citenum{10.1063/1.3463799} for the cases $S_{\mathrm{f}} = S_{\mathrm{i}} + 1$ and $S_{\mathrm{f}} = S_{\mathrm{i}}$, and on the theory presented in Ref. \citenum{10.1063/1.4937571} for $S_{\mathrm{f}} = S_{\mathrm{i}} - 1$. Spin-adapted RPA is briefly revisited in our Supporting Information (SI). Making use of the aforementioned RPA results, we translate the RPA Focks and kernels to DFT and TDDFT ones.

The DFT Fockian is expressed as:
\begin{equation}
f_{pq}^{\sigma} = h_{pq}^{\sigma} +
\big[(p_{\sigma} q_{\sigma} \mid r_{\tau} r_{\tau})
- c_{\mathrm{HF}}\delta_{\sigma\tau} (p_{\sigma} r_{\tau} \mid r_{\tau} q_{\sigma})\big]
\, n_{r_\tau}+v^{\sigma,\mathrm{xc}}_{pq}
\end{equation}

\[
\sigma,\tau = \alpha,\beta, \qquad n_{r_\tau} = 0,1.
\]
\begin{equation}
    v^{\sigma,\mathrm{xc}}_{pq}=\int d \mathbf{r} p^\dagger_{\sigma} \frac{\delta E_{\mathrm{xc}}}{\delta \rho^\sigma} q_{\sigma}
\end{equation}

And the TDDFT kernel is expressed as:
\begin{equation}
\begin{split}
    \bigl[K^{\sigma\tau,\sigma' \tau'}\bigr]_{pq,rs}
 &= (p_{\sigma} q_{\tau} \mid s_{\tau'} r_{\sigma'}) 
   - c_{\mathrm{HF}}(p_{\sigma} r_{\sigma'} \mid s_{\tau'} q_{\tau}) \\
   &+ \frac{\partial^2 E_{\mathrm{xc}}}{\partial D^{\sigma\tau}_{pq}\partial D^{\tau^\prime \sigma^\prime}_{sr}}
\end{split}
\end{equation}
From a computational standpoint, the Fock matrix and the kernel are first constructed in the atomic orbital (AO) basis and are subsequently transformed into the molecular orbital (MO) basis.
\begin{equation}
    f^\sigma_{\mu \nu } = h_{\mu \nu} +(\mu \nu |\eta \theta)(D^{\alpha \alpha}_{\eta \theta}+D^{\beta \beta}_{\eta \theta})- c_{\mathrm{HF}} (\mu \theta|\eta \nu) D^{\sigma \sigma}_{\eta \theta} +\frac{\partial E_{\mathrm{xc}}}{\partial D_{\mu \nu}^{\sigma \sigma}}
\end{equation}
\begin{equation}
    [K]^{\sigma \sigma^\prime, \tau \tau^\prime}_{\mu \nu, \pi \theta} =(\mu \nu |\theta \pi) \delta_{\sigma \sigma^\prime} \delta_{\tau \tau^\prime} -c_{\mathrm{HF}} (\mu \pi|\theta \nu) \delta_{\sigma \tau} \delta_{\sigma^\prime \tau^\prime} +\frac{\partial^2 E_{\mathrm{xc}}}{\partial D^{\sigma \sigma^\prime}_{\mu \nu}\partial D^{\tau^\prime \tau}_{\theta \pi}}
\end{equation}
Among 16 kernels, we only have 6 non-zero ones: $[K]^{\alpha \alpha, \alpha \alpha}_{\mu \nu, \pi \theta}$, $[K]^{\alpha \alpha, \beta \beta}_{\mu \nu, \pi \theta}$, $[K]^{\beta \beta, \alpha \alpha}_{\mu \nu, \pi \theta}$, $[K]^{\beta \beta, \beta \beta}_{\mu \nu, \pi \theta}$, $[K]^{\alpha \beta, \alpha \beta}_{\mu \nu, \pi \theta}$, $[K]^{\beta \alpha, \beta \alpha}_{\mu \nu, \pi \theta}$. Among the 6 non-zero kernels, four are independent:
$[K]^{\alpha \alpha, \alpha \alpha}_{\mu \nu, \pi \theta}$, $[K]^{\alpha \alpha, \beta \beta}_{\mu \nu, \pi \theta}$ ($[K]^{\beta \beta, \alpha \alpha}_{\mu \nu, \pi \theta}=[K]^{\alpha \alpha, \beta \beta}_{\theta \pi, \nu \mu}$),
 $[K]^{\beta \beta,\beta \beta}_{\mu \nu, \pi \theta}$, $[K]^{\alpha \beta, \alpha \beta}_{\mu \nu, \pi \theta}$ ($[K]^{\beta \alpha, \beta \alpha}_{\mu \nu, \pi \theta}=[K]^{\alpha \beta, \alpha \beta}_{\theta \pi, \nu \mu}$)
Among the 4 independent kernels, three are spin-conserving:
\begin{equation}
    [K]^{\alpha \alpha, \alpha \alpha}_{\mu \nu, \pi \theta} = (\mu \nu |\theta \pi)  -c_{\mathrm{HF}} (\mu \pi|\theta \nu)  +\frac{\partial^2 E_{\mathrm{xc}}}{\partial D^{\alpha \alpha}_{\mu \nu}\partial D^{\alpha \alpha}_{\theta \pi}}
\end{equation}
\begin{equation}
    [K]^{\beta \beta, \beta \beta}_{\mu \nu, \pi \theta} = (\mu \nu |\theta \pi)  -c_{\mathrm{HF}} (\mu \pi|\theta \nu)  +\frac{\partial^2 E_{\mathrm{xc}}}{\partial D^{\beta \beta}_{\mu \nu}\partial D^{\beta \beta}_{\theta \pi}}
\end{equation}
\begin{equation}
    [K]^{\alpha \alpha, \beta \beta}_{\mu \nu, \pi \theta} = (\mu \nu |\theta \pi) +\frac{\partial^2 E_{\mathrm{xc}}}{\partial D^{\alpha \alpha}_{\mu \nu}\partial D^{\beta \beta}_{\theta \pi}}
\end{equation}
The one left is spin-flip:
\begin{equation}
    [K]^{\alpha \beta, \alpha \beta}_{\mu \nu, \pi \theta} =  -c_{\mathrm{HF}} (\mu \pi|\theta \nu)  +\frac{\partial^2 E_{\mathrm{xc}}}{\partial D^{\alpha \beta}_{\mu \nu}\partial D^{\beta \alpha}_{\theta \pi}}
\end{equation}
We now focus on the pure xc part, and we denote:
\begin{equation}
    \bigl[K_{\mathrm{xc}}\bigr]^{\sigma\tau,\sigma' \tau'}_{\mu \nu,\pi \theta}
 = \frac{\partial^2 E_{\mathrm{xc}}}{\partial D^{\sigma\tau}_{\mu \nu}\partial D^{\tau^\prime \sigma^\prime}_{\theta \pi}}=\int d\mathbf{r} \frac{\partial^2 e_{\mathrm{xc}}}{\partial \gamma_i \gamma_j} \frac{\partial \gamma_i}{\partial D^{\sigma\tau}_{\mu \nu}} \frac{\partial \gamma_j}{\partial D^{\tau^\prime \sigma^\prime}_{\theta \pi}}
\end{equation}
where \(e_{\mathrm{xc}}\) denotes the integrand for \(E_{\mathrm{xc}}\), and \(\gamma\) represents a set of input variables of \(e_{\mathrm{xc}}\), such as the electron density \(n\), the spin magnetization component \(m_z\), and their spatial gradients \(\nabla n\) and \(\nabla m_z\). Obviously, the three xc kernels only need collinear xc energy functionals $E_{\mathrm{xc}}[n,m_z]$, because noncollinear xc functionals $E_{\mathrm{xc}}[n,\mathbf{m}]$ have correct collinear limits. However, $   [K_{\mathrm{xc}}]^{\alpha \beta, \alpha \beta}_{\mu \nu, \pi \theta}$ should use noncollinear xc functionals. Making use of the multicollinear approach, we can construct the spin-flip xc kernel from the spin-conserving xc kernels.
We denote:
\begin{equation}
    D^0_{\mu \nu} = D^{\alpha \alpha}_{\mu \nu} +D^{\beta \beta}_{\mu \nu}
\end{equation}
\begin{equation}
    D^z_{\mu \nu} = D^{\alpha \alpha}_{\mu \nu} -D^{\beta \beta}_{\mu \nu}
\end{equation}
These kernels can be written as functions of $D^0$ and $D^z$. We construct the spin-flip xc kernel by:
\begin{equation}
     \bigl[K_{\mathrm{xc}}\bigr]^{\beta \alpha, \beta \alpha}_{\mu \nu,\pi \theta}(D^0,D^z) = \int_0^1 [K^T]_{\mu \nu,\pi \theta}(D^0,tD^z) dt
\end{equation}
By doing this, we avoid an explicit implementation of a noncollinear xc functional.

Finally, we transform the Fock and kernel under AO representation to MO representation:
\begin{equation}
    f^\sigma_{pq}= (c^{\sigma}_{\mu p})^* f^\sigma_{\mu \nu} c^\sigma_{\nu q}
\end{equation}
\begin{equation}
    [K]^{\sigma \sigma^\prime, \tau \tau^\prime}_{pq,rs} = (c^\sigma_{\mu p})^* c^{\sigma^\prime}_{\nu q} [K]^{\sigma \sigma^\prime, \tau \tau^\prime}_{\mu \nu,\theta \pi} c^\tau_{\theta r} (c^{\tau^\prime}_{\pi s})^*
\end{equation}

Intuitively, spin-adapted TDDFT incorporates spin-flip excitations and, therefore, necessitates a noncollinear exchange–correlation (xc) kernel, which can be constructed within the multicollinear formalism. However, the introduction of rigorously defined noncollinear pure xc functionals does not, in general, preserve spin degeneracy. The spin degeneracy condition (cf. eqs. 117-130 in ref. \citenum{10.1063/1.3463799} for detailed derivation) is 
\begin{equation}
    K^T=K^{\alpha \beta,\alpha \beta} = K^{\beta \alpha, \beta \alpha},
      \label{eq:sdc}
\end{equation}
where the short notation $K^T$ is mentioned in our Supporting Information.
This limitation arises because the kernels of noncollinear pure xc functionals still depend explicitly on the spin density, thereby breaking the spin symmetry that is maintained at the random-phase approximation (RPA) level, where the kernels are independent of the spin density. Another problem of directly applying noncollinear functionals is the multi-counting of spin correlation in the pure part, which will lead to some unphysically low roots.

To address the two problems, we formulate a sufficiently stringent criterion for preserving spin degeneracy, namely the use of spin-unpolarized pure xc functionals. To retain a physically meaningful description of spin-dependent interactions under this constraint, we employ hybrid functionals in which the Hartree–Fock (HF) exchange component reintroduces the requisite spin interactions. 

\subsection{Sanity Check on $\langle \hat{S}^2 \rangle$}

We present a sanity check of the spin-adaptation procedure by replacing the electronic Hamiltonian operator, $\hat{H}$, with the total spin operator, $\hat{S}^2$, expressed in second quantization. Equivalently, this corresponds to replacing the Fock and kernel with their spin counterparts. \cite{doi:10.1021/acs.jctc.6c00314}
By implementing this consistency assessment, we computationally corroborate the spin purity of the calculated electronic states. Interestingly, we can see that kernels of 'spin' also satisfy the spin degeneracy condition (eq. \ref{eq:sdc}).

We define:
\begin{subequations}
    \begin{equation}
    S_{pq} = \sum_{\mu \nu} c^*_{\mu p} c_{\nu q} \langle g_\mu | g_\nu  \rangle = \sum_{\mu \nu} c^*_{\mu p} c_{\nu q} S_{\mu \nu}
\end{equation}
\begin{equation}
    S_{\bar{p}q} = \sum_{\mu \nu} c^*_{\bar{\mu} \bar{p}} c_{\nu q} \langle g_\mu | g_\nu  \rangle = \sum_{\mu \nu} c^*_{\bar{\mu} \bar{p}} c_{\nu q} S_{\mu \nu}
\end{equation}
 \begin{equation}
    S_{p\bar{q}} = \sum_{\mu \nu} c^*_{\mu p} c_{\bar{\nu} \bar{q}} \langle g_\mu | g_\nu  \rangle = \sum_{\mu \nu} c^*_{\mu p} c_{\bar{\nu} \bar{q}} S_{\mu \nu}
\end{equation}
 \begin{equation}
    S_{\bar{p}\bar{q}} = \sum_{\mu \nu} c^*_{\bar{\mu} \bar{p}} c_{\bar{\nu} \bar{q}} \langle g_\mu | g_\nu  \rangle = \sum_{\mu \nu} c^*_{\bar{\mu} \bar{p}} c_{\bar{\nu} \bar{q}} S_{\mu \nu}
\end{equation}
\label{eq:smat}
\end{subequations}
Then, the Fock of 'spin' is written as:
\begin{equation}
\begin{split}
     (f_s)_{pq}^\alpha &= \frac{1}{2} S_{pq} \sum_j (S_{jj}-S_{\bar{j}\bar{j}}) -\frac{1}{2} \sum_j S_{pj}S_{jq} -\sum_j S_{p\bar{j}}S_{\bar{j}q}  \\
     &= \frac{1}{2} \delta_{pq}\sum_j (S_{jj}-S_{\bar{j}\bar{j}}) - \frac{1}{2} n_p\delta_{pq}-\sum_j S_{p\bar{j}}S_{\bar{j}q}
\end{split}
\end{equation}
\begin{equation}
\begin{split}
     (f_s)_{pq}^\beta &= -\frac{1}{2} S_{\bar{p} \bar{q}} \sum_j (S_{jj}-S_{\bar{j}\bar{j}}) -\frac{1}{2} \sum_j S_{\bar{p}\bar{j}}S_{\bar{j}\bar{q}} -\sum_j S_{\bar{p}j}S_{j\bar{q}}\\
     &= -\frac{1}{2} \delta_{pq} \sum_j (S_{jj}-S_{\bar{j}\bar{j}}) -\frac{1}{2}n_{\bar{p}} \delta_{pq}-\sum_j S_{\bar{p}j}S_{j\bar{q}}
\end{split}
\end{equation}
The left two terms are zero.

The kernel of 'spin' is written as:
\begin{equation}
(K_s)_{pq,rs}^{\alpha \alpha, \alpha \alpha}
= \frac{1}{2}\Big( S_{pq}\,S_{sr}-S_{pr}\,S_{sq}\Big) = \frac{1}{2}(\delta_{pq} \delta_{sr}- \delta_{pr} \delta_{sq})
\end{equation}

\begin{equation}
(K_s)_{pq,rs}^{\beta \beta, \beta \beta}
= \frac{1}{2}\Big( S_{\bar{p}\bar{q}}\,S_{\bar{s}\bar{r}}-S_{\bar{p}\bar{r}}\,S_{\bar{s}\bar{q}}\Big)
= \frac{1}{2}(\delta_{pq} \delta_{sr}- \delta_{pr} \delta_{sq})
\end{equation}

\begin{equation}
(K_s)_{pq,rs}^{\alpha \alpha, \beta \beta}
= -\frac{1}{2}\,S_{pq}\,S_{\bar{s}\bar{r}} \;-\; S_{p\bar{r}}\,S_{\bar{s}q}
= -\frac{1}{2}\delta_{pq} \delta_{sr}- S_{p\bar{r}}\,S_{\bar{s}q}
\end{equation}

\begin{equation}
(K_s)_{pq,rs}^{\beta \beta, \alpha \alpha}
= -\frac{1}{2}\,S_{\bar{p}\bar{q}}\,S_{sr} \;-\; S_{\bar{p}r}\,S_{s\bar{q}}
= -\frac{1}{2}\delta_{pq} \delta_{sr} - S_{\bar{p}r}\,S_{s\bar{q}}
\end{equation}

\begin{equation}
(K_s)_{pq,rs}^{\alpha \beta, \alpha \beta}
= \frac{1}{2}\,S_{pr}\,S_{\bar{s}\bar{q}} \;+\; S_{p\bar{q}}\,S_{\bar{s}r}
= \frac{1}{2}\delta_{pr} \delta_{sq} + S_{p\bar{q}}\,S_{\bar{s}r}
\end{equation}

\begin{equation}
(K_s)_{pq,rs}^{\beta \alpha, \beta \alpha}
= \frac{1}{2}\,S_{\bar{p}\bar{r}}\,S_{sq} \;+\; S_{\bar{p}q}\,S_{s\bar{r}}
= \frac{1}{2}\delta_{pr} \delta_{sq} + S_{\bar{p}q}\,S_{s\bar{r}}
\end{equation}

The left ten terms are zero. 

Making use of the aforementioned equations, we know that $M_{S_\mathrm{f}}$ can be expressed by the Fock and kernel. We translate those Focks and kernels to Focks of 'spin' and kernels of 'spin', which gives $(M_s)_{S_\mathrm{f}}$. The spin change by spin-adapted TDDFT can be calculated via:
\begin{equation}
    \Delta \langle \hat{S}^2 \rangle = \frac{X^\dagger M_s X}{X^\dagger N X}
\end{equation}
The spin of excited states is:
\begin{equation}
    \langle \hat{S}^2 \rangle_\lambda =\langle \hat{S}^2 \rangle_0 +\Delta \langle \hat{S}^2 \rangle_\lambda
\end{equation}

\subsection{Single-particle density matrix}
\label{sec:single-particle-density}

The aim of this subsection is to calculate the unrelaxed spin-free
one-particle density matrix of an excited state obtained from SA-TDDFT or
SA-TDA.  The second-quantized one-particle operators are
\begin{equation}
 \hat\gamma_{p_\sigma q_{\sigma'}}
 =\hat a_{p_\sigma}^\dagger\hat a_{q_{\sigma'}},
 \end{equation}
 
 \begin{equation}
 \hat E_{xy}=\hat\gamma_{x_\alpha y_\alpha}
             +\hat\gamma_{x_\beta y_\beta},
\end{equation}
\begin{equation}
 \hat E_{xy}^\dagger=\hat E_{yx}.
 \label{eq:spdm-density-operator}
\end{equation}
Uppercase labels denote spin orbitals and lowercase labels denote spatial
orbitals.  For the high-spin RO reference,
\begin{equation}
 |0\rangle=|S_{\mathrm i}S_{\mathrm i}\rangle,
 \qquad
 \langle0|\hat\gamma_{PQ}|0\rangle=\delta_{PQ}n_P,
 \qquad
 n_P\in\{0,1\}.
 \label{eq:spdm-reference-contraction}
\end{equation}
We evaluate the $S_{\mathrm f}=S_{\mathrm i}+1$ and $S_{\mathrm f}=S_{\mathrm i}$ sectors with EOM3 and the $S_{\mathrm f}=S_{\mathrm i}-1$ sector with
EOM1.  The construction is obtained through the direct replacement
\begin{equation}
 \hat H\longrightarrow\hat E_{xy},
 \qquad
 M_{S_{\mathrm f}}\longrightarrow[M_e^{xy}]_{S_{\mathrm f}},
 \label{eq:spdm-operator-replacement}
\end{equation}
within the corresponding equations in the Supporting
Information.  At the spin-orbital level, the one-body matrix is
\begin{equation}
 [M_e^{xy}]_{PQ,RS}
 =\langle S_{\mathrm i}S_{\mathrm i}|
 \{\hat\gamma_{QP},\hat E_{xy},\hat\gamma_{RS}\}
 |S_{\mathrm i}S_{\mathrm i}\rangle.
 \label{eq:spdm-general-matrix}
\end{equation}
Every $[M_e^{xy}]_{S_{\mathrm f}}$ below is represented in exactly the same
final response basis as $M_{S_{\mathrm f}}$ and $N_{S_{\mathrm f}}$,
including the same normalization, excitation--deexcitation pair reversal,
phase convention, and zero-mode projection where required.  Consequently,
the Hamiltonian eigenvectors contract directly with the one-body matrices
given below.

For $S_{\mathrm f}=S_{\mathrm i}+1$, the complete matrix in the ordered
basis $(\mathrm{CV}(1)\mid\mathrm{VC}(1))$ is
\begin{equation}
 [M_e^{xy}]_{S_{\mathrm f}}
 =\begin{pmatrix}
 \delta_{ij}\delta_{ax}\delta_{by}
 -\delta_{ab}\delta_{jx}\delta_{iy}&0\\
 0&\delta_{ij}\delta_{bx}\delta_{ay}
 -\delta_{ab}\delta_{ix}\delta_{jy}.
 \end{pmatrix}
\end{equation}

For an open-shell reference, the $S_\mathrm{f}=S_\mathrm{i}$ matrix is written in
the normalized order
\begin{equation}
 (\mathrm{CV}(0),\mathrm{CO}(0),\mathrm{OV}(0),\mathrm{CV}(1)
 \mid\mathrm{VC}(0),\mathrm{OC}(0),\mathrm{VO}(0),\mathrm{VC}(1)).
 \label{eq:spdm-equal-order}
\end{equation}
Its explicit working form is
\begin{widetext}
\begin{equation}
\begingroup
\setlength{\arraycolsep}{2pt}
\renewcommand{\arraystretch}{1.05}
\resizebox{\linewidth}{!}{$
[M_e^{xy}]_{S_{\mathrm f}}=
\left(
\begin{array}{cccc|cccc}
\delta_{ij}\delta_{ax}\delta_{by}-\delta_{ab}\delta_{jx}\delta_{iy}
&\dfrac{3\delta_{ij}\delta_{ax}\delta_{vy}}{2\sqrt2}
&-\dfrac{3\delta_{ab}\delta_{vx}\delta_{iy}}{2\sqrt2}
&0&0&0&0&0\\[1.2ex]

\dfrac{3\delta_{ij}\delta_{ux}\delta_{by}}{2\sqrt2}
&\delta_{ij}\delta_{ux}\delta_{vy}-\delta_{uv}\delta_{jx}\delta_{iy}
&0
&\dfrac{\sqrt{(S_{\mathrm i}+1)/S_{\mathrm i}}\,
\delta_{ij}\delta_{ux}\delta_{by}}{2\sqrt2}
&0&0&0&0\\[1.2ex]

-\dfrac{3\delta_{ab}\delta_{jx}\delta_{uy}}{2\sqrt2}
&0
&\delta_{uv}\delta_{ax}\delta_{by}-\delta_{ab}\delta_{vx}\delta_{uy}
&\dfrac{\sqrt{(S_{\mathrm i}+1)/S_{\mathrm i}}\,
\delta_{ab}\delta_{jx}\delta_{uy}}{2\sqrt2}
&0&0&0&0\\[1.2ex]

0
&\dfrac{\sqrt{(S_{\mathrm i}+1)/S_{\mathrm i}}\,
\delta_{ij}\delta_{ax}\delta_{vy}}{2\sqrt2}
&\dfrac{\sqrt{(S_{\mathrm i}+1)/S_{\mathrm i}}\,
\delta_{ab}\delta_{vx}\delta_{iy}}{2\sqrt2}
&\delta_{ij}\delta_{ax}\delta_{by}-\delta_{ab}\delta_{jx}\delta_{iy}
&0&0&0&0\\
\hline
0&0&0&0
&\delta_{ij}\delta_{bx}\delta_{ay}-\delta_{ab}\delta_{ix}\delta_{jy}
&\dfrac{3\delta_{ij}\delta_{vx}\delta_{ay}}{2\sqrt2}
&-\dfrac{3\delta_{ab}\delta_{ix}\delta_{vy}}{2\sqrt2}
&0\\[1.2ex]

0&0&0&0
&\dfrac{3\delta_{ij}\delta_{bx}\delta_{uy}}{2\sqrt2}
&\delta_{ij}\delta_{vx}\delta_{uy}-\delta_{uv}\delta_{ix}\delta_{jy}
&0
&\dfrac{\sqrt{(S_{\mathrm i}+1)/S_{\mathrm i}}\,
\delta_{ij}\delta_{bx}\delta_{uy}}{2\sqrt2}\\[1.2ex]

0&0&0&0
&-\dfrac{3\delta_{ab}\delta_{ux}\delta_{jy}}{2\sqrt2}
&0
&\delta_{uv}\delta_{bx}\delta_{ay}-\delta_{ab}\delta_{ux}\delta_{vy}
&\dfrac{\sqrt{(S_{\mathrm i}+1)/S_{\mathrm i}}\,
\delta_{ab}\delta_{ux}\delta_{jy}}{2\sqrt2}\\[1.2ex]

0&0&0&0
&0
&\dfrac{\sqrt{(S_{\mathrm i}+1)/S_{\mathrm i}}\,
\delta_{ij}\delta_{vx}\delta_{ay}}{2\sqrt2}
&\dfrac{\sqrt{(S_{\mathrm i}+1)/S_{\mathrm i}}\,
\delta_{ab}\delta_{ix}\delta_{vy}}{2\sqrt2}
&\delta_{ij}\delta_{bx}\delta_{ay}-\delta_{ab}\delta_{ix}\delta_{jy}
\end{array}
\right)
$}
\endgroup
\label{eq:spdm-equal-working-matrix}
\end{equation}
\end{widetext}
For a closed-shell reference, the $S_{\mathrm{f}}=S_{\mathrm{i}} $ matrix must instead be
derived directly in the order $(\mathrm{CV}(0)\mid\mathrm{VC}(0))$:
\begin{equation}
 [M_e^{xy}]_{S_{\mathrm f}}
 =\begin{pmatrix}
 \delta_{ij}\delta_{ax}\delta_{by}
 -\delta_{ab}\delta_{jx}\delta_{iy}&0\\
 0&\delta_{ij}\delta_{bx}\delta_{ay}
 -\delta_{ab}\delta_{ix}\delta_{jy}
 \end{pmatrix}.
 \label{eq:spdm-closed-equal-working-matrix}
\end{equation}
The open-shell expression contains
$\sqrt{(S_{\mathrm i}+1)/S_{\mathrm i}}$ and therefore cannot be reduced to
the closed-shell expression by simply substituting $S_{\mathrm i}=0$.

For $S_{\mathrm f}=S_{\mathrm i}-1$, EOM1 gives
\begin{equation}
\footnotesize
 [\widetilde M_e^{xy}]
 =\frac{1}{S_{\mathrm i}(2S_{\mathrm i}-1)}[M_e^{xy}]_+
 -\frac{2S_{\mathrm i}+1}{S_{\mathrm i}(2S_{\mathrm i}-1)}[M_e^{xy}]_0
 +\frac{2S_{\mathrm i}+1}{2S_{\mathrm i}-1}[M_e^{xy}]_-.
 \label{eq:spdm-minus-block-combination}
\end{equation}
The ordered basis is
\begin{equation}
 (\mathrm{CV(1)},\mathrm{CO(1)},\mathrm{OV(1)},\mathrm{OO(1)},\mathrm{TT(1)}
 \mid\mathrm{VC(1)},\mathrm{OC(1)},\mathrm{VO(1)}),
 \label{eq:spdm-minus-order}
\end{equation}
where OO contains the ordered pairs $t\ne u$ and TT contains the diagonal
pairs $t=u$.  The matrix is:
\begin{widetext}
\begin{equation}
\begingroup
\setlength{\arraycolsep}{2pt}
\renewcommand{\arraystretch}{1.05}
\scriptsize
\resizebox{\linewidth}{!}{$
[M_e^{xy}]_{S_{\mathrm f}}=
\left(
\begin{array}{cccc|c|ccc}
[\widetilde A_e^{xy}]_{ai,bj} &[\widetilde A_e^{xy}]_{ai,vj}
&[\widetilde A_e^{xy}]_{ai,bv}&[\widetilde A_e^{xy}]_{ai,vw}
&[\widetilde L_e^{xy}]_{ai,vv}
&[\widetilde B_e^{xy}]_{ai,jb}&[\widetilde B_e^{xy}]_{ai,jv}&[\widetilde B_e^{xy}]_{ai,vb}\\

[\widetilde A_e^{xy}]_{ui,bj} &[\widetilde A_e^{xy}]_{ui,vj}
&[\widetilde A_e^{xy}]_{ui,bv}&[\widetilde A_e^{xy}]_{ui,vw}
&[\widetilde L_e^{xy}]_{ui,vv}
&[\widetilde B_e^{xy}]_{ui,jb}&[\widetilde B_e^{xy}]_{ui,jv}&[\widetilde B_e^{xy}]_{ui,vb}\\

[\widetilde A_e^{xy}]_{au,bj} &[\widetilde A_e^{xy}]_{au,vj}
&[\widetilde A_e^{xy}]_{au,bv}&[\widetilde A_e^{xy}]_{au,vw}
&[\widetilde L_e^{xy}]_{au,vv}
&[\widetilde B_e^{xy}]_{au,jb}&[\widetilde B_e^{xy}]_{au,jv}&[\widetilde B_e^{xy}]_{au,vb}\\

[\widetilde A_e^{xy}]_{tu,bj} &[\widetilde A_e^{xy}]_{tu,vj}
&[\widetilde A_e^{xy}]_{tu,bv}&[\widetilde A_e^{xy}]_{tu,vw}
&[\widetilde L_e^{xy}]_{tu,vv}
&[\widetilde B_e^{xy}]_{tu,jb}&[\widetilde B_e^{xy}]_{tu,jv}&[\widetilde B_e^{xy}]_{tu,vb}\\
\hline
[\widetilde L_e^{xy}]_{tt,bj}&[\widetilde L_e^{xy}]_{tt,vj}
&[\widetilde L_e^{xy}]_{tt,bv}&[\widetilde L_e^{xy}]_{tt,vw}
&[\widetilde D_e^{xy}]_{tt,vv}
&[\widetilde L_e^{xy}]_{tt,jb}&[\widetilde L_e^{xy}]_{tt,jv}&[\widetilde L_e^{xy}]_{tt,vb}\\
\hline
[\widetilde B_e^{xy}]_{ia,bj}&[\widetilde B_e^{xy}]_{ia,vj}
&[\widetilde B_e^{xy}]_{ia,bv}&[\widetilde B_e^{xy}]_{ia,vw}
&[\widetilde L_e^{xy}]_{ia,vv}
&[\widetilde A_e^{xy}]_{ia,jb}&[\widetilde A_e^{xy}]_{ia,jv}&[\widetilde A_e^{xy}]_{ia,vb}\\

[\widetilde B_e^{xy}]_{iu,bj}&[\widetilde B_e^{xy}]_{iu,vj}
&[\widetilde B_e^{xy}]_{iu,bv}&[\widetilde B_e^{xy}]_{iu,vw}
&[\widetilde L_e^{xy}]_{iu,vv}
&[\widetilde A_e^{xy}]_{iu,jb}&[\widetilde A_e^{xy}]_{iu,jv}&[\widetilde A_e^{xy}]_{iu,vb}\\

[\widetilde B_e^{xy}]_{ua,bj}&[\widetilde B_e^{xy}]_{ua,vj}
&[\widetilde B_e^{xy}]_{ua,bv}&[\widetilde B_e^{xy}]_{ua,vw}
&[\widetilde L_e^{xy}]_{ua,vv}
&[\widetilde A_e^{xy}]_{ua,jb}&[\widetilde A_e^{xy}]_{ua,jv}&[\widetilde A_e^{xy}]_{ua,vb}
\end{array}
\right),
$}
\endgroup
\label{eq:spdm-minus-full-matrix}
\end{equation}
\end{widetext}
where the last three block rows vanish.  The entries
in Eq.~\eqref{eq:spdm-minus-full-matrix} are obtained from
Eq.~\eqref{eq:spdm-minus-block-combination} and the tables shown in the Appendix.

For the $S_{\mathrm{f}}=S_{\mathrm{i}}-1 $ sector, the same TT zero-mode projection is applied
simultaneously to $M_{S_{\mathrm f}}$, $N_{S_{\mathrm f}}$, and every TT
connected block of $[M_e^{xy}]_{S_{\mathrm f}}$, so all three matrices remain
in the same final projected basis.  

When SA-TDA is used, $Z_\lambda$, $[M_e^{xy}]_{S_{\mathrm f}}$,
and $N_{S_{\mathrm f}}$ in the following equation are all restricted
to the same positive principal subspace of the final response basis,
after the zero-mode projection where required.
Let $Z_\lambda$ denote the Hamiltonian eigenvector in the response
space of the selected method.  For full SA-TDDFT in the
$S_{\mathrm f}=S_{\mathrm i}-1$ sector, $Z_\lambda$ is the canonical
finite-root representative defined in the Supporting Information.
The SA-TDDFT or SA-TDA density change is
\begin{equation}
 \Delta D_{xy}^{(\lambda)}
 =\frac{Z_\lambda^\dagger[M_e^{xy}]_{S_{\mathrm f}}Z_\lambda}
 {Z_\lambda^\dagger N_{S_{\mathrm f}}Z_\lambda}.
 \label{eq:spdm-density-change-tddft}
\end{equation}

The total spin-free density is
\begin{equation}
 D_{xy}^{(\lambda)}=D_{xy}^{(0)}+\Delta D_{xy}^{(\lambda)}.
 \label{eq:spdm-total-density}
\end{equation}
In an orthonormal RO spatial-orbital basis, the reference density is
\begin{equation}
 D_{xy}^{(0)}
 =\delta_{xy}\left(n_{x_\alpha}+n_{x_\beta}\right).
 \label{eq:spdm-reference-density}
\end{equation}
Because the EOM1 state one-particle density matrix is not necessarily
Hermitian, its exact orbital representation is generally
biorthogonal. \cite{10.1063/1.464746} Assuming that $D^{(\lambda)}$ is
diagonalizable, the coefficients of the right and left orbitals are
defined by 
\begin{align}
 \sum_y
 D_{yx}^{(\lambda)}
 R_{yk}^{(\lambda)}
 &=
 n_k^{(\lambda)}
 R_{xk}^{(\lambda)},
 \\
 \sum_x
 \left[L_{xk}^{(\lambda)}\right]^*
 D_{yx}^{(\lambda)}
 &=
 n_k^{(\lambda)}
 \left[L_{yk}^{(\lambda)}\right]^* ,
\end{align}
with the biorthonormality condition
\begin{equation}
 \sum_x
 \left[L_{xk}^{(\lambda)}\right]^*
 R_{xl}^{(\lambda)}
 =
 \delta_{kl}.
\end{equation}

The corresponding right and left natural orbitals are
\begin{align}
 \psi_k^{R,(\lambda)}(\mathbf r)
 &=
 \sum_x
 \phi_x(\mathbf r)
 R_{xk}^{(\lambda)},
 \\
 \psi_k^{L,(\lambda)}(\mathbf r)
 &=
 \sum_x
 \phi_x(\mathbf r)
 L_{xk}^{(\lambda)} .
\end{align}
They satisfy
\begin{equation}
 \int
 \left[
 \psi_k^{L,(\lambda)}(\mathbf r)
 \right]^*
 \psi_l^{R,(\lambda)}(\mathbf r)
 \,d\mathbf r
 =
 \delta_{kl}.
\end{equation}

The density matrix has the biorthogonal spectral representation
\begin{equation}
 D_{xy}^{(\lambda)}
 =
 \sum_k
 n_k^{(\lambda)}
 R_{yk}^{(\lambda)}
 \left[
 L_{xk}^{(\lambda)}
 \right]^* .
\end{equation}
Consequently, the diagonal real-space density is 
\begin{align}
 \rho_\lambda(\mathbf r)
 &=
 \sum_{xy}
 \phi_x^*(\mathbf r)
 D_{xy}^{(\lambda)}
 \phi_y(\mathbf r)
 \nonumber\\
 &=
 \sum_k
 n_k^{(\lambda)}
 \psi_k^{R,(\lambda)}(\mathbf r)
 \left[
 \psi_k^{L,(\lambda)}(\mathbf r)
 \right]^* .
\end{align}

For a non-Hermitian density matrix, the spectral values
$n_k^{(\lambda)}$ are not guaranteed to be real and should therefore
be regarded as biorthogonal spectral weights rather than conventional
orbital occupation numbers.

In the Hermitian limit, the left and right eigenvectors can be chosen
such that
\begin{equation}
 L^{(\lambda)}
 =
 R^{(\lambda)}
 =
 U^{(\lambda)},
\end{equation}
and the above expression reduces to the conventional
natural-orbital expansion \cite{PhysRev.97.1474}
\begin{equation}
 \rho_\lambda(\mathbf r)
 =
 \sum_k
 n_k^{(\lambda)}
 \left|
 \psi_k^{(\lambda)}(\mathbf r)
 \right|^2 ,
\end{equation}
where
\begin{equation}
 \psi_k^{(\lambda)}(\mathbf r)
 =
 \sum_x
 \phi_x(\mathbf r)
 U_{xk}^{(\lambda)} .
\end{equation}

Real occupation numbers are required for occupation-based measures
of radical character. We therefore define the Hermitian part of the
state density matrix as
\begin{equation}
 \overline{D}_{xy}^{(\lambda)}
 =
 \frac{1}{2}
 \left[
 D_{xy}^{(\lambda)}
 +
 \left(
 D_{yx}^{(\lambda)}
 \right)^*
 \right]
\end{equation}
and determine its natural orbitals from
\begin{equation}
 \sum_y
 \overline{D}_{yx}^{(\lambda)}
 U_{yk}^{(\lambda)}
 =
 \overline{n}_k^{(\lambda)}
 U_{xk}^{(\lambda)} .
\end{equation}
The corresponding spectral representation is
\begin{equation}
 \overline{D}_{xy}^{(\lambda)}
 =
 \sum_k
 \overline{n}_k^{(\lambda)}
 U_{yk}^{(\lambda)}
 \left[
 U_{xk}^{(\lambda)}
 \right]^* .
\end{equation}

The Hermitian natural orbitals are
\begin{equation}
 \overline{\psi}_k^{(\lambda)}(\mathbf r)
 =
 \sum_x
 \phi_x(\mathbf r)
 U_{xk}^{(\lambda)} .
\end{equation}
The associated real-space density is
\begin{align}
 \overline{\rho}_\lambda(\mathbf r)
 &=
 \sum_{xy}
 \phi_x^*(\mathbf r)
 \overline{D}_{xy}^{(\lambda)}
 \phi_y(\mathbf r)
 \nonumber\\
 &=
 \sum_k
 \overline{n}_k^{(\lambda)}
 \left|
 \overline{\psi}_k^{(\lambda)}(\mathbf r)
 \right|^2
 \nonumber\\
 &=
 \operatorname{Re}
 \rho_\lambda(\mathbf r).
\end{align}
For the real-valued nonrelativistic calculations considered here,
the diagonal EOM density is real within numerical precision, and
therefore
\begin{equation}
 \overline{\rho}_\lambda(\mathbf r)
 =
 \rho_\lambda(\mathbf r).
\end{equation}

Natural occupations are expected to satisfy \cite{RevModPhys.35.668}
\begin{equation}
 0
 \le
 \overline{n}_k^{(\lambda)}
 \le
 2
\end{equation}
and obey the electron-number sum rule
\begin{equation}
 \sum_k
 \overline{n}_k^{(\lambda)}
 =
 \operatorname{Tr}
 \overline{D}^{(\lambda)}
 =
 N_{\mathrm e}.
\end{equation}

The effective numbers of unpaired electrons are evaluated from the
real occupations $\overline{n}_k^{(\lambda)}$ using the linear and
nonlinear Head–Gordon indices \cite{HEADGORDON2003508},
\begin{align}
 n_u^{(\lambda)}
 &=
 \sum_k
 \min
 \left[
 \overline{n}_k^{(\lambda)},
 2-\overline{n}_k^{(\lambda)}
 \right],
 \\
 n_{u,\mathrm{nl}}^{(\lambda)}
 &=
 \sum_k
 \left[
 \overline{n}_k^{(\lambda)}
 \right]^2
 \left[
 2-\overline{n}_k^{(\lambda)}
 \right]^2 .
\end{align}
Equivalently, the contribution of the $k$th natural orbital to the
nonlinear index is
\begin{equation}
 u_k^{(\lambda)}
 =
 \left\{
 \overline{n}_k^{(\lambda)}
 \left[
 2-\overline{n}_k^{(\lambda)}
 \right]
 \right\}^2,
\end{equation}
such that
\begin{equation}
 n_{u,\mathrm{nl}}^{(\lambda)}
 =
 \sum_k
 u_k^{(\lambda)}.
\end{equation}

The nonlinear index suppresses contributions from small fractional occupations
associated with dynamical correlation and is therefore used as the
primary measure of radical character. \cite{doi:10.1021/acs.jctc.7b01012}

\section{Numerical Results}

\subsection{Computational Details}

All calculations on benchmarking the dataset QUEST and bond dissociation were carried out using the aug-cc-pVTZ basis set. All calculations on calculating diradicals and triradicals were carried out using the cc-pVTZ basis set. Our theoretical framework is implemented in our JAX-based software package IQC. IQC (v1.0.1) can be obtained from our github release \cite{iqc_user_v101}.

\subsection{Benchmark on Dataset}

 Computational results for the systems listed in Table I of Ref. \citenum{10.1063/5.0275059} are summarized in Table~\ref{tab:lucas} of the Supporting Information (SI). We use SA-TDDFT to obtain that the total electronic spin of the target state is $S-1$, while the spin of the chosen reference state is $S$. 
The primary objective of the table is to elucidate the relationship between unphysically negative excitation energies and noncollinear functionals. All fully spin-adapted methods employing certain noncollinear exchange–correlation functionals exhibit such pseudoroots, which we attribute to the breaking of spin degeneracy induced by the noncollinear nature of these functionals and the multiple counting of spin correlation in pure xc functionals after spin-adaptation. However, in the present work, this limitation is eliminated by the use of spin-unpolarized pure exchange–correlation functionals.
\begin{table}[t]
\centering
\small
\renewcommand{\arraystretch}{1.08}
\caption{Error statistics for QUEST~1 vertical transition energies. Errors are computed as
$E_{\mathrm{calc}}-\mathrm{TBE}$.}
\label{tab:quest1_error}

\begin{tabular*}{\columnwidth}{@{\extracolsep{\fill}}llrrr@{}}
\hline\hline
Functional & Set & ME & MAE & RMSE \\
\hline

\multicolumn{5}{@{}l}{\textit{SA-TDDFT}} \\
HFLYP  & S   & -0.06 & 0.71 & 0.82 \\
       & T   & -0.10 & 0.34 & 0.51 \\
       & All & -0.07 & 0.59 & 0.73 \\

HF     & S   & -0.47 & 0.76 & 0.91 \\
       & T   & -0.16 & 0.44 & 0.64 \\
       & All & -0.37 & 0.66 & 0.83 \\

BHHLYP & S   & -0.06 & 0.31 & 0.47 \\
       & T   &  0.52 & 0.52 & 0.55 \\
       & All &  0.13 & 0.38 & 0.50 \\

M06-HF & S   & -0.66 & 0.86 & 1.08 \\
       & T   & -0.02 & 0.50 & 0.56 \\
       & All & -0.45 & 0.74 & 0.94 \\

M06-2X & S   & -0.30 & 0.46 & 0.63 \\
       & T   &  0.43 & 0.43 & 0.47 \\
       & All & -0.06 & 0.45 & 0.58 \\

\hline
\multicolumn{5}{@{}l}{\textit{X-SF-TDA}} \\
SVWN5  & S   & -0.19 & 0.71 & 0.91 \\
       & T   &  0.38 & 0.40 & 0.60 \\
       & All &  0.01 & 0.60 & 0.82 \\

BLYP   & S   & -0.52 & 0.73 & 0.91 \\
       & T   &  0.29 & 0.29 & 0.43 \\
       & All & -0.25 & 0.58 & 0.78 \\

B3LYP  & S   & -0.39 & 0.47 & 0.64 \\
       & T   &  0.12 & 0.15 & 0.22 \\
       & All & -0.22 & 0.37 & 0.54 \\

BHHLYP & S   & -0.38 & 0.56 & 0.69 \\
       & T   &  0.15 & 0.48 & 1.09 \\
       & All & -0.21 & 0.53 & 0.84 \\

HF     & S   & -0.50 & 0.83 & 0.97 \\
       & T   & -0.07 & 0.43 & 0.65 \\
       & All & -0.36 & 0.70 & 0.88 \\

\hline
\multicolumn{5}{@{}l}{\textit{SA-SF-DFT}} \\
BHHLYP & S   & -0.11 & 0.28 & 0.35 \\
       & T   & -0.07 & 0.18 & 0.19 \\
       & All & -0.10 & 0.25 & 0.31 \\

\hline
\multicolumn{5}{@{}l}{\textit{MRSF-TDDFT}} \\
BHHLYP & S   & -0.36 & 0.41 & 0.52 \\
       & T   & -0.46 & 0.47 & 0.51 \\
       & All & -0.39 & 0.43 & 0.52 \\

\hline\hline
\end{tabular*}

\vspace{0.35em}
\begin{minipage}{\columnwidth}
\footnotesize
ME, MAE, and RMSE are reported in eV. S and T denote singlet and triplet states,
respectively. Only available entries were included in the statistics; failed calculations
and unavailable entries denoted by ``--'' in the energy table were skipped.
\end{minipage}
\end{table}

The excitation energies of the molecules in the QUEST1 set \cite{Loos2018Mountaineering} were computed. The lowest triplet state was employed as the reference state for performing spin-adapted TDDFT calculations, and singlets were obtained. We compare several spin-adapted approaches, including X-SF-TDA, MRSF-TDDFT and SA-SF-DFT. X-SF-TDA uses noncollinear kernels provided by the multicollinear approach. Codes for X-SF-TDA and X-TDA are taken from the open-source package XTDDFT. \cite{xtddft_commit_02efd408} Codes for MRSF-TDDFT are taken from the open-source package OpenQP \cite{openqp_github}. The state-resolved numerical values underlying the error analysis are reported in SI Table~\ref{tab:quest1}, and the associated statistical error measures are summarized in Table~\ref{tab:quest1_error}. For X-SF-TDA, the B3LYP functional yields the highest accuracy, with an RMSE of 0.54 eV. For SA-TDDFT, the BHHLYP functional provides the best performance, with an RMSE of 0.50 eV. These two error metrics are comparable to each other and are also similar to that obtained with MRSF-TDDFT. Among all methods considered, SA-SF-DFT exhibits the highest overall accuracy, with an RMSE of 0.31.

\begin{table}[t]
\centering
\footnotesize
\renewcommand{\arraystretch}{1.02}
\caption{Error statistics for the lowest doublet and quartet vertical transition energies
of nonlinear open-shell molecules in the QUEST data set. Errors are computed as
$E_{\mathrm{calc}}-\mathrm{TBE}$.}
\label{tab:quest_rad_error}

\begin{tabular*}{\columnwidth}{@{\extracolsep{\fill}}llrrr@{}}
\hline\hline
Functional & Set & ME & MAE & RMSE \\
\hline

\multicolumn{5}{@{}l}{\textit{SA-TDDFT}} \\
HFLYP  & D   & -0.40 & 1.22 & 1.51 \\
       & Q   & -1.35 (-0.66) & 1.60 (1.19) & 2.08 (1.43) \\
       & All & -0.85 & 1.40 & 1.80 \\

HF     & D   & -0.42 & 1.00 & 1.22 \\
       & Q   & -1.33 & 1.54 & 2.02 \\
       & All & -0.85 & 1.26 & 1.65 \\

BHHLYP & D   & -0.41 & 0.41 & 0.49 \\
       & Q   &  0.06 (-0.48) & 0.27 (0.57) & 0.34 (0.71) \\
       & All & -0.19 & 0.35 & 0.42 \\

M06-HF & D   & -1.03 & 1.23 & 1.35 \\
       & Q   & -1.71 (-0.57) & 2.05 (0.58) & 2.24 (0.73) \\
       & All & -1.36 & 1.62 & 1.83 \\

M06-2X & D   & -0.50 & 0.64 & 0.71 \\
       & Q   & -0.53 (-0.28) & 0.56 (0.32) & 0.63 (0.40) \\
       & All & -0.51 & 0.60 & 0.67 \\

\hline
\multicolumn{5}{@{}l}{\textit{SA-TDA}} \\
HFLYP  & D   & -0.21 & 0.33 & 0.44 \\
       & Q   & -0.34 (0.04) & 0.69 (0.62) & 0.78 (0.68) \\
       & All & -0.27 & 0.50 & 0.62 \\

HF     & D   & -0.23 & 0.31 & 0.42 \\
       & Q   & -0.34 & 0.67 & 0.76 \\
       & All & -0.28 & 0.48 & 0.60 \\

BHHLYP & D   & -0.24 & 0.24 & 0.31 \\
       & Q   &  0.21 (-0.16) & 0.27 (0.30) & 0.34 (0.37) \\
       & All & -0.03 & 0.25 & 0.33 \\

M06-HF & D   & -0.70 & 0.76 & 0.80 \\
       & Q   & -1.94 (-0.36) & 1.94 (0.44) & 2.08 (0.54) \\
       & All & -1.29 & 1.32 & 1.54 \\

M06-2X & D   & -0.36 & 0.36 & 0.38 \\
       & Q   & -0.32 (-0.15) & 0.38 (0.23) & 0.41 (0.27) \\
       & All & -0.34 & 0.37 & 0.40 \\

\hline
\multicolumn{5}{@{}l}{\textit{X-TDA}} \\
SVWN5  & D   &  0.04 & 0.37 & 0.48 \\
       & Q   & -0.32 & 0.39 & 0.51 \\
       & All & -0.13 & 0.38 & 0.49 \\

BLYP   & D   &  0.24 & 0.45 & 0.58 \\
       & Q   & -0.35 & 0.37 & 0.49 \\
       & All & -0.04 & 0.41 & 0.54 \\

BHHLYP & D   &  0.22 & 0.24 & 0.35 \\
       & Q   & -0.16 & 0.30 & 0.37 \\
       & All &  0.04 & 0.27 & 0.36 \\

B3LYP  & D   &  0.24 & 0.31 & 0.40 \\
       & Q   & -0.24 & 0.25 & 0.31 \\
       & All &  0.01 & 0.28 & 0.36 \\

HF     & D   &  0.23 & 0.47 & 0.67 \\
       & Q   & -0.34 & 0.67 & 0.76 \\
       & All & -0.04 & 0.57 & 0.72 \\

\hline\hline
\end{tabular*}

\vspace{0.35em}
\begin{minipage}{\columnwidth}
\footnotesize
ME, MAE, and RMSE are reported in eV. D and Q denote doublet and quartet
states, respectively. Parenthetical statistics for SA-TDDFT and SA-TDA quartet
rows were computed from the parenthetical quartet values (noncollinear
functionals are applied) in Supporting Information.
\end{minipage}
\end{table}

We subsequently investigated the excitation energies of open‐shell molecules in the QUEST dataset, omitting linear species because of self-consistent field (SCF) convergence problems associated with spatial orbital degeneracies. The state-resolved numerical excitation energies are compiled in SI Table~\ref{tab:quest_rad}, and the corresponding statistical error metrics are reported in Table~\ref{tab:quest_rad_error}. The lowest-energy doublet state was adopted as the reference state, and both $S_{\mathrm{f}} = S_{\mathrm{i}}$ and $S_{\mathrm{f}} = S_{\mathrm{i}} + 1$ spin manifolds were examined. Because SA-TDA is formally equivalent to the conventional spin–flip-up TDA formulation when spin-unpolarized functionals are employed, we additionally tested noncollinear exchange–correlation functionals for the $S_{\mathrm{f}} = S_{\mathrm{i}} + 1$ excitations. Note that the spin-degeneracy condition in eq. \ref{eq:sdc} is not satisfied because noncollinear exchange–correlation functionals are employed. Consequently, the working equations cannot be cast into a single unified formalism (cf. eqs. 95–98 in ref. \citenum{10.1063/1.3463799}). Considering only the non-parenthetical entries, the most accurate functional for SA-TDDFT is BHHLYP, yielding mean absolute errors (MAEs) of 0.41 eV for doublets, 0.27 eV for quartets, and 0.35 eV overall. For SA-TDA, BHHLYP again provides the best performance, with reduced MAEs of 0.24 eV for doublets, 0.27 eV for quartets, and 0.25 eV overall, rendering it the most accurate method–functional combination in this comparison. For X-TDA, BHHLYP is likewise the best overall functional, with an overall MAE/root-mean-square error (RMSE) of 0.27/0.36 eV, although B3LYP affords the smallest quartet MAE of 0.25 eV and exhibits nearly equivalent overall accuracy. Consequently, among the three approaches, SA-TDA/BHHLYP delivers the most accurate and balanced description of the excitation energies, whereas X-TDA/BHHLYP and X-TDA/B3LYP perform comparably closely.

\subsection{Bond Dissociation}

\begin{figure}
    \centering
    \includegraphics[width=1\linewidth]{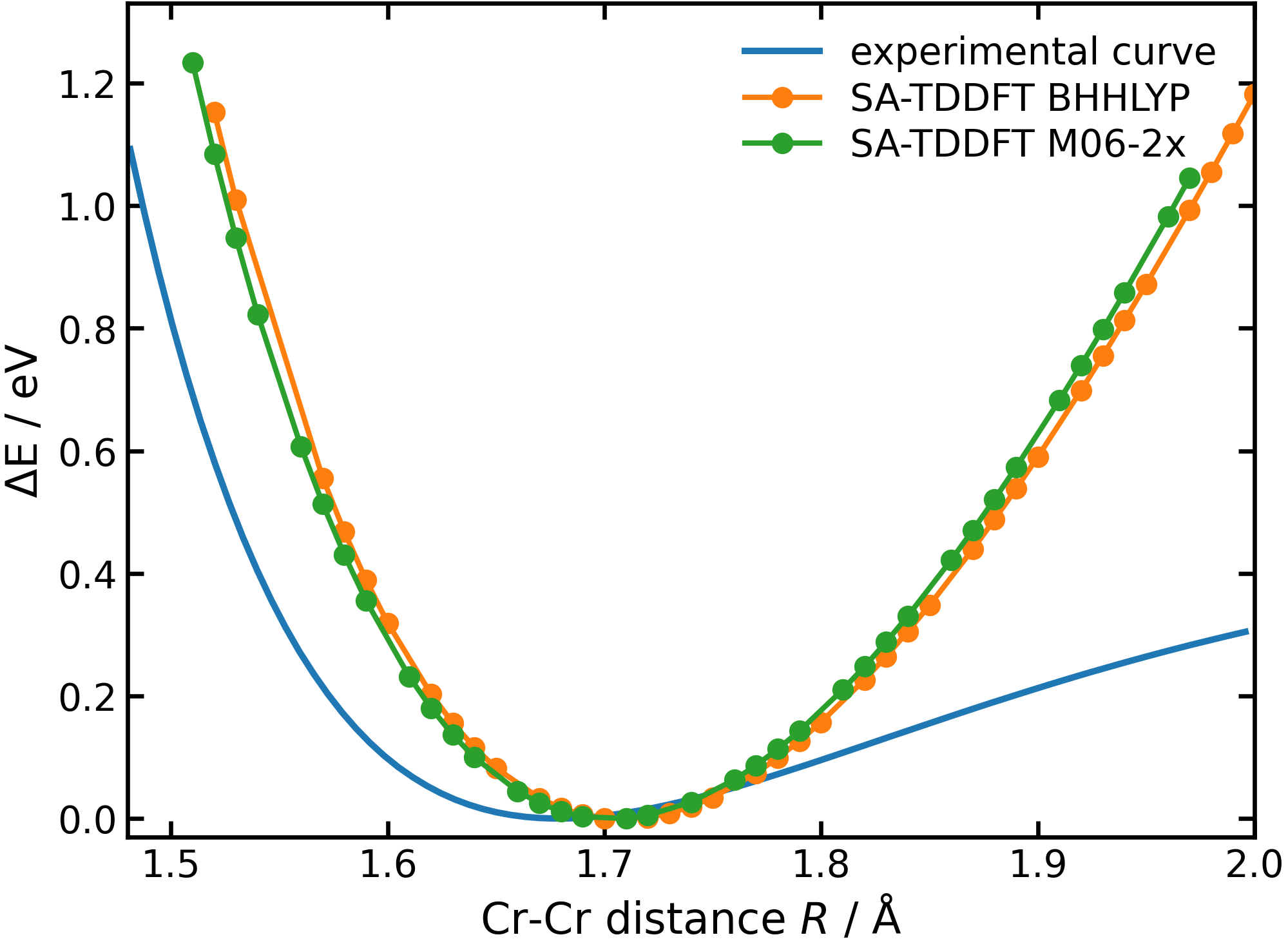}
    \caption{Near-equilibrium potential-energy curves of $\mathrm{Cr}_2$ shown as relative energies $\Delta E(R) = E(R) - E_{\mathrm{min}}$.}
    \label{fig:cr2}
\end{figure}

A central challenge for quantum-chemical methods is accurately describing strong correlation; the bond dissociation of $\mathrm{Cr}_2$ is a canonical example. \cite{doi:10.1021/jacs.2c06357} Here we take the triplet state as the reference and perform $S_\mathrm{f}=S_{\mathrm{i}}-1$ spin-adapted TDDFT calculations to recover the energy of the singlet ground state. Spin-free exact-two-component (X2C) within the one-electron approximation \cite{10.1063/1.2137315,10.1063/1.1436462,10.1063/1.4758987} is applied for special relativity. For each functional, we shift the potential-energy curve so that its minimum is set to zero; the resulting relative energies are listed in SI Table~\ref{tab:cr2} and plotted in Fig.~\ref{fig:cr2}. The experimental curve is taken from Ref.~\citenum{doi:10.1021/jacs.2c06357} (a fitted potential), which is based on the original measurements of Casey and Leopold. \cite{doi:10.1021/j100106a005} Both SA-TDDFT potential energy curves reproduce the position of the minimum (1.7~\AA), in close agreement with the experimental equilibrium bond length (also 1.7~\AA), but they remain systematically steeper than the experimental curve at geometries displaced from this minimum.

\begin{figure}[t]
\centering

\includegraphics[width=\columnwidth]{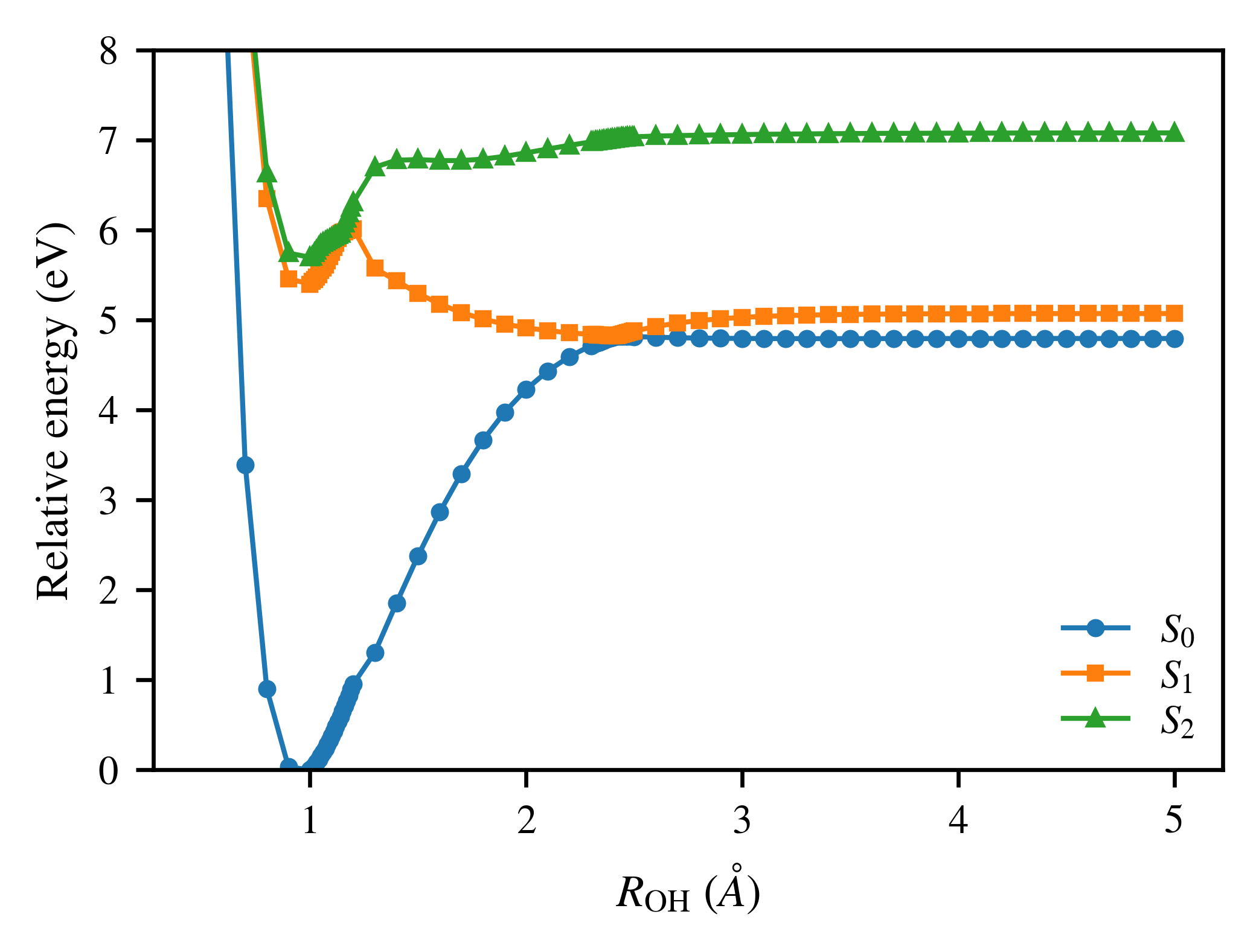}

\vspace{0.6em}

\begin{minipage}{0.49\columnwidth}
    \centering
    \includegraphics[width=\linewidth]{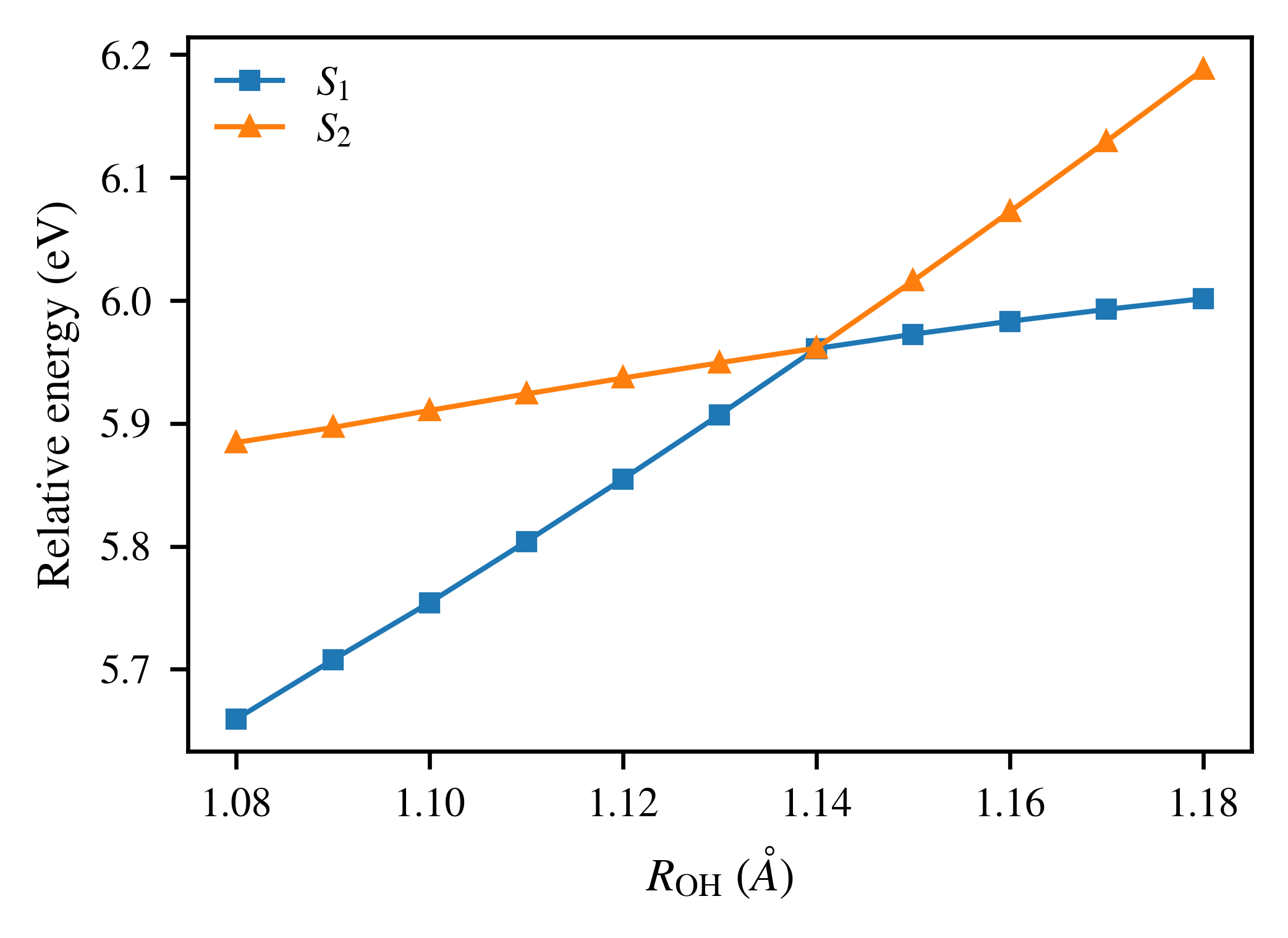}
\end{minipage}
\hfill
\begin{minipage}{0.49\columnwidth}
    \centering
    \includegraphics[width=\linewidth]{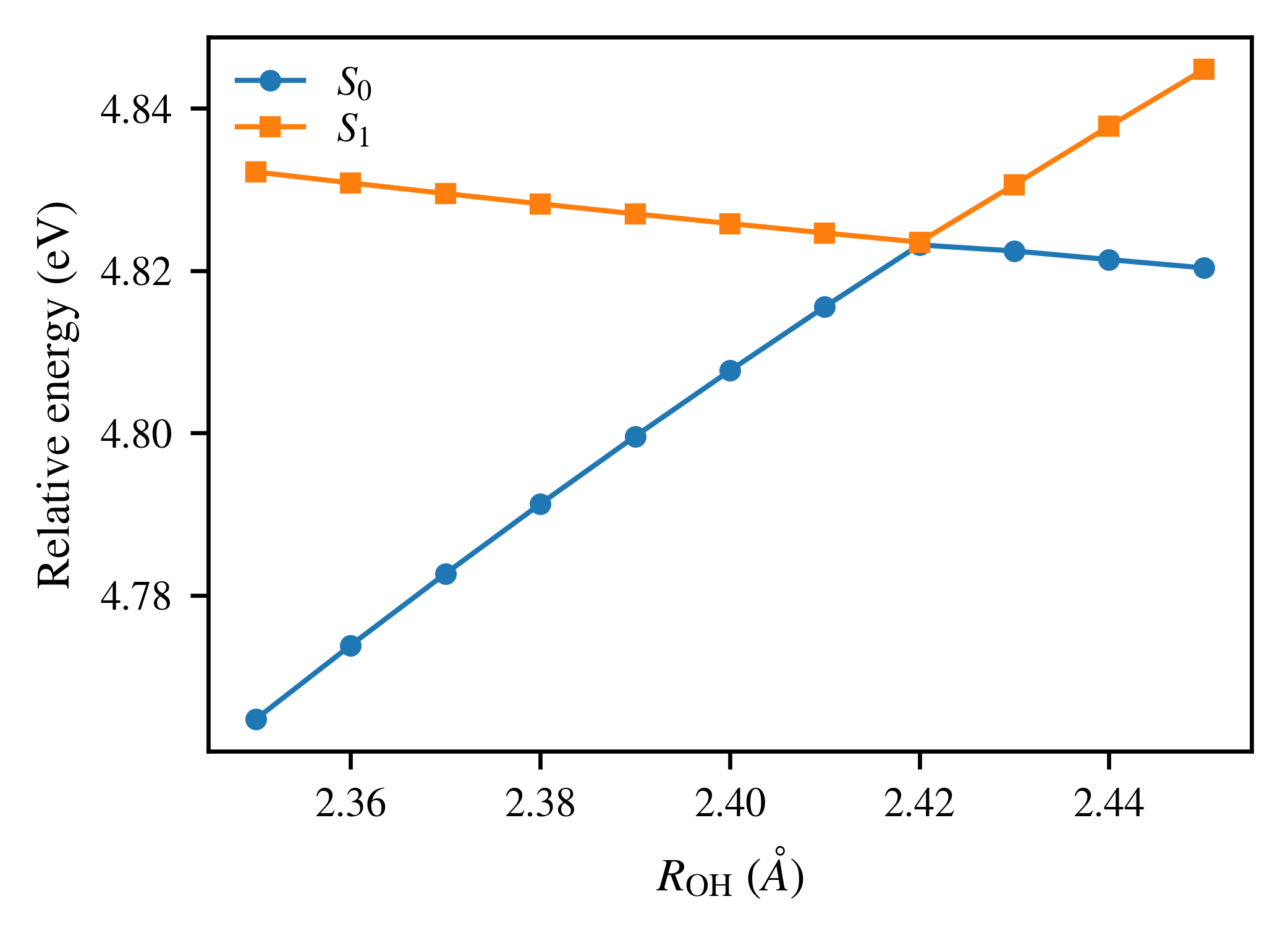}
\end{minipage}

\caption{
Relative energies of the three lowest singlet states of phenol along the O--H dissociation coordinate, computed with SA-TDDFT/BHHLYP.
Top: full scan of $S_0$, $S_1$, and $S_2$ as functions of $R_{\mathrm{OH}}$.
Bottom left: enlarged view of the $S_1/S_2$ near-degeneracy region around $R_{\mathrm{OH}}\approx 1.14$~\AA.
Bottom right: enlarged view of the $S_0/S_1$ near-degeneracy region around $R_{\mathrm{OH}}\approx 2.42$~\AA.
}
\label{fig:phenol_scan}
\end{figure}

Another challenging and practically relevant test case in the field of nonadiabatic dynamics is the O–H bond dissociation in phenol \cite{C4SC01967A,doi:10.1021/ja509016a}. We employ the triplet state as the reference and perform spin-adapted TDDFT calculations with $S_\mathrm{f}=S_{\mathrm{i}}-1$ to obtain the energies of the three lowest singlet states. The geometry is taken from ref.~\citenum{C4SC01967A} (cf. S-3 of their supporting information). The values plotted in Fig.~\ref{fig:phenol_scan} are tabulated in SI Table~\ref{tab:phenol}. The resulting potential energy profiles closely reproduce those reported in Figure 2 of ref.~\citenum{C4SC01967A}. The one-dimensional scan along the O–H dissociation coordinate reveals two pronounced near-degeneracies at 1.14~\AA~ and 2.42~\AA, in good agreement with the previously reported conical intersection geometries at 1.316~\AA~ and 2.231~\AA~ in ref.~\citenum{C4SC01967A}.

\subsection{Diradicals and Triradicals}
Diradical and triradical systems constitute prototypical cases in which spin-flip TDDFT is known to outperform conventional spin-conserving TDDFT. In this subsection, we assess the performance of SA-TDDFT/SA-TDA for such challenging open-shell species. To this end, we investigate the set of systems reported in ref. \citenum{doi:10.1021/acs.jctc.7b01012} and compare the resulting results with previously published spin-flip TDDFT data as well as with benchmark values obtained from high-level wavefunction-based methods presented in that reference. Geometries are taken from that reference and are also provided in the raw input files of our Supporting Information. In this subsection, we choose triplet or quartet as the reference state and then conduct SA-TDA $S_\mathrm{f}=S_\mathrm{i}-1$ calculations.

\begin{figure*}[!t]
\centering
\begin{minipage}[t]{0.49\textwidth}
    \centering
    \includegraphics[width=\linewidth]{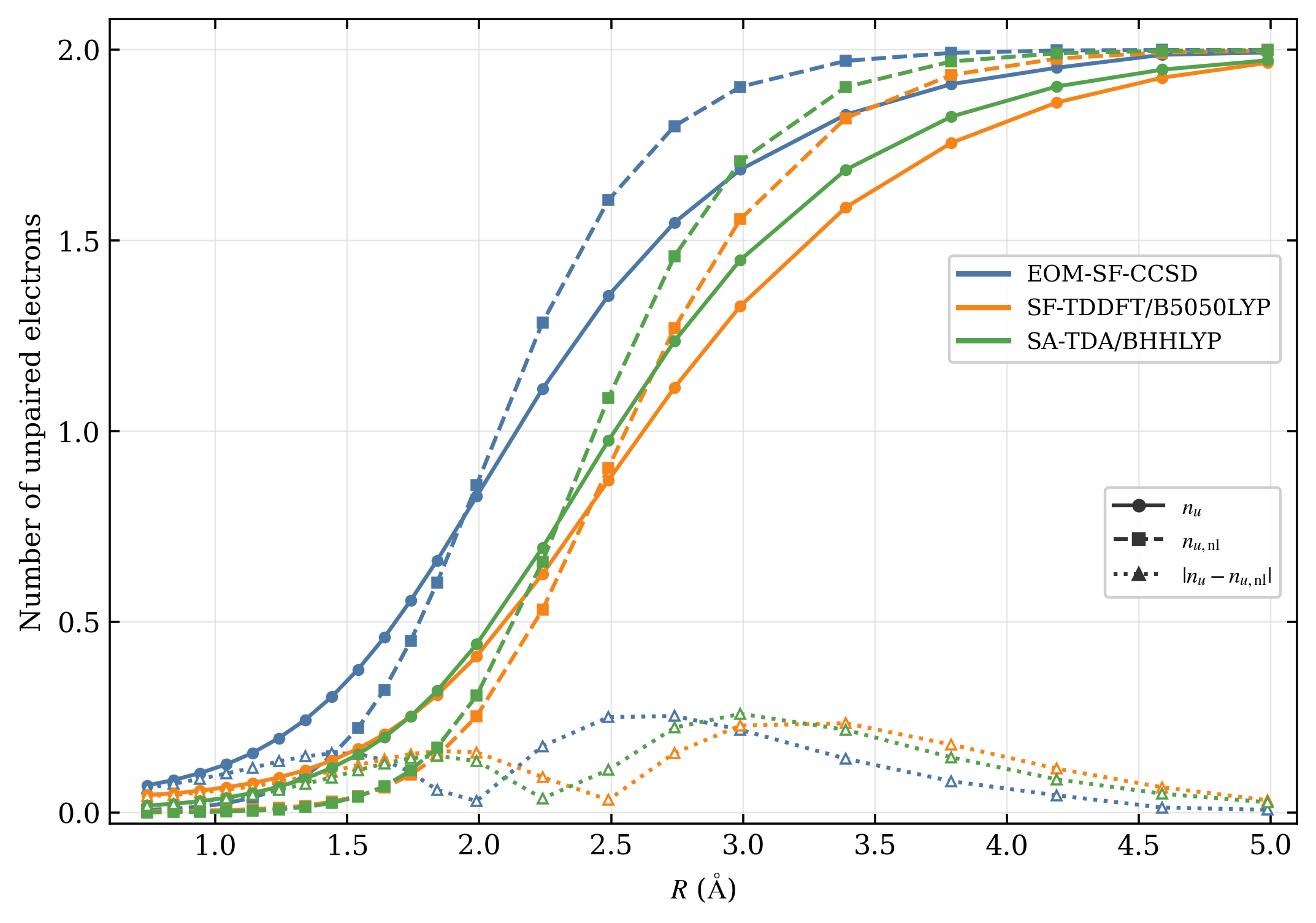}
    \\[-0.5ex]{\scriptsize (a)}
\end{minipage}
\hfill
\begin{minipage}[t]{0.49\textwidth}
    \centering
    \includegraphics[width=\linewidth]{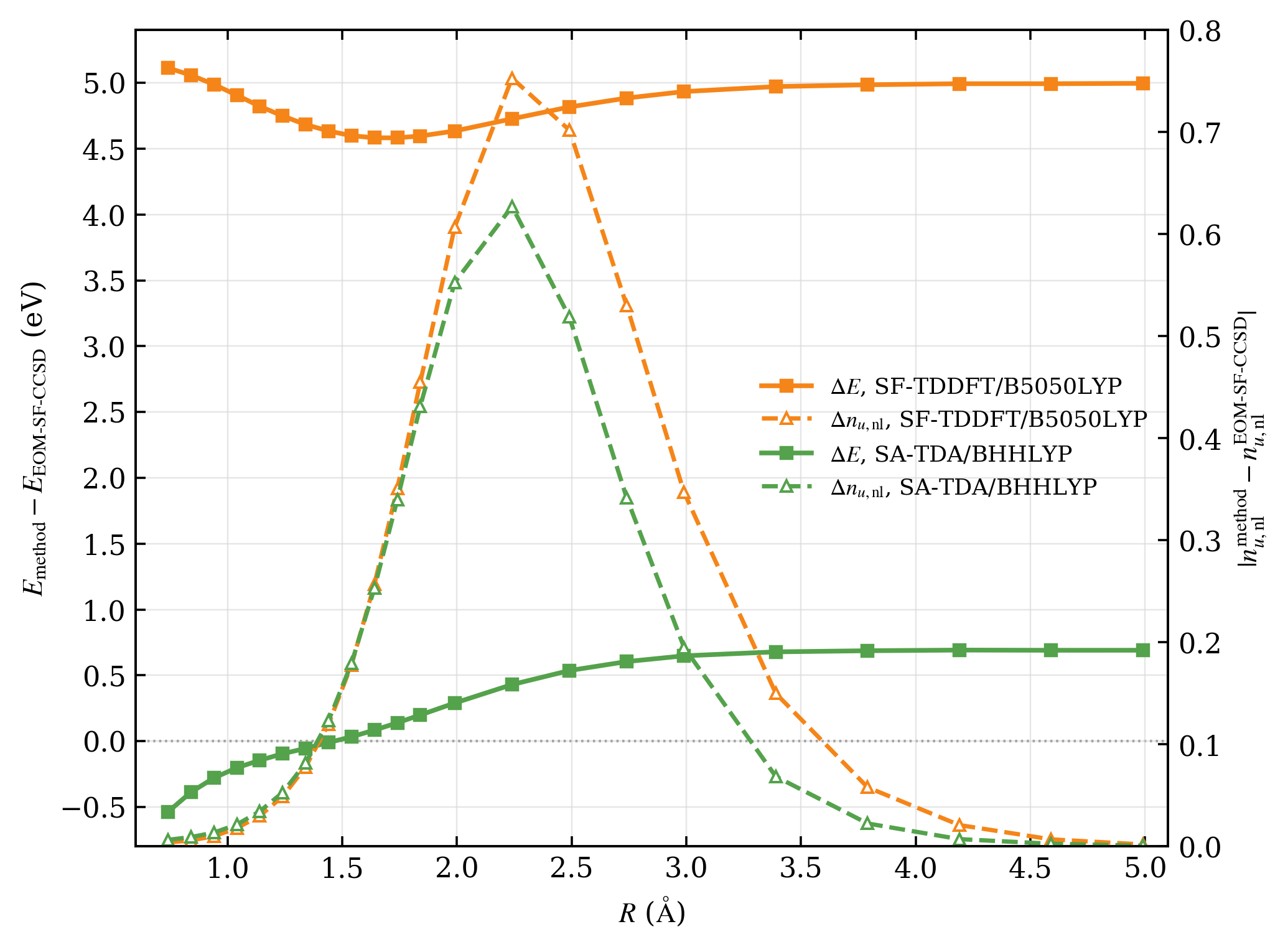}
    \\[-0.5ex]{\scriptsize (b)}
\end{minipage}

\caption{Natural-orbital diagnostics for the lowest singlet state of H$_2$ along the H--H bond-stretching coordinate. (a) Linear and nonlinear Head--Gordon indices, $n_u$ and $n_{u,\mathrm{nl}}$, together with their absolute difference, calculated using EOM--SF--CCSD, SF--TDDFT/B5050LYP, and SA--TDA/BHHLYP. (b) Deviations of SF--TDDFT/B5050LYP and SA--TDA/BHHLYP from EOM--SF--CCSD, defined as $\Delta E^M=E_{S_0}^M-E_{S_0}^{\mathrm{EOM\text{-}SF\text{-}CCSD}}$ and $\Delta n_{u,\mathrm{nl}}^M=\left|n_{u,\mathrm{nl}}^M-n_{u,\mathrm{nl}}^{\mathrm{EOM\text{-}SF\text{-}CCSD}}\right|$. The total energies are compared directly without shifting or re-zeroing the potential-energy curves. Energy deviations are in eV.}
\label{fig:h2}
\end{figure*}

We start by calculating the bond dissociation of $\mathrm{H}_2$. Figure~\ref{fig:h2} compares the density-based radical-character diagnostics and total-energy deviations along the H$_2$ dissociation coordinate. The values plotted in Fig.~\ref{fig:h2} are tabulated in SI Table~\ref{tab:h2_natural_orbitals}. Consistent with the analysis of Orms \textit{et al.},\cite{doi:10.1021/acs.jctc.7b01012} the linear index $n_u$ is larger than the nonlinear index $n_{u,\mathrm{nl}}$ in the weakly correlated region, whereas this ordering is reversed as the bond is stretched. Both indices ultimately approach two at the dissociation limit, corresponding to two effectively unpaired electrons. EOM--SF--CCSD predicts an earlier onset of diradical character than either density-functional method. Nevertheless, SA--TDA/BHHLYP generally lies between EOM--SF--CCSD and SF--TDDFT/B5050LYP in the intermediate bond-breaking region and shifts the crossing of $n_u$ and $n_{u,\mathrm{nl}}$ from approximately 2.42~\AA{} for SF--TDDFT/B5050LYP to 2.30~\AA{}, closer to the EOM--SF--CCSD value of 1.94~\AA{}. Relative to EOM--SF--CCSD, SA--TDA/BHHLYP reduces the mean absolute errors in $n_u$ and $n_{u,\mathrm{nl}}$ from 0.206 and 0.215 to 0.184 and 0.177, respectively. The maximum deviation in the nonlinear index is also reduced from 0.753 for SF--TDDFT/B5050LYP to 0.627 for SA--TDA/BHHLYP, with both maxima occurring near $R=2.24$~\AA{}. The unshifted total-energy comparison shows a large, nearly constant offset of approximately 4.6--5.1~eV for SF--TDDFT/B5050LYP, whereas the SA--TDA/BHHLYP deviation is much smaller, ranging from $-0.54$ to $0.69$~eV. Thus, SA--TDA/BHHLYP provides improved absolute agreement and natural-orbital diagnostics for this example.

\begin{table*}[!t]
\centering
\scriptsize
\setlength{\tabcolsep}{3pt}
\renewcommand{\arraystretch}{1.03}
\caption{
Vertical energy splittings and effective numbers of
unpaired electrons for the organic diradicals and triradicals. The energy
splitting is defined as $\Delta E=E(S_\mathrm{f})-E(S_\mathrm{i})$, and the
Head--Gordon indices refer to the target state. Energy splittings are in
eV. 
}
\label{tab:radical_natural_orbitals}
\begin{ruledtabular}
\begin{tabular}{l rcc rcc rcc}
& \multicolumn{3}{c}{EOM--SF--CCSD}
& \multicolumn{3}{c}{SF--TDDFT/B5050LYP}
& \multicolumn{3}{c}{SA--TDA/BHHLYP} \\
System & $\Delta E$ & $n_u$ & $n_{u,\mathrm{nl}}$
& $\Delta E$ & $n_u$ & $n_{u,\mathrm{nl}}$
& $\Delta E$ & $n_u$ & $n_{u,\mathrm{nl}}$ \\
\multicolumn{10}{c}{Diradicals: singlet/triplet} \\
CH$_2$ &  0.94 & 0.454 & 0.253 &  0.30 & 0.217 & 0.079 &  0.13 & 0.211 & 0.078 \\
A       & -2.46 & 0.414 & 0.156 & -2.66 & 0.171 & 0.044 & -2.79 & 0.154 & 0.039 \\
B       & -1.90 & 0.522 & 0.258 & -2.04 & 0.302 & 0.116 & -2.20 & 0.241 & 0.096 \\
C       & -0.24 & 1.337 & 1.452 & -0.23 & 1.129 & 1.286 & -0.34 & 1.068 & 1.214 \\
D       &  0.24 & 2.135 & 2.003 &  0.16 & 2.118 & 2.010 & -0.04 & 2.011 & 2.000 \\
\multicolumn{10}{c}{Triradicals: doublet/quartet} \\
F       & -2.21 & 1.699 & 1.278 & -2.44 & 1.300 & 1.095 & -3.51 & 1.061 & 1.004 \\
G       & -2.00 & 1.516 & 1.164 & -2.82 & 1.200 & 1.056 & -3.01 & 1.160 & 1.044 \\
H       &  0.41 & 3.211 & 3.015 &  0.23 & 3.218 & 3.033 &  0.06 & 3.022 & 3.000 \\
I       & -0.14 & 3.211 & 3.014 & -0.14 & 3.129 & 3.012 & -0.15 & 3.022 & 3.000 \\
\hline
\multicolumn{4}{l}{ME}   & -0.253 & -0.191 & -0.096 & -0.499 & -0.283 & -0.124 \\
\multicolumn{4}{l}{MAE}  &  0.256 &  0.192 &  0.101 &  0.499 &  0.283 &  0.124 \\
\multicolumn{4}{l}{RMSE} &  0.370 &  0.229 &  0.123 &  0.649 &  0.316 &  0.155 \\
\end{tabular}
\end{ruledtabular}
\end{table*}

Table~\ref{tab:radical_natural_orbitals} presents the vertical energy splittings and effective numbers of unpaired electrons for nine representative organic radicals spanning several open-shell bonding patterns. The diradical subset comprises the same-center diradical methylene (CH$_2$), the $\sigma\sigma$-type ortho-, meta-, and para-benzynes (A--C), and the $\sigma\pi$-type 1-(2-dehydroisopropyl)-4-dehydrobenzyne (D). The triradical subset includes the all-$\sigma$ 1,3,5- and 1,2,4-tridehydrobenzynes (F and G), together with the $\sigma\pi$-type 2- and 5-dehydro-meta-xylylenes (H and I). Structures can be found in figure 2 of ref. \citenum{doi:10.1021/acs.jctc.7b01012}. With $\Delta E=E(S_\mathrm{f})-E(S_\mathrm{i})$, SA--TDA/BHHLYP reproduces the EOM--SF--CCSD spin ordering for eight of the nine systems. The exception is D, for which the EOM--SF--CCSD splitting of $+0.24$ eV is predicted to be $-0.041$ eV. The SA--TDA energy errors are negative for all nine systems, giving ME, MAE, and RMSE values of $-0.499$, $0.499$, and $0.649$ eV, respectively, and indicating a systematic over-stabilization of the $S_\mathrm{f}$ states relative to their $S_\mathrm{i}$ counterparts. The largest deviations occur for F, G, and CH$_2$, with errors of $-1.305$, $-1.007$, and $-0.813$ eV, respectively. In comparison, SF--TDDFT/B5050LYP yields smaller energy-gap errors, with an MAE of $0.256$ eV and an RMSE of $0.370$ eV. 

Despite this systematic energetic bias, SA--TDA preserves the qualitative classification of the multiradical states. The nonlinear Head--Gordon index identifies CH$_2$, A, and B as weak diradicals, with $n_{u,\mathrm{nl}}=0.078$, $0.039$, and $0.096$, respectively; C exhibits substantial but intermediate diradical character ($n_{u,\mathrm{nl}}=1.214$), whereas D approaches the ideal diradical limit ($n_{u,\mathrm{nl}}=2.000$). For the triradicals, the values of $1.004$ and $1.044$ obtained for F and G are characteristic of closed-shell-like doublets dominated by one unpaired electron, while the values of approximately $3.000$ for H and I correctly describe open-shell triradical doublets. Relative to EOM--SF--CCSD, the MAE/RMSE of the linear index $n_u$ are $0.283/0.316$, whereas those of $n_{u,\mathrm{nl}}$ decrease to $0.124/0.155$. SA-TDA closely aligns with SF-TDDFT in this situation.

\FloatBarrier
\section{Conclusion}
In conclusion, we have developed a rigorously defined, spin-adapted time-dependent density functional theory (TDDFT) in this work. In practice, we present promising applications in strongly correlated systems, conical intersections, and diradicals/triradicals. Several promising directions for future research can be identified as follows.
\begin{enumerate}
    \item The development of more accurate hybrid exchange–correlation functionals, in which the pure exchange–correlation contribution remains spin-unpolarized.
    \item The integration of the present spin-adapted TDDFT framework with spin–orbit coupling (SOC). \cite{Li01122013}
    \item The derivation and implementation of analytic energy gradients within the spin-adapted TDDFT formalism. \cite{10.1063/5.0025428} Such gradients would enable the subsequent development of nonadiabatic coupling schemes. \cite{doi:10.1021/acs.accounts.1c00312}
\end{enumerate}

\section{Acknowledgement}
We gratefully acknowledge the helpful discussions provided by Yunlong Xiao. This work was supported by the Smart Grid–National Science and Technology Major Project (Grant No. 2025ZD0808601), whose contribution is sincerely appreciated.

\section{Data and Software Availability}
The data underpinning this study are fully reported within the published article. Our package IQC (v1.0.1) can be obtained from our github release \cite{iqc_user_v101}. Raw input files and output files are presented in our Supporting Information.                                                

\bibliography{main}
\section*{Appendix: \\
Translating Spin-Adapted RPA to Spin-Adapted TDDFT}

\setcounter{secnumdepth}{2}
\renewcommand\thesubsection{\Roman{subsection}}
\setcounter{subsection}{0}

\author{Xiaoyu Zhang}
\email{zhangxiaoyu@stu.pku.edu.cn}

\affiliation{College of Chemistry and Molecular Engineering, Peking University, Beijing 100871, the People's Republic of China}

\newif\ifsuppfirstsubsection
\suppfirstsubsectiontrue
\AddToHook{cmd/subsection/before}[supp-subsection-clearpage]{%
  \ifsuppfirstsubsection
    \global\suppfirstsubsectionfalse
  \else
    \clearpage
  \fi
}

\subsection{Matrix Elements for 1RDM}

\begin{table}[p]
\centering
\caption{Matrix elements of $[M_e^{xy}]_+$, $[M_e^{xy}]_0$, and
$[M_e^{xy}]_-$ without OO transitions, based on EOM1.}
\label{tab:spdm-minus1}
\begin{ruledtabular}
\begin{tabular}{@{}l@{}}
\textit{CV--CV}\\
$\begin{aligned}[t]
\bigl[[M_e^{xy}]_{+}\bigr]_{ai,bj}
 &=\delta_{ij}\delta_{ax}\delta_{by}
 -\delta_{ab}\delta_{jx}\delta_{iy}\\
\bigl[[M_e^{xy}]_{0}\bigr]_{ai,bj}
 &=\delta_{ij}\delta_{ax}\delta_{by}-\delta_{ab}\delta_{jx}\delta_{iy}\\
\bigl[[M_e^{xy}]_{-}\bigr]_{ai,bj}
 &=\delta_{ij}\delta_{ax}\delta_{by}-\delta_{ab}\delta_{jx}\delta_{iy}
\end{aligned}$\\

\textit{VC--VC}\\
$\begin{aligned}[t]
\bigl[[M_e^{xy}]_{+}\bigr]_{ia,jb}&=0\\
\bigl[[M_e^{xy}]_{0}\bigr]_{ia,jb}&=0\\
\bigl[[M_e^{xy}]_{-}\bigr]_{ia,jb}&=0
\end{aligned}$\\

\textit{CV--CO}\\
$\begin{aligned}[t]
\bigl[[M_e^{xy}]_{+}\bigr]_{ai,vj}&=\delta_{ij}\delta_{ax}\delta_{vy}\\
\bigl[[M_e^{xy}]_{0}\bigr]_{ai,vj}&=\delta_{ij}\delta_{ax}\delta_{vy}\\
\bigl[[M_e^{xy}]_{-}\bigr]_{ai,vj}&=\delta_{ij}\delta_{ax}\delta_{vy}
\end{aligned}$\\

\textit{VC--OC}\\
$\begin{aligned}[t]
\bigl[[M_e^{xy}]_{+}\bigr]_{ia,jv}&=0\\
\bigl[[M_e^{xy}]_{0}\bigr]_{ia,jv}&=0\\
\bigl[[M_e^{xy}]_{-}\bigr]_{ia,jv}&=0
\end{aligned}$\\

\textit{CO--CV}\\
$\begin{aligned}[t]
\bigl[[M_e^{xy}]_{+}\bigr]_{ui,bj}&=0\\
\bigl[[M_e^{xy}]_{0}\bigr]_{ui,bj}&=\tfrac12\delta_{ij}\delta_{ux}\delta_{by}\\
\bigl[[M_e^{xy}]_{-}\bigr]_{ui,bj}&=\delta_{ij}\delta_{ux}\delta_{by}
\end{aligned}$\\

\textit{OC--VC}\\
$\begin{aligned}[t]
\bigl[[M_e^{xy}]_{+}\bigr]_{iu,jb}&=0\\
\bigl[[M_e^{xy}]_{0}\bigr]_{iu,jb}&=0\\
\bigl[[M_e^{xy}]_{-}\bigr]_{iu,jb}&=0
\end{aligned}$\\

\textit{CV--OV}\\
$\begin{aligned}[t]
\bigl[[M_e^{xy}]_{+}\bigr]_{ai,bv}&=-\delta_{ab}\delta_{vx}\delta_{iy}\\
\bigl[[M_e^{xy}]_{0}\bigr]_{ai,bv}&=-\delta_{ab}\delta_{vx}\delta_{iy}\\
\bigl[[M_e^{xy}]_{-}\bigr]_{ai,bv}&=-\delta_{ab}\delta_{vx}\delta_{iy}
\end{aligned}$\\

\textit{VC--VO}\\
$\begin{aligned}[t]
\bigl[[M_e^{xy}]_{+}\bigr]_{ia,vb}&=0\\
\bigl[[M_e^{xy}]_{0}\bigr]_{ia,vb}&=0\\
\bigl[[M_e^{xy}]_{-}\bigr]_{ia,vb}&=0
\end{aligned}$\\

\textit{OV--CV}\\
$\begin{aligned}[t]
\bigl[[M_e^{xy}]_{+}\bigr]_{au,bj}&=0\\
\bigl[[M_e^{xy}]_{0}\bigr]_{au,bj}&=-\tfrac12\delta_{ab}\delta_{jx}\delta_{uy}\\
\bigl[[M_e^{xy}]_{-}\bigr]_{au,bj}&=-\delta_{ab}\delta_{jx}\delta_{uy}
\end{aligned}$\\

\end{tabular}
\end{ruledtabular}
\end{table}

\begin{table}[p]
\centering
\caption{Continuation of Table~\ref{tab:spdm-minus1}.}
\label{tab:spdm-minus1-continued}
\begin{ruledtabular}
\begin{tabular}{@{}l@{}}

\textit{VO--VC}\\
$\begin{aligned}[t]
\bigl[[M_e^{xy}]_{+}\bigr]_{ua,jb}&=0\\
\bigl[[M_e^{xy}]_{0}\bigr]_{ua,jb}&=0\\
\bigl[[M_e^{xy}]_{-}\bigr]_{ua,jb}&=0
\end{aligned}$\\

\textit{CO--CO}\\
$\begin{aligned}[t]
\bigl[[M_e^{xy}]_{+}\bigr]_{ui,vj}&=0\\
\bigl[[M_e^{xy}]_{0}\bigr]_{ui,vj}
 &=\tfrac12\Bigl(
 \delta_{ij}\delta_{ux}\delta_{vy}-\delta_{uv}\delta_{jx}\delta_{iy}\Bigr)\\
\bigl[[M_e^{xy}]_{-}\bigr]_{ui,vj}
 &=\delta_{ij}\delta_{ux}\delta_{vy}-\delta_{uv}\delta_{jx}\delta_{iy}
\end{aligned}$\\

\textit{OC--OC}\\
$\begin{aligned}[t]
\bigl[[M_e^{xy}]_{+}\bigr]_{iu,jv}&=0\\
\bigl[[M_e^{xy}]_{0}\bigr]_{iu,jv}&=0\\
\bigl[[M_e^{xy}]_{-}\bigr]_{iu,jv}&=0
\end{aligned}$\\

\textit{CO--OV}\\
$\begin{aligned}[t]
\bigl[[M_e^{xy}]_{+}\bigr]_{ui,bv}&=0\\
\bigl[[M_e^{xy}]_{0}\bigr]_{ui,bv}&=0\\
\bigl[[M_e^{xy}]_{-}\bigr]_{ui,bv}&=0
\end{aligned}$\\

\textit{OC--VO}\\
$\begin{aligned}[t]
\bigl[[M_e^{xy}]_{+}\bigr]_{iu,vb}&=0\\
\bigl[[M_e^{xy}]_{0}\bigr]_{iu,vb}&=0\\
\bigl[[M_e^{xy}]_{-}\bigr]_{iu,vb}&=0
\end{aligned}$\\

\textit{OV--CO}\\
$\begin{aligned}[t]
\bigl[[M_e^{xy}]_{+}\bigr]_{au,vj}&=0\\
\bigl[[M_e^{xy}]_{0}\bigr]_{au,vj}&=0\\
\bigl[[M_e^{xy}]_{-}\bigr]_{au,vj}&=0
\end{aligned}$\\

\textit{VO--OC}\\
$\begin{aligned}[t]
\bigl[[M_e^{xy}]_{+}\bigr]_{ua,jv}&=0\\
\bigl[[M_e^{xy}]_{0}\bigr]_{ua,jv}&=0\\
\bigl[[M_e^{xy}]_{-}\bigr]_{ua,jv}&=0
\end{aligned}$\\

\textit{OV--OV}\\
$\begin{aligned}[t]
\bigl[[M_e^{xy}]_{+}\bigr]_{au,bv}&=0\\
\bigl[[M_e^{xy}]_{0}\bigr]_{au,bv}
 &=\tfrac12\Bigl(
 \delta_{uv}\delta_{ax}\delta_{by}-\delta_{ab}\delta_{vx}\delta_{uy}\Bigr)\\
\bigl[[M_e^{xy}]_{-}\bigr]_{au,bv}
 &=\delta_{uv}\delta_{ax}\delta_{by}-\delta_{ab}\delta_{vx}\delta_{uy}
\end{aligned}$\\

\textit{VO--VO}\\
$\begin{aligned}[t]
\bigl[[M_e^{xy}]_{+}\bigr]_{ua,vb}&=0\\
\bigl[[M_e^{xy}]_{0}\bigr]_{ua,vb}&=0\\
\bigl[[M_e^{xy}]_{-}\bigr]_{ua,vb}&=0
\end{aligned}$
\end{tabular}
\end{ruledtabular}
\end{table}

\begin{table}[p]
\centering
\caption{Matrix elements of $[M_e^{xy}]_+$, $[M_e^{xy}]_0$, and
$[M_e^{xy}]_-$ between excitation and deexcitation blocks without OO
transitions, based on EOM1.}
\label{tab:spdm-minus2}
\begin{ruledtabular}
\begin{tabular}{@{}l@{}}
\textit{CV--VC}\\
$\begin{aligned}[t]
\bigl[[M_e^{xy}]_{+}\bigr]_{ai,jb}&=0\\
\bigl[[M_e^{xy}]_{0}\bigr]_{ai,jb}&=0\\
\bigl[[M_e^{xy}]_{-}\bigr]_{ai,jb}&=0
\end{aligned}$\\

\textit{VC--CV}\\
$\begin{aligned}[t]
\bigl[[M_e^{xy}]_{+}\bigr]_{ia,bj}&=0\\
\bigl[[M_e^{xy}]_{0}\bigr]_{ia,bj}&=0\\
\bigl[[M_e^{xy}]_{-}\bigr]_{ia,bj}&=0
\end{aligned}$\\

\textit{CV--OC}\\
$\begin{aligned}[t]
\bigl[[M_e^{xy}]_{+}\bigr]_{ai,jv}&=0\\
\bigl[[M_e^{xy}]_{0}\bigr]_{ai,jv}&=0\\
\bigl[[M_e^{xy}]_{-}\bigr]_{ai,jv}&=0
\end{aligned}$\\

\textit{VC--CO}\\
$\begin{aligned}[t]
\bigl[[M_e^{xy}]_{+}\bigr]_{ia,vj}&=0\\
\bigl[[M_e^{xy}]_{0}\bigr]_{ia,vj}&=0\\
\bigl[[M_e^{xy}]_{-}\bigr]_{ia,vj}&=0
\end{aligned}$\\

\textit{OC--CV}\\
$\begin{aligned}[t]
\bigl[[M_e^{xy}]_{+}\bigr]_{iu,bj}&=0\\
\bigl[[M_e^{xy}]_{0}\bigr]_{iu,bj}&=0\\
\bigl[[M_e^{xy}]_{-}\bigr]_{iu,bj}&=0
\end{aligned}$\\

\textit{CO--VC}\\
$\begin{aligned}[t]
\bigl[[M_e^{xy}]_{+}\bigr]_{ui,jb}&=0\\
\bigl[[M_e^{xy}]_{0}\bigr]_{ui,jb}&=0\\
\bigl[[M_e^{xy}]_{-}\bigr]_{ui,jb}&=0
\end{aligned}$\\

\textit{CO--OC}\\
$\begin{aligned}[t]
\bigl[[M_e^{xy}]_{+}\bigr]_{ui,jv}&=0\\
\bigl[[M_e^{xy}]_{0}\bigr]_{ui,jv}&=0\\
\bigl[[M_e^{xy}]_{-}\bigr]_{ui,jv}&=0
\end{aligned}$\\

\textit{OC--CO}\\
$\begin{aligned}[t]
\bigl[[M_e^{xy}]_{+}\bigr]_{iu,vj}&=0\\
\bigl[[M_e^{xy}]_{0}\bigr]_{iu,vj}&=0\\
\bigl[[M_e^{xy}]_{-}\bigr]_{iu,vj}&=0
\end{aligned}$\\

\textit{CV--VO}\\
$\begin{aligned}[t]
\bigl[[M_e^{xy}]_{+}\bigr]_{ai,vb}&=0\\
\bigl[[M_e^{xy}]_{0}\bigr]_{ai,vb}&=0\\
\bigl[[M_e^{xy}]_{-}\bigr]_{ai,vb}&=0
\end{aligned}$\\

\end{tabular}
\end{ruledtabular}
\end{table}

\begin{table}[p]
\centering
\caption{Continuation of Table~\ref{tab:spdm-minus2}.}
\label{tab:spdm-minus2-continued}
\begin{ruledtabular}
\begin{tabular}{@{}l@{}}

\textit{VC--OV}\\
$\begin{aligned}[t]
\bigl[[M_e^{xy}]_{+}\bigr]_{ia,bv}&=0\\
\bigl[[M_e^{xy}]_{0}\bigr]_{ia,bv}&=0\\
\bigl[[M_e^{xy}]_{-}\bigr]_{ia,bv}&=0
\end{aligned}$\\

\textit{VO--CV}\\
$\begin{aligned}[t]
\bigl[[M_e^{xy}]_{+}\bigr]_{ua,bj}&=0\\
\bigl[[M_e^{xy}]_{0}\bigr]_{ua,bj}&=0\\
\bigl[[M_e^{xy}]_{-}\bigr]_{ua,bj}&=0
\end{aligned}$\\

\textit{OV--VC}\\
$\begin{aligned}[t]
\bigl[[M_e^{xy}]_{+}\bigr]_{au,jb}&=0\\
\bigl[[M_e^{xy}]_{0}\bigr]_{au,jb}&=0\\
\bigl[[M_e^{xy}]_{-}\bigr]_{au,jb}&=0
\end{aligned}$\\

\textit{CO--VO}\\
$\begin{aligned}[t]
\bigl[[M_e^{xy}]_{+}\bigr]_{ui,vb}&=0\\
\bigl[[M_e^{xy}]_{0}\bigr]_{ui,vb}&=-\tfrac12\delta_{uv}\delta_{bx}\delta_{iy}\\
\bigl[[M_e^{xy}]_{-}\bigr]_{ui,vb}&=-\delta_{uv}\delta_{bx}\delta_{iy}
\end{aligned}$\\

\textit{OC--OV}\\
$\begin{aligned}[t]
\bigl[[M_e^{xy}]_{+}\bigr]_{iu,bv}&=0\\
\bigl[[M_e^{xy}]_{0}\bigr]_{iu,bv}&=0\\
\bigl[[M_e^{xy}]_{-}\bigr]_{iu,bv}&=0
\end{aligned}$\\

\textit{VO--CO}\\
$\begin{aligned}[t]
\bigl[[M_e^{xy}]_{+}\bigr]_{ua,vj}&=0\\
\bigl[[M_e^{xy}]_{0}\bigr]_{ua,vj}&=0\\
\bigl[[M_e^{xy}]_{-}\bigr]_{ua,vj}&=0
\end{aligned}$\\

\textit{OV--OC}\\
$\begin{aligned}[t]
\bigl[[M_e^{xy}]_{+}\bigr]_{au,jv}&=0\\
\bigl[[M_e^{xy}]_{0}\bigr]_{au,jv}&=\tfrac12\delta_{uv}\delta_{ax}\delta_{jy}\\
\bigl[[M_e^{xy}]_{-}\bigr]_{au,jv}&=\delta_{uv}\delta_{ax}\delta_{jy}
\end{aligned}$\\

\textit{OV--VO}\\
$\begin{aligned}[t]
\bigl[[M_e^{xy}]_{+}\bigr]_{au,vb}&=0\\
\bigl[[M_e^{xy}]_{0}\bigr]_{au,vb}&=0\\
\bigl[[M_e^{xy}]_{-}\bigr]_{au,vb}&=0
\end{aligned}$\\

\textit{VO--OV}\\
$\begin{aligned}[t]
\bigl[[M_e^{xy}]_{+}\bigr]_{ua,bv}&=0\\
\bigl[[M_e^{xy}]_{0}\bigr]_{ua,bv}&=0\\
\bigl[[M_e^{xy}]_{-}\bigr]_{ua,bv}&=0
\end{aligned}$
\end{tabular}
\end{ruledtabular}
\end{table}

\begin{table}[p]
\centering
\caption{Matrix elements of $[M_e^{xy}]_+$, $[M_e^{xy}]_0$, and
$[M_e^{xy}]_-$ involving OO transitions, based on EOM1.  OO denotes the
full ordered $(t,u)$ product space in this table, including TT.}
\scriptsize
\label{tab:spdm-minus3}
\begin{ruledtabular}
\begin{tabular}{@{}l@{}}
\textit{CV--OO}\\
$\begin{aligned}[t]
\bigl[[M_e^{xy}]_{+}\bigr]_{ai,vw}&=0\\
\bigl[[M_e^{xy}]_{0}\bigr]_{ai,vw}&=0\\
\bigl[[M_e^{xy}]_{-}\bigr]_{ai,vw}&=0
\end{aligned}$\\

\textit{VC--OO}\\
$\begin{aligned}[t]
\bigl[[M_e^{xy}]_{+}\bigr]_{ia,vw}&=0\\
\bigl[[M_e^{xy}]_{0}\bigr]_{ia,vw}&=0\\
\bigl[[M_e^{xy}]_{-}\bigr]_{ia,vw}&=0
\end{aligned}$\\

\textit{OO--CV}\\
$\begin{aligned}[t]
\bigl[[M_e^{xy}]_{+}\bigr]_{tu,bj}&=0\\
\bigl[[M_e^{xy}]_{0}\bigr]_{tu,bj}&=0\\
\bigl[[M_e^{xy}]_{-}\bigr]_{tu,bj}&=0
\end{aligned}$\\

\textit{OO--VC}\\
$\begin{aligned}[t]
\bigl[[M_e^{xy}]_{+}\bigr]_{tu,jb}&=0\\
\bigl[[M_e^{xy}]_{0}\bigr]_{tu,jb}&=0\\
\bigl[[M_e^{xy}]_{-}\bigr]_{tu,jb}&=0
\end{aligned}$\\

\textit{CO--OO}\\
$\begin{aligned}[t]
\bigl[[M_e^{xy}]_{+}\bigr]_{ui,vw}&=0\\
\bigl[[M_e^{xy}]_{0}\bigr]_{ui,vw}&=-\tfrac12\delta_{uv}\delta_{wx}\delta_{iy}\\
\bigl[[M_e^{xy}]_{-}\bigr]_{ui,vw}&=-\delta_{uv}\delta_{wx}\delta_{iy}
\end{aligned}$\\

\textit{OC--OO}\\
$\begin{aligned}[t]
\bigl[[M_e^{xy}]_{+}\bigr]_{iu,vw}&=0\\
\bigl[[M_e^{xy}]_{0}\bigr]_{iu,vw}&=0\\
\bigl[[M_e^{xy}]_{-}\bigr]_{iu,vw}&=0
\end{aligned}$\\

\textit{OO--CO}\\
$\begin{aligned}[t]
\bigl[[M_e^{xy}]_{+}\bigr]_{tu,vj}&=0\\
\bigl[[M_e^{xy}]_{0}\bigr]_{tu,vj}&=-\tfrac12\delta_{tu}\delta_{jx}\delta_{vy}\\
\bigl[[M_e^{xy}]_{-}\bigr]_{tu,vj}&=-\delta_{tv}\delta_{jx}\delta_{uy}
\end{aligned}$\\

\textit{OO--OC}\\
$\begin{aligned}[t]
\bigl[[M_e^{xy}]_{+}\bigr]_{tu,jv}&=0\\
\bigl[[M_e^{xy}]_{0}\bigr]_{tu,jv}&=\tfrac12\delta_{tu}\delta_{vx}\delta_{jy}\\
\bigl[[M_e^{xy}]_{-}\bigr]_{tu,jv}&=\delta_{uv}\delta_{tx}\delta_{jy}
\end{aligned}$\\

\textit{OV--OO}\\
$\begin{aligned}[t]
\bigl[[M_e^{xy}]_{+}\bigr]_{au,vw}&=0\\
\bigl[[M_e^{xy}]_{0}\bigr]_{au,vw}&=\tfrac12\delta_{uw}\delta_{ax}\delta_{vy}\\
\bigl[[M_e^{xy}]_{-}\bigr]_{au,vw}&=\delta_{uw}\delta_{ax}\delta_{vy}
\end{aligned}$\\

\textit{VO--OO}\\
$\begin{aligned}[t]
\bigl[[M_e^{xy}]_{+}\bigr]_{ua,vw}&=0\\
\bigl[[M_e^{xy}]_{0}\bigr]_{ua,vw}&=0\\
\bigl[[M_e^{xy}]_{-}\bigr]_{ua,vw}&=0
\end{aligned}$\\

\textit{OO--OV}\\
$\begin{aligned}[t]
\bigl[[M_e^{xy}]_{+}\bigr]_{tu,bv}&=0\\
\bigl[[M_e^{xy}]_{0}\bigr]_{tu,bv}&=\tfrac12\delta_{tu}\delta_{vx}\delta_{by}\\
\bigl[[M_e^{xy}]_{-}\bigr]_{tu,bv}&=\delta_{uv}\delta_{tx}\delta_{by}
\end{aligned}$\\

\textit{OO--VO}\\
$\begin{aligned}[t]
\bigl[[M_e^{xy}]_{+}\bigr]_{tu,vb}&=0\\
\bigl[[M_e^{xy}]_{0}\bigr]_{tu,vb}&=-\tfrac12\delta_{tu}\delta_{bx}\delta_{vy}\\
\bigl[[M_e^{xy}]_{-}\bigr]_{tu,vb}&=-\delta_{tv}\delta_{bx}\delta_{uy}
\end{aligned}$\\

\textit{OO--OO}\\
$\begin{aligned}[t]
\bigl[[M_e^{xy}]_{+}\bigr]_{tu,vw}&=0\\
\bigl[[M_e^{xy}]_{0}\bigr]_{tu,vw}&=0\\
\bigl[[M_e^{xy}]_{-}\bigr]_{tu,vw}
 &=\delta_{uw}\delta_{tx}\delta_{vy}-\delta_{tv}\delta_{wx}\delta_{uy}
\end{aligned}$
\end{tabular}
\end{ruledtabular}
\end{table}

\end{document}


\title{Supporting Information: Translating Spin-Adapted RPA to Spin-Adapted TDDFT }

\author{Xiaoyu Zhang}
\email{zhangxiaoyu@stu.pku.edu.cn}
\affiliation{College of Chemistry and Molecular Engineering, Peking University, Beijing 100871, China}

\author{Tai Wang}
\affiliation{College of Chemistry and Molecular Engineering, Peking University, Beijing 100871, China}

\date{\today}

\maketitle

\setcounter{equation}{0}
\renewcommand{\theequation}{S\arabic{equation}}
\setcounter{figure}{0}
\renewcommand{\thefigure}{S\arabic{figure}}
\setcounter{table}{0}
\renewcommand{\thetable}{S\arabic{table}}
\setcounter{section}{0}
\renewcommand{\thesection}{S\arabic{section}}
\setcounter{subsection}{0}
\renewcommand{\thesubsection}{\thesection.\arabic{subsection}}

\section{Spin-Adapted RPA}
\subsection{Operator Representation}

We rewrite single excitation operators as tensor operators \cite{PhysRevA.2.2208}:
\begin{equation}
    O^\dagger_{pq}(0,0) =\frac{1}{\sqrt{2}}(\hat{a}_{p_{\alpha}}^\dagger \hat{a}_{q_\alpha}+\hat{a}_{p_{\beta}}^\dagger \hat{a}_{q_\beta})
\end{equation}
\begin{equation}
    O^\dagger_{pq}(1,1)=-\hat{a}_{p_\alpha}^\dagger \hat{a}_{q_\beta}
\end{equation}
\begin{equation}
    O^\dagger_{pq}(1,0) =\frac{1}{\sqrt{2}}(\hat{a}_{p_{\alpha}}^\dagger \hat{a}_{q_\alpha}-\hat{a}_{p_{\beta}}^\dagger \hat{a}_{q_\beta})
\end{equation}
\begin{equation}
    O^\dagger_{pq}(1,-1)=\hat{a}_{p_\beta}^\dagger \hat{a}_{q_\alpha}
\end{equation}
where $\hat{a}^\dagger$ and $\hat{a}$ are the creation and annihilation operators, and the subscripts $\alpha$ and $\beta$ represent spin-up and spin-down, respectively.

The formalism of the equation-of-motion (EOM) focuses on a transition operator $O_\lambda^\dagger$ \cite{RevModPhys.40.153}, which changes an eigenstate $|0\rangle$ of the Hamiltonian to another eigenstate $|\lambda \rangle$:
\begin{equation}
    O_\lambda^\dagger |0\rangle =|\lambda \rangle
\end{equation}
Also, it satisfies the killer condition:
\begin{equation}
    O_\lambda |0 \rangle = 0
\end{equation}
$O_\lambda^\dagger$ is determined by one of the following three conditions:
\begin{equation}
    \langle 0| \delta O_\lambda [H,O_\lambda^\dagger]|0\rangle = \omega_\lambda   \langle 0| \delta O_\lambda O_\lambda^\dagger|0\rangle 
\label{eq:eom1}
\end{equation}
\begin{equation}
    \langle 0| [\delta O_\lambda, [H,O_\lambda^\dagger]]|0\rangle = \omega_\lambda   \langle 0| [\delta O_\lambda ,O_\lambda^\dagger]|0\rangle 
    \label{eq:eom2}
\end{equation}
\begin{equation}
    \langle 0| [\delta O_\lambda, H,O_\lambda^\dagger]|0\rangle = \omega_\lambda   \langle 0| [\delta O_\lambda, O_\lambda^\dagger]|0\rangle 
    \label{eq:eom3}
\end{equation}
Here, $\omega_\lambda$ is the excitation energy. It is noteworthy that the three conditions are equivalent only if $|0 \rangle$ is exact, the killer condition is satisfied, and $O_\lambda$ is expanded on a complete basis. Thus, it will be tricky to decide which condition should be used in the following derivation. We preserve the transferability of our derivation and only adhere to specific conditions when deriving explicit working equations. We use $\{\}$ here to represent the commutators in EOM conditions. For example, if we are using the first EOM condition, we have $\{A,B,C\}= A[B,C]$ and $\{A,B\}= AB$.
In the framework of equation-of-motion (EOM), given that the high spin open-shell reference state can be expressed by a single slater determinant and is acquired via ROHF, RPA can be expressed as:
\begin{equation}
    MX=\omega NX
    \label{eq:eom}
\end{equation}
\begin{equation}
    (M)_{(pq)\Gamma^\prime \mu^\prime,(rs)\Gamma\mu} =  \langle S_{\mathrm{i}} S_{\mathrm{i}}|\{O_{pq}(\Gamma^\prime,\mu^\prime),H,O^\dagger_{rs}(\Gamma,\mu)\}|S_{\mathrm{i}}S_{\mathrm{i}} \rangle
    \label{eq:Mdef}
\end{equation}
\begin{equation}
    (N)_{(pq)\Gamma^\prime \mu^\prime,(rs)\Gamma \mu} = \langle S_{\mathrm{i}} S_i|\{O_{pq}(\Gamma^\prime,\mu^\prime),O^\dagger_{rs}(\Gamma,\mu)\}|S_{\mathrm{i}}S_i \rangle
    \label{eq:Ndef}
\end{equation}
where $\Gamma$ is the rank of a tensor operator, and $\mu$ is one component of the tensor operator. $S_{\mathrm{i}}$ is the initial spin and $S_{\mathrm{f}}$ is the final spin. The second-quantized Hamiltonian can be expressed as:
\begin{equation}
    \hat{H}= h^\sigma_{pq} \hat{a}^\dagger_{p_\sigma} \hat{a}_{q_\sigma} +\frac{1}{2} (p_\sigma q_\sigma |r_{\sigma^\prime} s_{\sigma^{\prime}}) \hat{a}^\dagger_{p_\sigma} \hat{a}^\dagger_{r_{\sigma^\prime}} \hat{a}_{s_{\sigma^\prime}} \hat{a}_{q_{\sigma}} 
\end{equation}
Note that there are some additional phase factors for a tensor de-activation operator and a tensor bra state:
\begin{equation}
    \bar{O}_{pq}(\Gamma,\mu)=(-1)^{\Gamma+\mu}O_{pq}(\Gamma,-\mu)
\end{equation}
\begin{equation}
    \langle \langle S_{\mathrm{i}}|_{M_i} =(-1)^{S_{\mathrm{i}}+M_i} \langle S_{\mathrm{i}},-M_i|
\end{equation}

By directly using eq. (3.10) in ref. \citenum{RevModPhys.47.471}, we have expressions in tensor equation-of-motion (TEOM):
\begin{equation}
    \begin{split}
          &(M_{S_{\mathrm{f}}})_{(pq)\Gamma^\prime,(rs)\Gamma}=  \sum_K (-1)^{S_{\mathrm{i}}-S_{\mathrm{f}}-\Gamma^\prime+K} \sqrt{2K+1}\\
          &W(\Gamma^\prime \Gamma S_{\mathrm{i}} S_{\mathrm{i}},KS_{\mathrm{f}}) \times \langle S_{\mathrm{i}}||\{[\bar{O}_{pq}(\Gamma^\prime),H,O^\dagger_{rs}(\Gamma)]^K\}||S_{\mathrm{i}} \rangle
    \end{split}
\end{equation}
\begin{equation}
    \begin{split}
          &(N_{S_{\mathrm{f}}})_{(pq)\Gamma^\prime,(rs)\Gamma}=  \sum_K (-1)^{S_{\mathrm{i}}-S_{\mathrm{f}}-\Gamma^\prime+K} \sqrt{2K+1}\\
          &W(\Gamma^\prime \Gamma S_{\mathrm{i}} S_{\mathrm{i}},KS_{\mathrm{f}}) \times \langle S_{\mathrm{i}}||\{[\bar{O}_{pq}(\Gamma^\prime),O^\dagger_{rs}(\Gamma)]^K\}||S_{\mathrm{i}} \rangle
    \end{split}
\end{equation}
where $W(\Gamma^\prime \Gamma S_{\mathrm{i}} S_{\mathrm{i}},KS_{\mathrm{f}})$ are Racah coefficients.
By using the Wigner-Eckart theorem and tensor decoupling, we have more explicit expressions:
\begin{equation}
      (M_{S_{\mathrm{f}}})_{(pq)\Gamma^\prime,(rs)\Gamma} = \sum_K (-1)^K [M_{\Gamma^\prime \Gamma}^K(S_{\mathrm{f}})]_{pq,rs}
      \label{eq:Mfinal1}
\end{equation}
\begin{equation}
\begin{split}
     [M_{\Gamma^\prime \Gamma}^K(S_{\mathrm{f}})]_{pq,rs} &= \sum_\mu (-1)^{\Gamma^\prime-\mu} \beta_{\Gamma^\prime \Gamma}^K(\mu;S_{\mathrm{f}})\\
     &\langle S_{\mathrm{i}} S_{\mathrm{i}}|  \{O_{pq}(\Gamma^\prime,\mu),H,O_{rs}^\dagger(\Gamma,\mu)\} |S_{\mathrm{i}} S_{\mathrm{i}}\rangle
     \end{split}
     \label{eq:Mfinal2}
\end{equation}
\begin{equation}
      (N_{S_{\mathrm{f}}})_{(pq)\Gamma^\prime,(rs)\Gamma} = \sum_K (-1)^K [N_{\Gamma^\prime \Gamma}^K(S_{\mathrm{f}})]_{pq,rs}
      \label{eq:Nfinal1}
\end{equation}
\begin{equation}
\begin{split}
     [N_{\Gamma^\prime \Gamma}^K(S_{\mathrm{f}})]_{pq,rs} &= \sum_\mu (-1)^{\Gamma^\prime-\mu} \beta_{\Gamma^\prime \Gamma}^K(\mu;S_{\mathrm{f}})\\
     &\langle S_{\mathrm{i}} S_{\mathrm{i}}|  \{O_{pq}(\Gamma^\prime,\mu),O_{rs}^\dagger(\Gamma,\mu)\} |S_{\mathrm{i}} S_{\mathrm{i}}\rangle
     \end{split}
     \label{eq:Nfinal2}
\end{equation}
where we use a mathematical intermediate
\begin{equation}
\begin{split}
    \beta_{\Gamma^\prime \Gamma}^K(\mu;S_{\mathrm{f}}) &= (-1)^{S_{\mathrm{i}}-S_{\mathrm{f}}-\Gamma^\prime}(2K+1)W(\Gamma^\prime \Gamma S_{\mathrm{i}} S_{\mathrm{i}},K S_{\mathrm{f}}) \\
    &\frac{\langle \Gamma^\prime(-\mu),\Gamma \mu|K0\rangle}{\langle S_{\mathrm{i}} S_{\mathrm{i}},S_{\mathrm{i}}(-S_{\mathrm{i}})|K0 \rangle}
    \end{split}
\end{equation}

\subsection{Matrix Representation}

Due to the non-zero condition of $W(\Gamma^\prime \Gamma S_{\mathrm{i}} S_{\mathrm{i}},K S_{\mathrm{f}})$, we have four triangular conditions to satisfy: $\Delta (\Gamma S_{\mathrm{i}}S_{\mathrm{f}})$, $\Delta (\Gamma^\prime S_{\mathrm{i}}S_{\mathrm{f}})$, $\Delta (\Gamma \Gamma^\prime K)$, and $\Delta (S_{\mathrm{i}} S_{\mathrm{i}}K)$. Thus, given $S_{\mathrm{i}}$ and $S_{\mathrm{f}}$, the range of $\Gamma$, $\Gamma^\prime$, and $K$ can be determined. We list all possible expressions of $M_{S_{\mathrm{f}}}$ in terms of $M_{\Gamma^\prime \Gamma}^K$ in table \ref{table:M}. All $M_{\Gamma^\prime \Gamma}^K$ can be evaluated by introducing five skeleton matrices, the details of which are shown in table \ref{table:Msub}. $N_{S_{\mathrm{f}}}$ is very similar here, and the only difference is that there will be no $H$ in table \ref{table:Msub}.
\begin{table}[H]
\caption{All possible expressions of $M_{S_{\mathrm{f}}}$ in terms of $M_{\Gamma^\prime \Gamma}^K$. This table was originally proposed by Li and Liu in ref. \citenum{10.1063/1.3463799}.}
\label{table:M}
\centering
\begin{tabular}{|c|c|c|c|}
\hline
 & $S_{\mathrm{f}}=S_{\mathrm{i}}+1$ & $S_{\mathrm{f}}=S_{\mathrm{i}}$ & $S_{\mathrm{f}}=S_{\mathrm{i}}-1$ \\
\hline
$S_{\mathrm{i}}=0$ & $\left[\mathbf{M}_{11}^{0}\right]$ & $\left[\mathbf{M}_{00}^{0}\right]$ & NP \\
\hline
$S_{\mathrm{i}}=\frac{1}{2}$ & $\left[\mathbf{M}_{11}^{0}-\mathbf{M}_{11}^{1}\right]$ & 
$\left[\!\!\begin{array}{cc}
\mathbf{M}_{00}^{0} & -\mathbf{M}_{01}^{1} \\
-\mathbf{M}_{10}^{1} & \mathbf{M}_{11}^{0}-\mathbf{M}_{11}^{1}
\end{array}\!\!\right]$ & NP \\
\hline
$S_{\mathrm{i}}\geq 1$ & $\left[\mathbf{M}_{11}^{0}-\mathbf{M}_{11}^{1}+\mathbf{M}_{11}^{2}\right]$ & 
$\left[\!\!\begin{array}{cc}
\mathbf{M}_{00}^{0} & -\mathbf{M}_{01}^{1} \\
-\mathbf{M}_{10}^{1} & \mathbf{M}_{11}^{0}-\mathbf{M}_{11}^{1}+\mathbf{M}_{11}^{2}
\end{array}\!\!\right]$ & 
$\left[\mathbf{M}_{11}^{0}-\mathbf{M}_{11}^{1}+\mathbf{M}_{11}^{2}\right]$ \\
\hline
\end{tabular}
\end{table}

\begin{table}[H]
\caption{$ S_{\mathrm{f}} $-independent skeleton matrices $ M $ for the submatrices $ \mathbf{M}_{\Gamma^{\prime}\Gamma}^{K}(S_{\mathrm{f}}) $. This table was originally proposed by Li and Liu in ref. \citenum{10.1063/1.3463799}. }
\label{table:Msub}
\centering
\begin{tabular}{ll}
\hline
Blocks & Expressions \\
\hline
$ M_{00} $ & $ [M_{00}]_{pq,rs}=\langle S_{\mathrm{i}}S_{\mathrm{i}}|\{O_{pq}(0,0),H,O_{rs}^{\dagger}(0,0)\}|S_{\mathrm{i}}S_{\mathrm{i}}\rangle $ \\
$ M_{01} $ & $ [M_{01}]_{pq,rs}=\langle S_{\mathrm{i}}S_{\mathrm{i}}|\{O_{pq}(0,0),H,O_{rs}^{\dagger}(1,0)\}|S_{\mathrm{i}}S_{\mathrm{i}}\rangle $ \\
$ M_{+} $ & $ [M_{+}]_{pq,rs}=\langle S_{\mathrm{i}}S_{\mathrm{i}}|\{O_{pq}(1,1),H,O_{rs}^{\dagger}(1,1)\}|S_{\mathrm{i}}S_{\mathrm{i}}\rangle $ \\
$ M_{0} $ & $ [M_{0}]_{pq,rs}=\langle S_{\mathrm{i}}S_{\mathrm{i}}|\{O_{pq}(1,0),H,O_{rs}^{\dagger}(1,0)\}|S_{\mathrm{i}}S_{\mathrm{i}}\rangle $ \\
$ M_{-} $ & $ [M_{-}]_{pq,rs}=\langle S_{\mathrm{i}}S_{\mathrm{i}}|\{O_{pq}(1,-1),H,O_{rs}^{\dagger}(1,-1)\}|S_{\mathrm{i}}S_{\mathrm{i}}\rangle $ \\
$ \mathbf{M}_{00}^{0}(S_{\mathrm{f}}=S_{\mathrm{i}}) $ & $ M_{00} $ \\
$ \mathbf{M}_{01}^{1}(S_{\mathrm{f}}=S_{\mathrm{i}}) $ & $ \sqrt{\frac{S_{\mathrm{i}}+1}{S_{\mathrm{i}}}}M_{01} $ \\
$ \mathbf{M}_{10}^{1}(S_{\mathrm{f}}=S_{\mathrm{i}}) $ & $ \sqrt{\frac{S_{\mathrm{i}}+1}{S_{\mathrm{i}}}}M_{01}^{\dagger} $ \\
$ \mathbf{M}_{11}^{0}(S_{\mathrm{f}}=S_{\mathrm{i}}) $ & $ \frac{1}{3}(M_{+}+M_{0}+M_{-}) $ \\
$ \mathbf{M}_{11}^{1}(S_{\mathrm{f}}=S_{\mathrm{i}}) $ & $ -\frac{1}{2S_{\mathrm{i}}}(-M_{+}+M_{-}) $ \\
$ \mathbf{M}_{11}^{2}(S_{\mathrm{f}}=S_{\mathrm{i}}) $ & $ -\frac{2S_{\mathrm{i}}+3}{6S_{\mathrm{i}}}(M_{+}-2M_{0}+M_{-}) $ \\
$ \mathbf{M}_{11}^{0}(S_{\mathrm{f}}=S_{\mathrm{i}}+1) $ & $ \frac{1}{3}(M_{+}+M_{0}+M_{-}) $ \\
$ \mathbf{M}_{11}^{1}(S_{\mathrm{f}}=S_{\mathrm{i}}+1) $ & $ \frac{1}{2}(-M_{+}+M_{-}) $ \\
$ \mathbf{M}_{11}^{2}(S_{\mathrm{f}}=S_{\mathrm{i}}+1) $ & $ \frac{1}{6}(M_{+}-2M_{0}+M_{-}) $ \\
$ \mathbf{M}_{11}^{0}(S_{\mathrm{f}}=S_{\mathrm{i}}-1) $& $ \frac{1}{3}(M_{+}+M_{0}+M_{-}) $ \\
$ \mathbf{M}_{11}^{1}(S_{\mathrm{f}}=S_{\mathrm{i}}-1) $ & $ -\frac{S_{\mathrm{i}}+1}{2S_{\mathrm{i}}}(-M_{+}+M_{-}) $ \\
$ \mathbf{M}_{11}^{2}(S_{\mathrm{f}}=S_{\mathrm{i}}-1) $ & $ \frac{(S_{\mathrm{i}}+1)(2S_{\mathrm{i}}+3)}{6S_{\mathrm{i}}(2S_{\mathrm{i}}-1)}(M_{+}-2M_{0}+M_{-}) $ \\
\hline
\end{tabular}
\end{table}

We have three basic symmetries in eq. \ref{eq:Mfinal2} and eq. \ref{eq:Nfinal2}:
\begin{equation}
    O_{qp}(\Gamma,\mu)= (-1)^{\Gamma+\mu} O^\dagger_{pq}(\Gamma,-\mu)
\end{equation}
\begin{equation}
    \beta^K_{\Gamma^\prime \Gamma}(-\mu;S_\mathrm{f}) = (-1)^{\Gamma^\prime+\Gamma-K} \beta^K_{\Gamma^\prime \Gamma}(\mu;S_\mathrm{f})
\end{equation}
\begin{equation}
    \beta^K_{\Gamma \Gamma^\prime}(\mu;S_\mathrm{f}) = (-1)^{\Gamma^\prime-\Gamma} \beta^K_{\Gamma^\prime \Gamma}(\mu;S_\mathrm{f})
\end{equation}

Thus, by the definition of M and N, we have:
\begin{equation}
    \begin{split}
     [M_{\Gamma^\prime \Gamma}^K(S_{\mathrm{f}})]_{qp,sr} &= (-1)^{\Gamma^\prime+\Gamma+K}  \sum_\mu (-1)^{\Gamma^\prime-\mu} \beta_{\Gamma^\prime \Gamma}^K(\mu;S_{\mathrm{f}})\\
     &\langle S_{\mathrm{i}} S_{\mathrm{i}}|  \{O_{pq}^\dagger(\Gamma^\prime,\mu),H,O_{rs}(\Gamma,\mu)\} |S_{\mathrm{i}} S_{\mathrm{i}}\rangle
     \end{split}
\end{equation}
\begin{equation}
    \begin{split}
     [N_{\Gamma^\prime \Gamma}^K(S_{\mathrm{f}})]_{qp,sr} &= (-1)^{\Gamma^\prime+\Gamma+K}  \sum_\mu (-1)^{\Gamma^\prime-\mu} \beta_{\Gamma^\prime \Gamma}^K(\mu;S_{\mathrm{f}})\\
     &\langle S_{\mathrm{i}} S_{\mathrm{i}}|  \{O_{pq}^\dagger(\Gamma^\prime,\mu),O_{rs}(\Gamma,\mu)\} |S_{\mathrm{i}} S_{\mathrm{i}}\rangle
     \end{split}
\end{equation}
If we use the EOM condition eq. \ref{eq:eom2} or eq. \ref{eq:eom3}, we have:
\begin{equation}
      [M_{\Gamma^\prime \Gamma}^K(S_{\mathrm{f}})]_{qp,sr} = (-1)^{\Gamma^\prime+\Gamma+K}[M_{\Gamma^\prime \Gamma}^K(S_{\mathrm{f}})]_{pq,rs}
\end{equation}
\begin{equation}
      [N_{\Gamma^\prime \Gamma}^K(S_{\mathrm{f}})]_{qp,sr} = (-1)^{\Gamma^\prime+\Gamma+K+1}[N_{\Gamma^\prime \Gamma}^K(S_{\mathrm{f}})]_{pq,rs}
\end{equation}
If we use the EOM condition eq. \ref{eq:eom3}, we particularly have:
\begin{equation}
      [M_{\Gamma^\prime \Gamma}^K(S_{\mathrm{f}})]_{pq,rs} = [M_{\Gamma \Gamma^\prime}^K(S_{\mathrm{f}})]_{rs,pq}
\end{equation}
\begin{equation}
      [N_{\Gamma^\prime \Gamma}^K(S_{\mathrm{f}})]_{pq,rs} = [N_{\Gamma \Gamma^\prime}^K(S_{\mathrm{f}})]_{rs,pq}
\end{equation}

It is imperative to categorize various orbitals for the purpose of evaluating matrix elements. The following system of notation for molecular orbitals should be adopted: \textit{i}, \textit{j}, \textit{k}, \textit{l},... denote doubly occupied orbitals, also referred to as the closed-shell component C. Notations \textit{t}, \textit{u}, \textit{v}, \textit{w},... are assigned to singly occupied orbitals along with their time-reversed counterparts, corresponding to the open-shell component O. \textit{a}, \textit{b}, \textit{c}, \textit{d},... are designated for virtual orbitals, identified as the vacant-shell component V, and \textit{p}, \textit{q}, \textit{r}, \textit{s},... pertain to unspecified orbitals.

In order to be consistent with conventional notations in TDDFT, we use different notations onto $M_{S_{\mathrm{f}}}$ and $N_{S_{\mathrm{f}}}$ based on the sequence of orbital indices (activation, de-activation, or diagonal):
\begin{equation}
\begin{aligned}
&(M_{S_{\mathrm{f}}})_{(pq)\Gamma^\prime,(rs)\Gamma} = {}\\
&\left\{\begin{array}{@{}l@{\quad}l@{}}
(A)_{(pq)\Gamma^\prime,(rs)\Gamma}, & p>q,\ r>s\ \text{or } p<q,\ r<s,\\
(B)_{(pq)\Gamma^\prime,(rs)\Gamma}, & p>q,\ r<s\ \text{or } p<q,\ r>s,\\
(L)_{(pq)\Gamma^\prime,(rs)\Gamma}, & p\neq q,\ r=s\ \text{or } p=q,\ r\neq s,\\
(D)_{(pq)\Gamma^\prime,(rs)\Gamma}, & p=q,\ r=s.
\end{array}\right.
\end{aligned}
\end{equation}

\begin{equation}
\begin{aligned}
&(N_{S_{\mathrm{f}}})_{(pq)\Gamma^\prime,(rs)\Gamma} = {}\\
&\left\{\begin{array}{@{}l@{\quad}l@{}}
(\Sigma)_{(pq)\Gamma^\prime,(rs)\Gamma}, & p>q,\ r>s\ \text{or } p<q,\ r<s,\\
(\Delta)_{(pq)\Gamma^\prime,(rs)\Gamma}, & p>q,\ r<s\ \text{or } p<q,\ r>s,\\
(\theta)_{(pq)\Gamma^\prime,(rs)\Gamma}, & p\neq q,\ r=s\ \text{or } p=q,\ r\neq s,\\
(\epsilon)_{(pq)\Gamma^\prime,(rs)\Gamma}, & p=q,\ r=s.
\end{array}\right.
\end{aligned}
\end{equation}

The notational conventions for blocks involving OO transitions require special treatment. With the exception of the cases labeled $L$ and $D$, no additional constraints are imposed on OO transitions. For instance, in the expression $A_{O_1O_2,O_3O_4}$, no ordering relation is required between $O_1$ and $O_2$, nor between $O_3$ and $O_4$.
Explicit expressions for skeleton matrices are presented in the next section.

In order to further deduce the matrices, we now start considering the operator basis. Apparently, we have eight tensor single excitation operators: $O^\dagger_{CV}(0)$, $O^\dagger_{CO}(0)$, $O^\dagger_{OV}(0)$, $O^\dagger_{OO}(0)$, $O^\dagger_{CV}(1)$, $O^\dagger_{CO}(1)$, $O^\dagger_{OV}(1)$, $O^\dagger_{OO}(1)$. Because of the triangular condition $\Delta (\Gamma S_{\mathrm{i}}S_{\mathrm{f}})$, the excitation operator basis is constructed as follows:
\begin{equation}
    q^\dagger_{S_{\mathrm{f}}=S_{\mathrm{i}}+1}=\{ O^\dagger_{CV}(1), O^\dagger_{CO}(1), O^\dagger_{OV}(1), O^\dagger_{OO}(1) \}
\end{equation}
\begin{equation}
\begin{split}
     q^\dagger_{S_{\mathrm{f}}=S_{\mathrm{i}}}=\{ O^\dagger_{CV}(0), O^\dagger_{CO}(0), O^\dagger_{OV}(0), O^\dagger_{OO}(0),\\ O^\dagger_{CV}(1), O^\dagger_{CO}(1), O^\dagger_{OV}(1), O^\dagger_{OO}(1)  \}
\end{split}
\end{equation}
\begin{equation}
    q_{S_{\mathrm{f}}=S_{\mathrm{i}}-1}^\dagger = \{  O^\dagger_{CV}(1), O^\dagger_{CO}(1), O^\dagger_{OV}(1), O^\dagger_{OO}(1)  \}
\end{equation}
However, there are two problems: redundancies and linear dependency. Redundancy is defined as follows:
\begin{equation}
    [O^\dagger(\Gamma)\times |S_{\mathrm{i}}\rangle \rangle]^{S_{\mathrm{f}}}=0
\end{equation}
or 
\begin{equation}
     [O^\dagger(\Gamma)\times |S_{\mathrm{i}}\rangle \rangle]^{S_{\mathrm{f}}}=|S_{\mathrm{i}} \rangle \rangle
\end{equation}
Linear dependency is described by:
\begin{equation}
     [O_j^\dagger(\Gamma)\times |S_{\mathrm{i}}\rangle \rangle]^{S_{\mathrm{f}}}=const \times   [O_k^\dagger(\Gamma^\prime)\times |S_{\mathrm{i}}\rangle \rangle]^{S_{\mathrm{f}}}
\end{equation}
After eliminating redundancy and linear dependency, the two bases are changed to:
\begin{equation}
    q_{S_{\mathrm{f}}=S_{\mathrm{i}}+1}^\dagger=\{ O^\dagger_{CV}(1)  \}
\end{equation}
\begin{equation}
    q^\dagger_{S_{\mathrm{f}}=S_{\mathrm{i}}} = \{   O^\dagger_{CV}(0), O^\dagger_{CO}(0), O^\dagger_{OV}(0),O^\dagger_{CV}(1) \}
\end{equation}
\begin{widetext}
For the three different cases, we now start to derive explicit matrix forms. 
For HF exchange and spin-unpolarized pure xc functionals, we have (cf. eq. \ref{eq:kt}):
\begin{equation}
      K^{\alpha\beta,\alpha\beta} = K^{\beta\alpha,\beta\alpha}=K^T:=K^{TC}
\end{equation}

The orbital-resolved Hamiltonian working equations below also use the ROHF
Brillouin/stationarity conditions
\begin{equation}
 f_{\rm CV}^{\alpha}+f_{\rm CV}^{\beta}=0,
 \qquad
 f_{\rm CO}^{\beta}=0,
 \qquad
 f_{\rm OV}^{\alpha}=0.
 \label{eq:si-rohf-brillouin}
\end{equation}
Together with $f^\sigma_{pq}=f^{\sigma *}_{qp}$, these conditions are
applied only to the Hamiltonian working equations.  They must not be used
for an unconstrained one-body insertion.

The choice of the equation-of-motion (EOM) condition (eq. \ref{eq:eom1}–eq. \ref{eq:eom3}) requires careful consideration. Among these, eq. \ref{eq:eom3} is generally preferred because the commutator of two operators has a lower particle rank than their direct product, rendering its matrix elements less sensitive to deficiencies in the reference state $|0 \rangle$. Consequently, we adopt eq. \ref{eq:eom3} for cases with $S_{\mathrm{f}} = S_{\mathrm{i}} + 1$ and $S_{\mathrm{f}} = S_{\mathrm{i}}$. 

However, this EOM condition leads to spin contamination when $S_{\mathrm{f}} = S_{\mathrm{i}} - 1$. This issue can be understood by examining the rank of the OO–OO block of the $N$ matrix, which was observed by Zhang and Herbert \cite{10.1063/1.4937571}. For $S_{\mathrm{i}} = 1$, we obtain
\begin{equation}
    N^{\mathrm{OO-OO}} = 
    \begin{pmatrix}
        2 & 0 & 0 & 0\\
        0 & 2 & 0 & 0\\
        0 & 0 & 2 & 0\\
        0 & 0 & 0 & 2\\
    \end{pmatrix}
\end{equation}
This matrix is problematic because it has rank 4, whereas only three singlet states are physically targeted. 

A viable remedy is to instead employ eq. \ref{eq:eom1}, which yields
\begin{equation}
    N^{\mathrm{OO-OO}} = 
    \begin{pmatrix}
        \frac{3}{2} & 0 & 0 & -\frac{3}{2}\\
        0 & 3 & 0 & 0\\
        0 & 0 & 3 & 0\\
        -\frac{3}{2} & 0 & 0 & \frac{3}{2}\\
    \end{pmatrix}
\end{equation}
This matrix has rank 3, thereby removing the spurious solution and eliminating spin contamination.

When $S_{\mathrm{f}}=S_{\mathrm{i}}+1$, we choose EOM condition Eq. \ref{eq:eom3}. We obtain that:
\begin{equation}
N_{S_{\mathrm{f}}}=diag\{ \delta_{ab}\delta_{ij},-\delta_{ab}\delta_{ij} \}
\end{equation}

\begin{equation}
     M_{S_{\mathrm{f}}}=
     \begin{pmatrix}
         \delta_{ij}f^\alpha_{ab}-\delta_{ab} f^\beta_{ji} +(K^{TC})_{ai,bj}& -(K^{TC})_{ai,jb}\\
         -(K^{TC})_{ai,jb} & \delta_{ij}f^\beta_{ab}-\delta_{ab} f^\alpha_{ji} +(K^{TC})_{ai,bj}
     \end{pmatrix}
\end{equation} 
Note that SA-TDA here works exactly the same as spin-flip-up TDA, but SA-TDDFT doesn't work the same as spin-flip-up TDDFT.

When $S_{\mathrm{f}}=S_{\mathrm{i}}$, we choose the EOM condition Eq. \ref{eq:eom3}. We obtain that:
\begin{equation}
M_{S_{\mathrm{f}}}= 
    \begin{pmatrix}
       \mathbf{A} & \mathbf{B} \\
\mathbf{B}^{*} & \widetilde{\mathbf{A}}^{*}
    \end{pmatrix}
\label{eq:equal}
\end{equation}
\begin{equation}
N_{S_{\mathrm{f}}}= 
    \begin{pmatrix}
      I & 0 \\
0 & -I
    \end{pmatrix}
\end{equation}
The detailed expressions of elements are shown in table. \ref{table:equal}.

For closed-shell systems, equations are reduced to common RPA. 

When $S_{\mathrm{f}}=S_{\mathrm{i}}-1$, we choose the EOM condition Eq.\ref{eq:eom1}. We denote:
\begin{equation}
    \widetilde{M} = \frac{1}{S_{\mathrm{i}}(2S_{\mathrm{i}}-1)}M_+ - \frac{2S_{\mathrm{i}}+1}{S_{\mathrm{i}}(2S_{\mathrm{i}}-1)}  M_0 + \frac{2S_{\mathrm{i}}+1}{2S_{\mathrm{i}} -1} M_-
\end{equation}

\begin{equation}
\begingroup
\setlength{\arraycolsep}{2pt}
\renewcommand{\arraystretch}{1.05}
\scriptsize
\resizebox{\linewidth}{!}{$
M_{S_{\mathrm{f}}}=
\left(
\begin{array}{cccc|c|ccc}
\widetilde{A}_{ai,bj} & \widetilde{A}_{ai,vj} & \widetilde{A}_{ai,bv} & \widetilde{A}_{ai,vw} & \widetilde{L}_{ai,vv}
& \widetilde{B}_{ai,jb} & \widetilde{B}_{ai,jv} & \widetilde{B}_{ai,vb} \\

\widetilde{A}_{ui,bj} & \widetilde{A}_{ui,vj} & \widetilde{A}_{ui,bv} & \widetilde{A}_{ui,vw} & \widetilde{L}_{ui,vv}
& \widetilde{B}_{ui,jb} & \widetilde{B}_{ui,jv} & \widetilde{B}_{ui,vb} \\

\widetilde{A}_{au,bj} & \widetilde{A}_{au,vj} & \widetilde{A}_{au,bv} & \widetilde{A}_{au,vw} & \widetilde{L}_{au,vv}
& \widetilde{B}_{au,jb} & \widetilde{B}_{au,jv} & \widetilde{B}_{au,vb} \\

\widetilde{A}_{tu,bj} & \widetilde{A}_{tu,vj} & \widetilde{A}_{tu,bv} & \widetilde{A}_{tu,vw} & \widetilde{L}_{tu,vv}
& \widetilde{B}_{tu,jb} & \widetilde{B}_{tu,jv} & \widetilde{B}_{tu,vb} \\
\hline
\widetilde{L}_{tt,bj} & \widetilde{{L}}_{tt,vj} & \widetilde{L}_{tt,bv} & \widetilde{L}_{tt,vw} & \widetilde{D}_{tt,vv}
& \widetilde{L}_{tt,jb} & \widetilde{L}_{tt,jv} & \widetilde{L}_{tt,vb} \\
\hline
\widetilde{B}_{ia,bj} & \widetilde{B}_{ia,vj} & \widetilde{B}_{ia,bv} & \widetilde{B}_{ia,vw} & \widetilde{L}_{ia,vv}
& \widetilde{A}_{ia,jb} & \widetilde{A}_{ia,jv} & \widetilde{A}_{ia,vb} \\

\widetilde{B}_{iu,bj} & \widetilde{B}_{iu,vj} & \widetilde{B}_{iu,bv} & \widetilde{B}_{iu,vw} & \widetilde{L}_{iu,vv}
& \widetilde{A}_{iu,jb} & \widetilde{A}_{iu,jv} & \widetilde{A}_{iu,vb} \\

\widetilde{B}_{ua,bj} & \widetilde{B}_{ua,vj} & \widetilde{B}_{ua,bv} & \widetilde{B}_{ua,vw} & \widetilde{L}_{ua,vv}
& \widetilde{A}_{ua,jb} & \widetilde{A}_{ua,jv} & \widetilde{A}_{ua,vb}
\end{array}
\right)
$}
\endgroup
\end{equation}
The detailed expressions of matrix elements are shown in Tables~\ref{tab:minus1},~\ref{tab:minus2}, and~\ref{tab:minus3}.

$N_{S_{\mathrm{f}}}$ has a very similar structure.
\begin{equation}
    N_{S_{\mathrm{f}}} = diag\{ \delta_{ij}\delta_{ab}, (\frac{1}{2S_\mathrm{i}}+1)\delta_{uv}\delta_{ij},(\frac{1}{2S_\mathrm{i}}+1)\delta_{uv}\delta_{ab},\frac{2S_\mathrm{i}+1}{2S_\mathrm{i}-1} \delta_{tv}\delta_{uw},\frac{2S_\mathrm{i}+1}{2S_\mathrm{i}-1}(-\frac{1}{2S_\mathrm{i}}+\delta_{tv}),0,0,0 \}
\end{equation}
Note that $N_{S_{\mathrm{f}}}$ here is block-diagonal but not diagonal.
\end{widetext}

For $S_{\rm f}=S_{\rm i}-1$, separate the full ordered OO product space as
\begin{equation}
 \mathrm{OO}_{\rm full}
 =\mathrm{OO}_{t\ne u}\oplus\mathrm{TT}_{t=u}.
 \label{eq:si-oo-tt-split}
\end{equation}
Since $n_O=2S_{\rm i}$, the TT block of the metric is
\begin{equation}
 N_{\rm TT}
 =\frac{2S_{\rm i}+1}{2S_{\rm i}-1}
 \left(I_{n_O}-\frac{1}{n_O}
 \boldsymbol1\boldsymbol1^\dagger\right).
 \label{eq:si-tt-metric}
\end{equation}
Its uniform vector is a zero mode.  Let
$U\in\mathbb C^{n_O\times(n_O-1)}$ satisfy
\begin{equation}
 U^\dagger U=I_{n_O-1},
 \qquad U^\dagger\boldsymbol1=0,
 \qquad
 UU^\dagger=I_{n_O}-\frac{1}{n_O}
 \boldsymbol1\boldsymbol1^\dagger,
 \label{eq:si-tt-traceless-basis}
\end{equation}
and, in the block order
$[\mathrm{CV},\mathrm{CO},\mathrm{OV},\mathrm{OO},\mathrm{TT}\mid
\mathrm{VC},\mathrm{OC},\mathrm{VO}]$, define
\begin{equation}
 P_{-1}=\operatorname{diag}
 (I_{\rm CV},I_{\rm CO},I_{\rm OV},I_{\rm OO},U,
 I_{\rm VC},I_{\rm OC},I_{\rm VO}).
 \label{eq:si-minus-projector}
\end{equation}
The physical generalized eigenvalue problem is formed from
\begin{equation}
 M_{S_{\rm f}}^{\rm phys}=P_{-1}^\dagger M_{S_{\rm f}}P_{-1},
 \qquad
 N_{S_{\rm f}}^{\rm phys}=P_{-1}^\dagger N_{S_{\rm f}}P_{-1}.
 \label{eq:si-minus-projected-pencil}
\end{equation}
The projection acts on every row and column connected to TT, rather than
only on the TT–TT subblock. In SA-TDA, the corresponding projector
$P_{-1}^{\rm TDA}$ consists of the first five diagonal blocks of $P_{-1}$.

 Similar to common TDDFT, we have redundant deexcitation-like roots, and we use the following equations to select:
\begin{equation}
    n_k = X_k^\dagger N X_k
\end{equation}
$X_k$ denotes the $k$-th vector in the solutions. If $n_k>0 $, the root $X_k$ is regarded as an excitation-like root that will be preserved. Otherwise, the root will be discarded.

Within the spin-adapted framework, SA-TDA is defined as the
restriction of the zero-mode-projected response problem to its complete
positive principal subspace.  For
$S_{\mathrm f}=S_{\mathrm i}-1$, this subspace is
\[
 [\mathrm{CV},\mathrm{CO},\mathrm{OV},\mathrm{OO},\mathrm{TT}],
\]
where TT denotes the $(n_O-1)$-dimensional traceless space generated by
$U$.  Thus, SA-TDA retains all $A$, $L$, and $D$ blocks, including all
positive-space OO/TT-connected blocks, and removes the blocks coupling
to VC, OC, and VO.  The same restriction is applied to
$N_{S_{\mathrm f}}$ and to any property matrix represented in the
response basis.

For $S_{\mathrm f}=S_{\mathrm i}-1$, after applying the OO/TT
projection above, write the response vector as
\begin{equation}
 Z=
 \begin{pmatrix}
 X\\
 Y
 \end{pmatrix},
\end{equation}
where $X$ contains the positive-amplitude blocks and $Y$ contains the
remaining blocks.  The projected EOM1 equation has the block form
\begin{equation}
 MZ=\omega NZ,
\end{equation}
with
\begin{equation}
 M-\omega N
 =
 \begin{pmatrix}
 M_{\mathrm{TDA}}-\omega N_{\mathrm{TDA}} & C\\
 0 & 0
 \end{pmatrix}.
\end{equation}
Here, $C$ is the upper-right coupling block and need not vanish.
Because the lower block rows vanish, the full EOM1 pencil is singular.
We therefore define its physical finite roots through the regular
positive-amplitude restriction
\begin{equation}
 M_{\mathrm{TDA}}X_\lambda
 =
 \omega_\lambda N_{\mathrm{TDA}}X_\lambda.
 \label{eq:si-minus-physical-root}
\end{equation}

For every root of Eq.~\eqref{eq:si-minus-physical-root}, choose the
canonical full-space representative
\begin{equation}
 Z_\lambda=
 \begin{pmatrix}
 X_\lambda\\
 0
 \end{pmatrix}.
 \label{eq:si-minus-canonical-vector}
\end{equation}
It satisfies the full EOM1 equation because
\begin{equation}
 MZ_\lambda
 =
 \begin{pmatrix}
 M_{\mathrm{TDA}}X_\lambda\\
 0
 \end{pmatrix}
 =
 \omega_\lambda
 \begin{pmatrix}
 N_{\mathrm{TDA}}X_\lambda\\
 0
 \end{pmatrix}
 =
 \omega_\lambda NZ_\lambda.
\end{equation}
The coupling block $C$ does not contribute to this canonical
representative.  Components in the singular $Y$ sector are not used
for root selection or property contractions.  Under this physical
root-selection convention, the reported SA-TDDFT and SA-TDA
excitation energies are identical.

\subsection{Skeleton Matrices}

We use this short notation for double-electron integrals:
\begin{equation}
    (p_{\sigma}q_{\sigma^\prime}|r_{\tau}s_{\tau^\prime})=\int d \mathbf{r_1} d\mathbf{r_2} \frac{p_{\sigma}(\mathbf{r_1})q_{\sigma^\prime}(\mathbf{r_1})r_{\tau}(\mathbf{r_2})s_{\tau^\prime}(\mathbf{r_2})}{|\mathbf{r_1}-\mathbf{r_2}|}\\
\end{equation}
Obviously, the integral satisfies:
\begin{equation}
      (p_{\sigma}q_{\sigma^\prime}|r_{\tau}s_{\tau^\prime})=  (p_{\sigma}q_{\sigma}|r_{\tau}s_{\tau})\delta_{\sigma \sigma^\prime} \delta_{\tau \tau^\prime}
\end{equation}

Another short notation is defined as:
\begin{equation}
    \bra{ij}\ \ket{kl}=(ik|jl)-(il|jk)
\end{equation}

We first define the first-order reduced density matrix as:
\begin{equation}
    D_{pq}^{\sigma \sigma^\prime} = f_{i\tau} c^*_{p_\sigma i_\tau} c_{q_{\sigma^\prime} i_\tau}
\end{equation}

The Fockian is defined as:
\begin{equation}
    f^{\sigma \sigma^\prime}_{pq}= \frac{\partial E}{\partial D^{\sigma \sigma^\prime}_{pq}}
\end{equation}

Then, the Fockian satisfies:
\begin{equation}
    f^{\sigma}_{pq}= f^{\sigma \sigma}_{pq}, \sigma=\alpha,\beta
\end{equation}
\begin{equation}
    f^{\sigma \sigma^\prime}_{pq}=0,\sigma \neq \sigma^\prime
\end{equation}
\begin{equation}
    f^{\sigma}_{pq} = f^{\sigma}_{qp}
\end{equation}
More explicitly, the RPA Fockian is expressed as:
\begin{equation}
\begin{split}
    f_{pq}^{\sigma} &= h_{pq}^{\sigma} +
\big[(p_{\sigma} q_{\sigma} \mid r_{\tau} r_{\tau})
- \delta_{\sigma\tau} (p_{\sigma} r_{\tau} \mid r_{\tau} q_{\sigma})\big]
\, n_{r_\tau}\\
&= h_{pq}^\sigma + \sum_{r_\tau \in \Phi_0} \bra{p_\sigma r_\tau} \ \ket{q_\sigma r_\tau}
\end{split}
\end{equation}

\[
\sigma,\tau = \alpha,\beta, \qquad n_{r_\tau} = 0,1.
\]

Similarly, the kernel matrix elements are defined as:
\begin{equation}
    \bigl[K^{\sigma\tau,\sigma' \tau'}\bigr]_{pq,rs}= \frac{\partial^2 E}{\partial D^{\sigma\tau}_{pq}\partial D^{\tau^\prime \sigma^\prime}_{sr}}
\end{equation}
Apparently, we have a basic symmetry for the kernel:
\begin{equation}
      \bigl[K^{\sigma\tau,\sigma' \tau'}\bigr]_{pq,rs} = \bigl[K^{\tau' \sigma', \tau \sigma}\bigr]_{sr,qp}=\bigl[K^{\sigma' \tau',\sigma\tau}\bigr]_{rs,pq}  
\end{equation}
More explicitly, the RPA kernel is:
\begin{equation}
\begin{split}
    \bigl[K^{\sigma\tau,\sigma' \tau'}\bigr]_{pq,rs}
 &= (p_{\sigma} q_{\tau} \mid s_{\tau'} r_{\sigma'}) 
   - (p_{\sigma} r_{\sigma'} \mid s_{\tau'} q_{\tau}) \\
   &=\bra{p_\sigma s_{\tau^\prime}}\ \ket{q_\tau r_{\sigma^\prime}}
\end{split}
\end{equation}

We define the following short notations:
\begin{equation}
    K^S =\frac{1}{2}(K^{\alpha\alpha,\alpha \alpha}+K^{\alpha\alpha,\beta \beta}+K^{\beta \beta,\alpha \alpha}+K^{\beta \beta,\beta \beta})
\end{equation}
\begin{equation}
    K^P =\frac{1}{2}(K^{\alpha\alpha,\alpha \alpha}-K^{\alpha\alpha,\beta \beta}+K^{\beta \beta,\alpha \alpha}-K^{\beta \beta,\beta \beta})
\end{equation}
\begin{equation}
    K^T =\frac{1}{2}(K^{\alpha\alpha,\alpha \alpha}-K^{\alpha\alpha,\beta \beta}-K^{\beta \beta,\alpha \alpha}+K^{\beta \beta,\beta \beta})
    \label{eq:kt}
\end{equation}
\begingroup
\footnotesize
\setlength{\LTpre}{0.5em}
\setlength{\LTpost}{0.5em}
\begin{longtable}{@{}cl@{}}
\caption{Matrix elements of Eq.~\ref{eq:equal}. Only the upper triangular
parts of $\mathbf{A}$ and $\mathbf{B}$ as well as the CV(1)-CV(1) block of
$\widetilde{\mathbf{A}}$ are listed since $\mathbf{A}$ is Hermitian,
$\mathbf{B}$ is symmetric, and $\widetilde{\mathbf{A}}$ differs from
$\mathbf{A}$ only in the triplet-triplet block. This table was originally proposed by Li and Liu in ref. \citenum{10.1063/1.3463799}.}\label{table:equal}\\
\toprule
Blocks & Matrix elements \\
\midrule
\endfirsthead
\caption[]{Matrix elements of Eq.~\ref{eq:equal} (continued).}\\
\toprule
Blocks & Matrix elements \\
\midrule
\endhead
\bottomrule
\endfoot
CV(0)-CV(0) &
$\begin{aligned}[t]
[\mathbf{A}]_{ai,bj}
&=
\delta_{ij}\left(\frac{f^{\alpha}_{ab}+f^{\beta}_{ab}}{2}\right)
-\delta_{ab}\left(\frac{f^{\alpha}_{ji}+f^{\beta}_{ji}}{2}\right)
+[K^{S}]_{ai,bj}
\end{aligned}$ \\[2ex]

CV(0)-CO(0) &
$\begin{aligned}[t]
[\mathbf{A}]_{ai,vj}
&=
\frac{1}{\sqrt{2}}
\left(
\delta_{ij}f^{\beta}_{av}
+[K^{\alpha\alpha,\beta\beta}]_{ai,vj}
+[K^{\beta\beta,\beta\beta}]_{ai,vj}
\right)
\end{aligned}$ \\[2ex]

CV(0)-OV(0) &
$\begin{aligned}[t]
[\mathbf{A}]_{ai,bv}
&=
\frac{1}{\sqrt{2}}
\left(
-\delta_{ab}f^{\alpha}_{vi}
+[K^{\alpha\alpha,\alpha\alpha}]_{ai,bv}
+[K^{\beta\beta,\alpha\alpha}]_{ai,bv}
\right)
\end{aligned}$ \\[2ex]

CO(0)-CO(0) &
$\begin{aligned}[t]
[\mathbf{A}]_{ui,vj}
&=
\delta_{ij}f^{\beta}_{uv}
-\delta_{uv}f^{\beta}_{ji}
+[K^{\beta\beta,\beta\beta}]_{ui,vj}
\end{aligned}$ \\[2ex]

CO(0)-OV(0) &
$\begin{aligned}[t]
[\mathbf{A}]_{ui,bv}
&=
[K^{\beta\beta,\alpha\alpha}]_{ui,bv}
\end{aligned}$ \\[2ex]

OV(0)-OV(0) &
$\begin{aligned}[t]
[\mathbf{A}]_{au,bv}
&=
\delta_{uv}f^{\alpha}_{ab}
-\delta_{ab}f^{\alpha}_{vu}
+[K^{\alpha\alpha,\alpha\alpha}]_{au,bv}
\end{aligned}$ \\[2ex]

CV(0)-CV(1) &
$\begin{aligned}[t]
[\mathbf{A}]_{ai,bj}
&=
-\sqrt{\frac{S_{\mathrm{i}}+1}{S_{\mathrm{i}}}}
\left[
\delta_{ij}\left(\frac{f^{\alpha}_{ab}-f^{\beta}_{ab}}{2}\right)
-\delta_{ab}\left(\frac{f^{\alpha}_{ji}-f^{\beta}_{ji}}{2}\right)
+[K^{P}]_{ai,bj}
\right]
\end{aligned}$ \\[2ex]

CO(0)-CV(1) &
$\begin{aligned}[t]
[\mathbf{A}]_{ui,bj}
&=
-\sqrt{\frac{S_{\mathrm{i}}+1}{S_{\mathrm{i}}}}
\frac{1}{\sqrt{2}}
\left(
-\delta_{ij}f^{\beta}_{ub}
+[K^{\beta\beta,\alpha\alpha}]_{ui,bj}
-[K^{\beta\beta,\beta\beta}]_{ui,bj}
\right)
\end{aligned}$ \\[2ex]

OV(0)-CV(1) &
$\begin{aligned}[t]
[\mathbf{A}]_{au,bj}
&=
-\sqrt{\frac{S_{\mathrm{i}}+1}{S_{\mathrm{i}}}}
\frac{1}{\sqrt{2}}
\left(
-\delta_{ab}f^{\alpha}_{ju}
+[K^{\alpha\alpha,\alpha\alpha}]_{au,bj}
-[K^{\alpha\alpha,\beta\beta}]_{au,bj}
\right)
\end{aligned}$ \\[2ex]

CV(1)-CV(1) &
$\begin{aligned}[t]
[\mathbf{A}]_{ai,bj}
&=
\delta_{ij}
\left[
\frac{1}{2}\left(1-\frac{1}{S_{\mathrm{i}}}\right)f^{\alpha}_{ab}
+
\frac{1}{2}\left(1+\frac{1}{S_{\mathrm{i}}}\right)f^{\beta}_{ab}
\right]
\\
&\quad
-\delta_{ab}
\left[
\frac{1}{2}\left(1+\frac{1}{S_{\mathrm{i}}}\right)f^{\alpha}_{ji}
+
\frac{1}{2}\left(1-\frac{1}{S_{\mathrm{i}}}\right)f^{\beta}_{ji}
\right]
+[K^{TC}]_{ai,bj}
\end{aligned}$ \\[2ex]

CV(0)-VC(0) &
$\begin{aligned}[t]
[\mathbf{B}]_{ai,jb}
&=
[K^{S}]_{ai,jb}
\end{aligned}$ \\[2ex]

CV(0)-OC(0) &
$\begin{aligned}[t]
[\mathbf{B}]_{ai,jv}
&=
\frac{1}{\sqrt{2}}
\left(
[K^{\alpha\alpha,\beta\beta}]_{ai,jv}
+[K^{\beta\beta,\beta\beta}]_{ai,jv}
\right)
\end{aligned}$ \\[2ex]

CV(0)-VO(0) &
$\begin{aligned}[t]
[\mathbf{B}]_{ai,ub}
&=
\frac{1}{\sqrt{2}}
\left(
[K^{\alpha\alpha,\alpha\alpha}]_{ai,ub}
+[K^{\beta\beta,\alpha\alpha}]_{ai,ub}
\right)
\end{aligned}$ \\[2ex]

CO(0)-OC(0) &
$\begin{aligned}[t]
[\mathbf{B}]_{ui,jv}
&=
[K^{\beta\beta,\beta\beta}]_{ui,jv}
\end{aligned}$ \\[2ex]

CO(0)-VO(0) &
$\begin{aligned}[t]
[\mathbf{B}]_{ui,vb}
&=
\delta_{uv}f^{\beta}_{bi}
+[K^{\beta\beta,\alpha\alpha}]_{ui,vb}
\end{aligned}$ \\[2ex]

OV(0)-VO(0) &
$\begin{aligned}[t]
[\mathbf{B}]_{au,vb}
&=
[K^{\alpha\alpha,\alpha\alpha}]_{au,vb}
\end{aligned}$ \\[2ex]

CV(0)-VC(1) &
$\begin{aligned}[t]
[\mathbf{B}]_{ai,jb}
&=
-\sqrt{\frac{S_{\mathrm{i}}+1}{S_{\mathrm{i}}}}
[K^{P}]_{ai,jb}
\end{aligned}$ \\[2ex]

CO(0)-VC(1) &
$\begin{aligned}[t]
[\mathbf{B}]_{ui,jb}
&=
-\sqrt{\frac{S_{\mathrm{i}}+1}{S_{\mathrm{i}}}}
\frac{1}{\sqrt{2}}
\left(
[K^{\beta\beta,\alpha\alpha}]_{ui,jb}
-[K^{\beta\beta,\beta\beta}]_{ui,jb}
\right)
\end{aligned}$ \\[2ex]

OV(0)-VC(1) &
$\begin{aligned}[t]
[\mathbf{B}]_{au,jb}
&=
-\sqrt{\frac{S_{\mathrm{i}}+1}{S_{\mathrm{i}}}}
\frac{1}{\sqrt{2}}
\left(
[K^{\alpha\alpha,\alpha\alpha}]_{au,jb}
-[K^{\alpha\alpha,\beta\beta}]_{au,jb}
\right)
\end{aligned}$ \\[2ex]

CV(1)-VC(1) &
$\begin{aligned}[t]
[\mathbf{B}]_{ai,jb}
&=
[K^{TC}]_{ai,jb}
\end{aligned}$ \\[2ex]

VC(1)-VC(1) &
$\begin{aligned}[t]
[\widetilde{\mathbf{A}}]_{ai,bj}
&=
\delta_{ij}
\left[
\frac{1}{2}\left(1+\frac{1}{S_{\mathrm{i}}}\right)f^{\alpha}_{ab}
+
\frac{1}{2}\left(1-\frac{1}{S_{\mathrm{i}}}\right)f^{\beta}_{ab}
\right]
\\
&\quad
-\delta_{ab}
\left[
\frac{1}{2}\left(1-\frac{1}{S_{\mathrm{i}}}\right)f^{\alpha}_{ji}
+
\frac{1}{2}\left(1+\frac{1}{S_{\mathrm{i}}}\right)f^{\beta}_{ji}
\right]
+[K^{TC}]_{ai,bj}
\end{aligned}$\\
\end{longtable}
\endgroup

\begingroup
\footnotesize
\setlength{\LTpre}{0.5em}
\setlength{\LTpost}{0.5em}
\begin{longtable}{@{}ll@{}}
\caption{Matrix elements of $A_{+}$, $A_{0}$, and $A_{-}$ not including OO excitations based on EOM condition Eq. \ref{eq:eom1}. Some of the matrix elements were originally proposed by Zhang and Herbert in ref. \citenum{10.1063/1.4937571}.}\label{tab:minus1}\\
\toprule
Blocks & Matrix elements \\
\midrule
\endfirsthead
\caption[]{Matrix elements of $A_{+}$, $A_{0}$, and $A_{-}$ (continued).}\\
\toprule
Blocks & Matrix elements \\
\midrule
\endhead
\bottomrule
\endfoot
CV-CV &
$\begin{aligned}[t]
[A_{+}]_{ai,bj} &= \delta_{ij} f^{\alpha}_{ab}-\delta_{ab} f^{\beta}_{ij}+K^{\alpha\beta,\alpha\beta}_{ai,bj}\\
[A_{0}]_{ai,bj} &= \delta_{ij}\left(f^{\alpha}_{ab}+f^{\beta}_{ab}\right)/2-\delta_{ab}\left(f^{\alpha}_{ij}+f^{\beta}_{ij}\right)/2+K^{T}_{ai,bj}\\
[A_{-}]_{ai,bj} &= \delta_{ij} f^{\beta}_{ab}-\delta_{ab} f^{\alpha}_{ij}+K^{\beta\alpha,\beta\alpha}_{ai,bj}
\end{aligned}$\\[1ex]

VC-VC &
$\begin{aligned}[t]
[A_{+}]_{ia,jb} &= 0\\
[A_{0}]_{ia,jb} &=0\\
[A_{-}]_{ia,jb} &= 0
\end{aligned}$\\[1ex]

CV-CO &
$\begin{aligned}[t]
[A_{+}]_{ai,vj} &= \delta_{ij} f^{\alpha}_{av}\\
[A_{0}]_{ai,vj} &= \left(\delta_{ij} f^{\alpha}_{av}+\delta_{ij} f^{\beta}_{av}-K^{\alpha\alpha,\beta\beta}_{ai,vj}+K^{\beta\beta,\beta\beta}_{ai,vj}\right)/2\\
[A_{-}]_{ai,vj} &= \delta_{ij} f^{\beta}_{av}+K^{\beta\alpha,\beta\alpha}_{ai,vj}
\end{aligned}$\\[1ex]

VC-OC &
$\begin{aligned}[t]
[A_{+}]_{ia,jv} &= 0\\
[A_{0}]_{ia,jv} &= 0\\
[A_{-}]_{ia,jv} &= 0
\end{aligned}$\\[1ex]

CO-CV &
$\begin{aligned}[t]
[A_{+}]_{ui,bj} &= 0\\
[A_{0}]_{ui,bj} &= \left(\delta_{ij} f^{\beta}_{bu}-K^{\beta\beta,\alpha\alpha}_{ui,bj}+K^{\beta\beta,\beta\beta}_{ui,bj}\right)/2\\
[A_{-}]_{ui,bj} &= \delta_{ij} f^{\beta}_{bu}+K^{\beta\alpha,\beta\alpha}_{ui,bj}
\end{aligned}$\\[1ex]

OC-VC &
$\begin{aligned}[t]
[A_{+}]_{iu,jb} &=0 \\
[A_{0}]_{iu,jb} &= 0  \\
[A_{-}]_{iu,jb} &= 0
\end{aligned}$\\[1ex]

CV-OV &
$\begin{aligned}[t]
[A_{+}]_{ai,bv} &= -\delta_{ab} f^{\beta}_{iv}\\
[A_{0}]_{ai,bv} &= \left(-\delta_{ab} f^{\alpha}_{iv}-\delta_{ab} f^{\beta}_{iv}+K^{\alpha\alpha,\alpha\alpha}_{ai,bv}-K^{\beta\beta,\alpha\alpha}_{ai,bv}\right)/2\\
[A_{-}]_{ai,bv} &= -\delta_{ab} f^{\alpha}_{iv}+K^{\beta\alpha,\beta\alpha}_{ai,bv}
\end{aligned}$\\[1ex]

VC-VO &
$\begin{aligned}[t]
[A_{+}]_{ia,vb} &= 0\\
[A_{0}]_{ia,vb} &= 0\\
[A_{-}]_{ia,vb} &= 0
\end{aligned}$\\[1ex]

OV-CV &
$\begin{aligned}[t]
[A_{+}]_{au,bj} &= 0\\
[A_{0}]_{au,bj} &= \left(-\delta_{ab} f^{\alpha}_{ju}+K^{\alpha\alpha,\alpha\alpha}_{au,bj}-K^{\alpha\alpha,\beta\beta}_{au,bj}\right)/2\\
[A_{-}]_{au,bj} &= -\delta_{ab} f^{\alpha}_{ju}+K^{\beta\alpha,\beta\alpha}_{au,bj}
\end{aligned}$\\[1ex]

VO-VC &
$\begin{aligned}[t]
[A_{+}]_{ua,jb} &= 0\\
[A_{0}]_{ua,jb} &= 0\\
[A_{-}]_{ua,jb} &= 0
\end{aligned}$\\[1ex]

CO-CO &
$\begin{aligned}[t]
[A_{+}]_{ui,vj} &= 0\\
[A_{0}]_{ui,vj} &= \left(\delta_{ij} f^{\beta}_{uv}-\delta_{uv} f^{\beta}_{ij}+K^{\beta\beta,\beta\beta}_{ui,vj}\right)/2\\
[A_{-}]_{ui,vj} &= \delta_{ij} f^{\beta}_{uv}-\delta_{uv} f^{\alpha}_{ij}+K^{\beta\alpha,\beta\alpha}_{ui,vj}
\end{aligned}$\\[1ex]

OC-OC &
$\begin{aligned}[t]
[A_{+}]_{iu,jv} &= 0\\
[A_{0}]_{iu,jv} &= 0\\
[A_{-}]_{iu,jv} &=0
\end{aligned}$\\[1ex]

CO-OV &
$\begin{aligned}[t]
[A_{+}]_{ui,bv} &= 0\\
[A_{0}]_{ui,bv} &= -K^{\beta\beta,\alpha\alpha}_{ui,bv}/2\\
[A_{-}]_{ui,bv} &= K^{\beta\alpha,\beta\alpha}_{ui,bv}
\end{aligned}$\\[1ex]

OC-VO &
$\begin{aligned}[t]
[A_{+}]_{iu,vb} &= 0\\
[A_{0}]_{iu,vb} &=0 \\
[A_{-}]_{iu,vb} &= 0
\end{aligned}$\\[1ex]

OV-CO &
$\begin{aligned}[t]
[A_{+}]_{au,vj} &= 0\\
[A_{0}]_{au,vj} &= -K^{\alpha\alpha,\beta\beta}_{au,vj}/2\\
[A_{-}]_{au,vj} &= K^{\beta\alpha,\beta\alpha}_{au,vj}
\end{aligned}$\\[1ex]

VO-OC &
$\begin{aligned}[t]
[A_{+}]_{ua,jv} &= 0\\
[A_{0}]_{ua,jv} &= 0\\
[A_{-}]_{ua,jv} &=0
\end{aligned}$\\[1ex]

OV-OV &
$\begin{aligned}[t]
[A_{+}]_{au,bv} &= 0\\
[A_{0}]_{au,bv} &= \left(\delta_{uv} f^{\alpha}_{ab}-\delta_{ab} f^{\alpha}_{uv}+K^{\alpha\alpha,\alpha\alpha}_{au,bv}\right)/2\\
[A_{-}]_{au,bv} &= \delta_{uv} f^{\beta}_{ab}-\delta_{ab} f^{\alpha}_{uv}+K^{\beta\alpha,\beta\alpha}_{au,bv}
\end{aligned}$\\[1ex]

VO-VO &
$\begin{aligned}[t]
[A_{+}]_{ua,vb} &= 0\\
[A_{0}]_{ua,vb} &= 0\\
[A_{-}]_{ua,vb} &= 0
\end{aligned}$\\

\end{longtable}
\endgroup

\begingroup
\footnotesize
\setlength{\LTpre}{0.5em}
\setlength{\LTpost}{0.5em}
\begin{longtable}{@{}ll@{}}
\caption{Matrix elements of $B_{+}$, $B_{0}$, and $B_{-}$ not including OO excitations based on EOM condition Eq. \ref{eq:eom1}.}\label{tab:minus2}\\
\toprule
Blocks & Matrix elements \\
\midrule
\endfirsthead
\caption[]{Matrix elements of $B_{+}$, $B_{0}$, and $B_{-}$ (continued).}\\
\toprule
Blocks & Matrix elements \\
\midrule
\endhead
\bottomrule
\endfoot
CV-VC &
$\begin{aligned}[t]
[B_{+}]_{ai,jb} &= -[K^{\alpha\beta,\alpha\beta}]_{ai,jb}\\
[B_{0}]_{ai,jb} &=-[K^{T}]_{ai,jb} \\
[B_{-}]_{ai,jb} &=-[K^{\beta \alpha, \beta \alpha}]_{ai,jb} \\
\end{aligned}$\\[1ex]

VC-CV &
$\begin{aligned}[t]
[B_{+}]_{ia,bj} &= 0\\
[B_{0}]_{ia,bj} &= 0\\
[B_{-}]_{ia,bj} &= 0\\
\end{aligned}$\\[1ex]

CV-OC &
$\begin{aligned}[t]
[B_{+}]_{ai,jv} &= -[K^{\alpha\beta,\alpha\beta}]_{ai,jv}\\
[B_{0}]_{ai,jv} &=\frac{1}{2}[K^{\alpha \alpha, \beta \beta}-K^{\beta \beta, \beta \beta}]_{ai,jv} \\
[B_{-}]_{ai,jv} &= 0\\
\end{aligned}$\\[1ex]

VC-CO &
$\begin{aligned}[t]
[B_{+}]_{ia,vj} &= 0\\
[B_{0}]_{ia,vj} &= 0\\
[B_{-}]_{ia,vj} &= 0\\
\end{aligned}$\\[1ex]

OC-CV &
$\begin{aligned}[t]
[B_{+}]_{iu,bj} &=0 \\
[B_{0}]_{iu,bj} &=0 \\
[B_{-}]_{iu,bj} &= 0\\
\end{aligned}$\\[1ex]

CO-VC &
$\begin{aligned}[t]
[B_{+}]_{ui,jb} &= 0\\
[B_{0}]_{ui,jb} &= \tfrac{1}{2}[K^{\beta\beta,\alpha\alpha}]_{ui,jb}-\tfrac{1}{2}[K^{\beta\beta,\beta\beta}]_{ui,jb} \\
[B_{-}]_{ui,jb} &= -[K^{\beta\alpha,\beta\alpha}]_{ui,jb}\\
\end{aligned}$\\[1ex]

CO-OC &
$\begin{aligned}[t]
[B_{+}]_{ui,jv} &= 0\\
[B_{0}]_{ui,jv} &=-\tfrac{1}{2}[K^{\beta\beta,\beta\beta}]_{ui,jv} \\
[B_{-}]_{ui,jv} &= 0\\
\end{aligned}$\\[1ex]

OC-CO &
$\begin{aligned}[t]
[B_{+}]_{iu,vj} &= 0\\
[B_{0}]_{iu,vj} &= 0\\
[B_{-}]_{iu,vj} &= 0\\
\end{aligned}$\\[1ex]

CV-VO &
$\begin{aligned}[t]
[B_{+}]_{ai,vb} &=-[K^{\alpha\beta,\alpha\beta}]_{ai,vb} \\
[B_{0}]_{ai,vb} &=  -\tfrac{1}{2}\Bigl(
   [K^{\alpha\alpha,\alpha\alpha}]_{ai,vb}
   - [K^{\beta\beta,\alpha\alpha}]_{ai,vb}
   \Bigr)\\
[B_{-}]_{ai,vb} &= 0\\
\end{aligned}$\\[1ex]

VC-OV &
$\begin{aligned}[t]
[B_{+}]_{ia,bv} &=0 \\
[B_{0}]_{ia,bv} &= 0 \\
[B_{-}]_{ia,bv} &= 0\\
\end{aligned}$\\[1ex]

VO-CV &
$\begin{aligned}[t]
[B_{+}]_{ua,bj} &= 0\\
[B_{0}]_{ua,bj} &= 0\\
[B_{-}]_{ua,bj} &= 0\\
\end{aligned}$\\[1ex]

OV-VC &
$\begin{aligned}[t]
[B_{+}]_{au,jb} &= 0\\
[B_{0}]_{au,jb} &= \tfrac{1}{2}\Bigl(
   -[K^{\alpha\alpha,\alpha\alpha}]_{au,jb}
   + [K^{\alpha\alpha,\beta\beta}]_{au,jb}
   \Bigr)\\
[B_{-}]_{au,jb} &= -[K^{\beta\alpha,\beta\alpha}]_{au,jb}\\
\end{aligned}$\\[1ex]

CO-VO &
$\begin{aligned}[t]
[B_{+}]_{ui,vb} &= 0\\
[B_{0}]_{ui,vb} &= -\frac12\delta_{uv}f_{bi}^{\beta}
 +\frac12[K^{\beta\beta,\alpha\alpha}]_{ui,vb}\\
[B_{-}]_{ui,vb} &= -f^\alpha_{bi}\delta_{uv}\\
\end{aligned}$\\[1ex]

OC-OV &
$\begin{aligned}[t]
[B_{+}]_{iu,bv} &= 0\\
[B_{0}]_{iu,bv} &= 0\\
[B_{-}]_{iu,bv} &=0\\
\end{aligned}$\\[1ex]

VO-CO &
$\begin{aligned}[t]
[B_{+}]_{ua,vj} &= 0\\
[B_{0}]_{ua,vj} &=0 \\
[B_{-}]_{ua,vj} &= 0\\
\end{aligned}$\\[1ex]

OV-OC &
$\begin{aligned}[t]
[B_{+}]_{au,jv} &= 0\\
[B_{0}]_{au,jv} &= \frac12\delta_{uv}f_{aj}^{\alpha}
 +\frac12[K^{\alpha\alpha,\beta\beta}]_{au,jv}\\
[B_{-}]_{au,jv} &= \delta_{uv}f^\beta_{aj}\\
\end{aligned}$\\[1ex]

OV-VO &
$\begin{aligned}[t]
[B_{+}]_{au,vb} &= 0\\
[B_{0}]_{au,vb} &= -\frac{1}{2}[K^{\alpha \alpha, \alpha \alpha}]_{au,vb}\\
[B_{-}]_{au,vb} &= 0\\
\end{aligned}$\\[1ex]

VO-OV &
$\begin{aligned}[t]
[B_{+}]_{ua,bv} &= 0\\
[B_{0}]_{ua,bv} &= 0\\
[B_{-}]_{ua,bv} &=0 \\
\end{aligned}$\\[1ex]

\end{longtable}
\endgroup

\begingroup
\footnotesize
\setlength{\LTpre}{0.5em}
\setlength{\LTpost}{0.5em}
\begin{longtable}{@{}ll@{}}
\caption{Matrix elements of $M_{+}$, $M_{0}$, and $M_{-}$ including OO excitations based on EOM condition Eq. \ref{eq:eom1}. Some of the matrix elements were originally proposed by Zhang and Herbert in ref. \citenum{10.1063/1.4937571}.}\label{tab:minus3}\\
\toprule
Blocks & Matrix elements \\
\midrule
\endfirsthead
\caption[]{Matrix elements of $M_{+}$, $M_{0}$, and $M_{-}$ (continued).}\\
\toprule
Blocks & Matrix elements \\
\midrule
\endhead
\bottomrule
\endfoot
CV-OO &
$\begin{aligned}[t]
[M_{+}]_{ai,vw} &= -K^{\alpha\beta,\alpha\beta}_{ai,vw}\\
[M_{0}]_{ai,vw} &= 0\\
[M_{-}]_{ai,vw} &= K^{\beta\alpha,\beta\alpha}_{ai,vw}
\end{aligned}$\\[1ex]

VC-OO &
$\begin{aligned}[t]
[M_{+}]_{ia,vw} &= 0\\
[M_{0}]_{ia,vw} &= 0\\
[M_{-}]_{ia,vw} &= 0
\end{aligned}$\\[1ex]

OO-CV &
$\begin{aligned}[t]
[M_{+}]_{tu,bj} &= 0\\
[M_{0}]_{tu,bj} &= \delta_{tu}\left(f^{\alpha}_{bj}-f^{\beta}_{bj}\right)/2\\
[M_{-}]_{tu,bj} &= K^{\beta\alpha,\beta\alpha}_{tu,bj}
\end{aligned}$\\[1ex]

OO-VC &
$\begin{aligned}[t]
[M_{+}]_{tu,jb} &= 0\\
[M_{0}]_{tu,jb} &= \frac12\delta_{tu}(f_{bj}^{\beta}-f_{bj}^{\alpha})\\
[M_{-}]_{tu,jb} &=-[K^{\beta\alpha,\beta\alpha}]_{tu,jb}
\end{aligned}$\\[1ex]

CO-OO &
$\begin{aligned}[t]
[M_{+}]_{ui,vw} &= 0\\
[M_{0}]_{ui,vw} &= -\delta_{uv} f^{\beta}_{iw}/2\\
[M_{-}]_{ui,vw} &= -\delta_{uv} f^{\alpha}_{iw}+K^{\beta\alpha,\beta\alpha}_{ui,vw}
\end{aligned}$\\[1ex]

OC-OO &
$\begin{aligned}[t]
[M_{+}]_{iu,vw} &= 0\\
[M_{0}]_{iu,vw} &=0\\
[M_{-}]_{iu,vw} &= 0
\end{aligned}$\\[1ex]

OO-CO &
$\begin{aligned}[t]
[M_{+}]_{tu,vj} &= 0\\
[M_{0}]_{tu,vj} &= -\delta_{tu} f^{\beta}_{vj}/2\\
[M_{-}]_{tu,vj} &= -\delta_{tv} f^{\alpha}_{ju}+K^{\beta\alpha,\beta\alpha}_{tu,vj}
\end{aligned}$\\[1ex]

OO-OC &
$\begin{aligned}[t]
[M_{+}]_{tu,jv} &=0 \\
[M_{0}]_{tu,jv} &=0\\
[M_{-}]_{tu,jv} &= \delta_{uv} f^{\beta}_{tj}
\end{aligned}$\\[1ex]

OV-OO &
$\begin{aligned}[t]
[M_{+}]_{au,vw} &= 0\\
[M_{0}]_{au,vw} &= \delta_{uw} f^{\alpha}_{av}/2\\
[M_{-}]_{au,vw} &= \delta_{uw} f^{\beta}_{av}+K^{\beta\alpha,\beta\alpha}_{au,vw}
\end{aligned}$\\[1ex]

VO-OO &
$\begin{aligned}[t]
[M_{+}]_{ua,vw} &= 0\\
[M_{0}]_{ua,vw} &= 0\\
[M_{-}]_{ua,vw} &= 0
\end{aligned}$\\[1ex]

OO-OV &
$\begin{aligned}[t]
[M_{+}]_{tu,bv} &= 0\\
[M_{0}]_{tu,bv} &= \delta_{ut} f^{\alpha}_{bv}/2\\
[M_{-}]_{tu,bv} &= \delta_{uv} f^{\beta}_{bt}+K^{\beta\alpha,\beta\alpha}_{tu,bv}
\end{aligned}$\\[1ex]

OO-VO &
$\begin{aligned}[t]
[M_{+}]_{tu,vb} &= 0\\
[M_{0}]_{tu,vb} &= 0\\
[M_{-}]_{tu,vb} &= -\delta_{vt} f^{\alpha}_{bu}
\end{aligned}$\\[1ex]

OO-OO &
$\begin{aligned}[t]
[M_{+}]_{tu,vw} &= 0\\
[M_{0}]_{tu,vw} &= 0\\
[M_{-}]_{tu,vw} &= \delta_{uw} f^{\beta}_{tv}-\delta_{tv} f^{\alpha}_{uw}+K^{\beta\alpha,\beta\alpha}_{tu,vw}
\end{aligned}$\\
\end{longtable}
\endgroup

\section{Raw Data of Numerical Results}

\begin{table}[H]
\centering
\caption{Experimental excitation energies and calculated values from SA-TDDFT, SF-CIS, and SF-TDA. All energies are in eV. For HFLYP, BHHLYP, M06-HF, and M06-2X, values in parentheses are the corresponding results with noncollinear functionals.}
\label{tab:lucas}
\begin{tabular}{llrccccc cc}
\hline
System & State & Exp. & \multicolumn{5}{c}{SA-TDDFT} & SF-CIS\textsuperscript{d,f} & SF-TDA\textsuperscript{e,f} \\
\cline{4-8}
       &       &      & HF & HFLYP & BHHLYP & M06-HF & M06-2X &  &  \\
\hline
N      & a$\,{}^2D$           & 2.380\textsuperscript{a} & 2.347 & 2.365 (1.656) & 1.267 (1.440) & 2.346 (1.984)  & 1.362 (2.019) &  2.290 &  2.300 \\
N      & b$\,{}^2P$           & 3.580\textsuperscript{a} & 4.633 & 4.668 (3.250) & 2.298 (3.400) & 4.704 (-0.325) & 2.496 (2.168) &  3.900 &  3.900 \\
P      & a$\,{}^2D$           & 1.410\textsuperscript{a} & 1.334 & 1.347 (0.841) & 0.788 (0.777) & 1.258 (-1.179\textsuperscript{g}) & 0.838 (0.956) &  1.190 &  1.200 \\
P      & b$\,{}^2P$           & 2.320\textsuperscript{a} & 3.104 & 3.135 (1.944) & 1.551 (1.962) & 3.118 (-0.328\textsuperscript{g}) & 1.689 (0.888) &  2.140 &  2.140 \\
O$_2$  & a$\,{}^1\Delta_g$    & 0.980\textsuperscript{b} & 0.471 & 0.474 (0.360) & 0.415 (0.547) & 0.375 (0.309)  & 0.424 (0.800) &  0.890 &  0.920 \\
O$_2$  & b$\,{}^1\Sigma_g^+$  & 1.640\textsuperscript{b} & 2.202 & 2.212 (1.840) & 0.331 (0.874) & 2.171 (-0.424) & 0.321 (0.136) &  1.950 &  1.950 \\
O$_2$  & c$\,{}^1\Sigma_u^-$  & 4.100\textsuperscript{c} & 3.827 & 3.832 (3.553) & 1.166 (2.059) & 2.565 (8.462)  & 1.249 (1.130) & -5.690 & -5.690 \\
S$_2$  & a$\,{}^1\Delta_g$    & 0.540\textsuperscript{b} & 0.103 & 0.104 (0.040) & 0.214 (0.221) & -0.003 (-0.224) & 0.208 (0.388) &  0.420 &  0.440 \\
S$_2$  & b$\,{}^1\Sigma_g^+$  & 0.990\textsuperscript{b} & 1.434 & 1.443 (1.092) & 0.398 (0.571) & 1.374 (-0.183) & 0.324 (0.214) &  1.000 &  1.000 \\
S$_2$  & c$\,{}^1\Sigma_u^-$  & 2.610\textsuperscript{c} & 2.248 & 2.257 (2.029) & 0.757 (1.142) & 1.408 (4.142)  & 0.815 (0.563) & -2.410 & -2.410 \\
\hline
\end{tabular}

\vspace{0.5em}
\begin{flushleft}
\footnotesize
\textsuperscript{a} Experimental data taken from the NIST Atomic Spectra Database.\cite{kramida2023nist_asd}\\
\textsuperscript{b} Experimental data taken from the NIST Diatomic Spectral Database.\cite{lovas2001nist_diatomic}\\
\textsuperscript{c} Experimental data taken from Ref.~\citenum{10.1063/1.5080458}.\\
\textsuperscript{d} Calculated data taken from Ref.~\citenum{10.1063/5.0275059}. SF-CIS refers to a modified variant of TCIS\cite{Zhao18022026}.\\ 
\textsuperscript{e} Calculated data taken from Ref.~\citenum{10.1063/5.0275059}. SF-TDA refers to a modified variant of SA-SF-CIS.\cite{10.1063/1.4937571}\\
\textsuperscript{f} Noncollinear SVWN5 was used.\\
\textsuperscript{g} For P with noncollinear M06-HF, the lowest roots form several threefold near-degenerate groups and no clear fivefold $^2D$ manifold is found among the first ten roots; the assignment is therefore ambiguous.
\end{flushleft}
\end{table}

\clearpage

\begin{table}[H]
\centering
\scriptsize
\setlength{\tabcolsep}{2pt}
\caption{Vertical transition energies for the QUEST 1 data set. TBE means the best estimation. Four methods SA-TDDFT, X-SF-TDA, SA-SF-DFT, and MRSF-TDDFT are computed. The lowest triplet state is chosen as the reference. The output excitation energies were re-zeroed to the singlet ground-state root: $\Delta E(S_n)=\omega(S_n)-\omega(S_0)$; for the triplet-reference rows, $\Delta E(T)=-\omega(S_0)$. Singlet rows were assigned by matching the ascending TBE singlet-state order to the ascending calculated singlet-root order for each molecule. Failed SCF calculations are denoted by ``--''. These data underlie the QUEST1 error statistics reported in the main text.}
\label{tab:quest1}
\resizebox{\textwidth}{!}{%
\begin{tabular}{llr|*{5}{r}|*{5}{r}|r|r}
\hline\hline
Molecule & State\textsuperscript{a} & TBE (eV)\textsuperscript{b}
& \multicolumn{5}{c|}{SA-TDDFT (eV)}
& \multicolumn{5}{c|}{X-SF-TDA (eV)\textsuperscript{c}}
& \multicolumn{1}{c|}{SA-SF-DFT (eV)\textsuperscript{d}}
& \multicolumn{1}{c}{MRSF-TDDFT (eV)\textsuperscript{e}} \\
\cline{4-8}\cline{9-13}\cline{14-14}\cline{15-15}
 & & & HFLYP & HF & BHHLYP & M06-HF & M06-2X
 & SVWN5 & BLYP & B3LYP & BHHLYP & HF
 & BHHLYP & BHHLYP \\
\hline
acetaldehyde & ${}^{1}A^{\prime\prime}(\mathrm{V}; n \to \pi^\ast)$ & 4.31 & 3.65 & 3.62 & 4.54 & 3.97 & 4.45 & 4.42 & 4.14 & 4.08 & 3.74 & 3.43 & 4.60 & 4.18 \\
acetaldehyde & ${}^{3}A^{\prime\prime}(\mathrm{V}; n \to \pi^\ast)$ & 3.97 & 3.89 & 3.85 & 4.48 & 4.27 & 4.42 & 4.10 & 4.01 & 3.91 & 3.58 & 3.94 & 3.80 & 3.25 \\
acetylene & ${}^{1}\Sigma_u^{-}(\mathrm{V}; \pi \to \pi^\ast)$ & 7.10 & 6.69 & 5.98 & 6.85 & 5.14 & 6.06 & -- & 6.51 & 6.44 & 6.11 & 5.99 & 6.94 & 6.32 \\
acetylene & ${}^{1}\Delta_u(\mathrm{V}; \pi \to \pi^\ast)$ & 7.44 & 7.18 & 6.17 & 6.93 & 5.82 & 6.47 & -- & 7.34 & 7.10 & 6.65 & 6.12 & 7.08 & 6.53 \\
ammonia & ${}^{1}A_2(\mathrm{R}; n \to 3s)$ & 6.59 & 5.43 & 5.02 & 7.02 & 5.46 & 6.79 & 7.72 & 7.36 & 6.99 & 6.31 & 4.95 & 7.18 & 6.74 \\
ammonia & ${}^{1}E(\mathrm{R}; n \to 3p)$ & 8.16 & 7.13 & 6.56 & 8.40 & 6.50 & 7.45 & 9.28 & 8.72 & 8.38 & 7.72 & 6.58 & 7.60 & 7.70 \\
ammonia & ${}^{3}A_2(\mathrm{R}; n \to 3s)$ & 6.31 & 5.32 & 4.94 & 6.89 & 5.40 & 6.65 & 7.51 & 7.20 & 6.80 & 6.12 & 4.96 & 6.55 & 6.05 \\
carbon monoxide & ${}^{1}\Pi(\mathrm{V}; n \to \pi^\ast)$ & 8.49 & 8.59 & 8.57 & 8.28 & 6.54 & 7.64 & 6.41 & 6.61 & 7.28 & 7.29 & 8.51 & 8.24 & 7.85 \\
carbon monoxide & ${}^{1}\Sigma^{-}(\mathrm{V}; \pi \to \pi^\ast)$ & 9.92 & 8.87 & 8.90 & 8.31 & 8.06 & 8.23 & 7.86 & 7.76 & 7.74 & 8.25 & 8.97 & 9.44 & 8.37 \\
carbon monoxide & ${}^{3}\Pi(\mathrm{V}; n \to \pi^\ast)$ & 6.28 & 5.84 & 5.85 & 6.92 & 5.61 & 6.81 & 6.21 & 6.70 & 6.30 & 6.42 & 5.92 & 6.14 & 5.94 \\
cyclopropene & ${}^{1}B_1(\mathrm{V}; \sigma \to \pi^\ast)$ & 6.68 & 7.66 & 5.10 & 6.63 & 6.19 & 6.17 & 5.70 & 5.83 & 6.39 & 6.63 & 5.03 & 6.40 & 6.40 \\
cyclopropene & ${}^{1}B_2(\mathrm{V}; \pi \to \pi^\ast)$ & 6.79 & 8.18 & 5.69 & 6.70 & 6.34 & 6.60 & 6.26 & 5.98 & 6.39 & 6.97 & 5.70 & 6.50 & 6.51 \\
cyclopropene & ${}^{3}B_2(\mathrm{V}; \pi \to \pi^\ast)$ & 4.38 & 4.40 & 5.20 & 4.69 & 4.10 & 4.53 & 4.53 & 4.49 & 4.46 & 4.41 & 5.21 & 4.06 & 3.95 \\
diazomethane & ${}^{1}A_2(\mathrm{V}; \pi \to \pi^\ast)$ & 3.14 & 2.17 & 2.16 & 3.06 & 2.86 & 3.24 & 3.57 & 3.16 & 2.93 & 2.46 & 1.93 & 3.11 & 2.75 \\
diazomethane & ${}^{1}B_1(\mathrm{R}; \pi \to 3s)$ & 5.54 & 7.02 & 6.63 & 5.93 & 6.51 & 5.92 & 5.21 & 4.70 & 5.12 & 4.98 & 6.74 & 5.58 & 5.21 \\
diazomethane & ${}^{1}A_1(\mathrm{V}; \pi \to \pi^\ast)$ & 5.90 & 7.07 & 6.82 & 6.01 & 6.80 & 5.95 & 5.62 & 5.27 & 5.16 & 5.48 & 6.92 & 5.86 & 5.34 \\
diazomethane & ${}^{3}A_2(\mathrm{V}; \pi \to \pi^\ast)$ & 2.79 & 2.64 & 2.63 & 3.13 & 3.44 & 3.35 & 3.39 & 3.15 & 2.89 & 2.45 & 2.74 & 2.63 & 2.15 \\
dinitrogen & ${}^{1}\Pi_g(\mathrm{V}; n \to \pi^\ast)$ & 9.34 & 8.87 & 8.84 & 9.05 & 7.89 & 8.71 & -- & 7.85 & 8.49 & 8.66 & 9.05 & 9.13 & 8.88 \\
dinitrogen & ${}^{1}\Sigma_u^{-}(\mathrm{V}; \pi \to \pi^\ast)$ & 9.88 & 9.53 & 9.49 & 9.33 & 8.37 & 8.79 & -- & 8.63 & 8.79 & 8.75 & 9.36 & 9.37 & 9.06 \\
dinitrogen & ${}^{3}\Sigma_u^{+}(\mathrm{V}; \pi \to \pi^\ast)$ & 7.70 & 8.50 & 8.47 & 8.65 & 8.26 & 8.52 & -- & 7.96 & 7.89 & 8.13 & 8.69 & 7.85 & 7.76 \\
ethylene & ${}^{1}B_{3u}(\mathrm{R}; \pi \to 3s)$ & 7.39 & 8.09 & 7.64 & 7.19 & 7.37 & 7.09 & 6.51 & 6.55 & 6.87 & 7.15 & 7.75 & 6.72 & 6.71 \\
ethylene & ${}^{1}B_{1u}(\mathrm{V}; \pi \to \pi^\ast)$ & 7.93 & 8.83 & 8.32 & 7.47 & 7.62 & 7.56 & 7.01 & 6.56 & 7.14 & 7.77 & 8.43 & 7.31 & 7.34 \\
ethylene & ${}^{3}B_{1u}(\mathrm{V}; \pi \to \pi^\ast)$ & 4.54 & 4.49 & 4.47 & 4.92 & 4.23 & 4.73 & 4.96 & 4.87 & 4.77 & 4.65 & 4.58 & 4.29 & 4.17 \\
formaldehyde & ${}^{1}A_2(\mathrm{V}; n \to \pi^\ast)$ & 3.98 & 3.35 & 3.31 & 4.18 & 3.77 & 4.12 & 3.99 & 3.79 & 3.72 & 3.40 & 3.13 & 4.22 & 3.81 \\
formaldehyde & ${}^{1}B_2(\mathrm{R}; n \to 3s)$ & 7.23 & 8.28 & 7.81 & 7.49 & 7.76 & 7.33 & 6.50 & 6.17 & 6.58 & 6.80 & 7.90 & 6.88 & 6.49 \\
formaldehyde & ${}^{3}A_2(\mathrm{V}; n \to \pi^\ast)$ & 3.58 & 3.54 & 3.50 & 4.10 & 4.01 & 4.05 & 3.62 & 3.63 & 3.52 & 3.20 & 3.59 & 3.39 & 2.85 \\
formamide & ${}^{1}A^{\prime\prime}(\mathrm{V}; n \to \pi^\ast)$ & 5.65 & 5.90 & 5.42 & 5.58 & 5.67 & 5.00 & -- & -- & 5.52 & 6.73 & 5.22 & 5.38 & 5.47 \\
formamide & ${}^{1}A^{\prime}(\mathrm{R}; n \to 3s)$ & 6.81 & 7.17 & 6.72 & 7.33 & 5.70 & 7.13 & -- & -- & 6.30 & 8.17 & 6.76 & 6.87 & 6.11 \\
formamide & ${}^{3}A^{\prime\prime}(\mathrm{V}; n \to \pi^\ast)$ & 5.38 & 6.19 & 5.74 & 6.23 & 6.17 & 6.06 & -- & -- & 5.42 & 9.60 & 5.77 & 5.51 & 4.56 \\
hydrogen chloride & ${}^{1}\Pi(\mathrm{CT})$ & 7.84 & 7.42 & 7.13 & 7.78 & 6.50 & 7.29 & -- & -- & 7.85 & 7.49 & 7.03 & 7.81 & 7.80 \\
hydrogen sulfide & ${}^{1}A_2(\mathrm{R}; n \to 4p)$ & 6.18 & 5.70 & 5.33 & 6.28 & 5.24 & 5.81 & -- & -- & 6.07 & 5.90 & 5.23 & 5.90 & 6.01 \\
hydrogen sulfide & ${}^{1}B_1(\mathrm{R}; n \to 4s)$ & 6.24 & 6.02 & 5.91 & 6.49 & 5.54 & 6.31 & -- & -- & 6.40 & 6.26 & 5.93 & 6.45 & 6.35 \\
hydrogen sulfide & ${}^{3}A_2(\mathrm{R}; n \to 4p)$ & 5.81 & 5.59 & 5.28 & 6.32 & 5.39 & 6.13 & -- & -- & 6.11 & 5.95 & 5.30 & 6.05 & 5.70 \\
ketene & ${}^{1}A_2(\mathrm{V}; \pi \to \pi^\ast)$ & 3.85 & 3.21 & 3.20 & 4.04 & 3.52 & 4.10 & 4.25 & 3.84 & 3.74 & 3.47 & 3.01 & 4.13 & 3.86 \\
ketene & ${}^{1}B_1(\mathrm{R}; n \to 3s)$ & 6.01 & 7.05 & 6.66 & 6.24 & 6.68 & 6.22 & 5.71 & 5.16 & 5.54 & 5.81 & 6.76 & 5.93 & 5.69 \\
ketene & ${}^{3}A_2(\mathrm{V}; n \to \pi^\ast)$ & 3.77 & 3.57 & 3.55 & 4.10 & 4.05 & 4.20 & 4.12 & 3.87 & 3.74 & 3.47 & 3.65 & 3.65 & 3.28 \\
methanimine & ${}^{1}A^{\prime\prime}(\mathrm{V}; n \to \pi^\ast)$ & 5.23 & 4.80 & 4.78 & 5.39 & 5.07 & 5.34 & 5.05 & 4.94 & 4.94 & 4.79 & 4.66 & 5.42 & 5.08 \\
methanimine & ${}^{3}A^{\prime\prime}(\mathrm{V}; n \to \pi^\ast)$ & 4.65 & 4.60 & 4.58 & 5.15 & 4.90 & 5.09 & 4.58 & 4.66 & 4.58 & 4.38 & 4.65 & 4.50 & 4.02 \\
nitrosomethane & ${}^{1}A^{\prime\prime}(\mathrm{V}; n \to \pi^\ast)$ & 1.96 & 1.59 & 1.58 & 1.93 & 1.64 & 1.87 & 1.87 & 1.67 & 1.70 & 1.64 & 1.41 & 1.99 & 1.89 \\
nitrosomethane & ${}^{1}A^{\prime}(\mathrm{V}; n,n \to \pi^\ast,\pi^\ast)$ & 4.76 & 4.96 & 4.94 & 3.16 & 4.67 & 3.01 & 4.69 & 4.61 & 4.77 & 4.86 & 4.98 & 4.75 & 4.77 \\
nitrosomethane & ${}^{3}A^{\prime\prime}(\mathrm{V}; n \to \pi^\ast)$ & 1.16 & 1.48 & 1.47 & 1.61 & 1.58 & 1.56 & 1.20 & 1.22 & 1.15 & 1.03 & 1.62 & 0.89 & 0.65 \\
streptocyanine-C1 & ${}^{1}B_2(\mathrm{V}; \pi \to \pi^\ast)$ & 7.13 & 7.22 & 7.19 & 7.15 & 6.53 & 7.00 & 6.43 & 6.26 & 6.71 & 7.31 & 7.20 & 7.13 & 7.59 \\
streptocyanine-C1 & ${}^{3}B_2(\mathrm{V}; \pi \to \pi^\ast)$ & 5.52 & 5.44 & 5.42 & 5.94 & 5.09 & 5.75 & 5.82 & 5.63 & 5.57 & 5.44 & 5.54 & 5.28 & 5.13 \\
thioformaldehyde & ${}^{1}B_1(\mathrm{R}; n \to 3s)$ & 2.22 & 1.65 & 1.65 & 2.38 & 1.68 & 2.32 & 2.43 & 2.25 & 2.11 & 1.84 & 1.45 & 2.46 & 2.09 \\
thioformaldehyde & ${}^{1}B_2(\mathrm{R}; n \to 4s)$ & 5.96 & 6.89 & 6.53 & 6.18 & 6.15 & 6.05 & 5.34 & 5.23 & 4.98 & 4.87 & 6.60 & 5.86 & 5.52 \\
thioformaldehyde & ${}^{3}A_2(\mathrm{V}; n \to \pi^\ast)$ & 1.95 & 1.95 & 1.95 & 2.36 & 2.08 & 2.31 & 2.21 & 2.18 & 2.01 & 1.75 & 2.03 & 1.93 & 1.46 \\
water & ${}^{1}B_1(\mathrm{R}; n \to 3s)$ & 7.62 & 6.18 & 5.74 & 8.00 & 6.23 & 7.98 & 9.10 & 8.55 & 8.05 & 7.15 & 5.65 & 8.14 & 7.67 \\
water & ${}^{1}A_2(\mathrm{R}; n \to 3p)$ & 9.41 & 8.20 & 7.63 & 9.41 & 6.98 & 8.90 & 10.74 & 10.04 & 9.59 & 8.75 & 7.66 & 8.53 & 8.74 \\
water & ${}^{3}B_1(\mathrm{R}; n \to 3s)$ & 7.25 & 6.06 & 5.65 & 7.84 & 6.13 & 7.81 & 8.80 & 8.34 & 7.81 & 6.90 & 5.68 & 7.33 & 6.78 \\
\hline\hline
\end{tabular}}

\vspace{0.5em}
\begin{minipage}{0.98\textwidth}
\scriptsize
\textsuperscript{a} Characterized as valence (V), Rydberg (R), or charge transfer (CT), according to Ref.~\cite{Loos2018Mountaineering}.\\
\textsuperscript{b} TBE values are from Ref.~\cite{Loos2018Mountaineering}.\\
\textsuperscript{c} X-SF-TDA uses the aug-cc-pVTZ basis set.\\
\textsuperscript{d} The SA-SF-DFT calculations use the BHHLYP functional and the aug-cc-pVTZ basis set. Results are directly taken from Ref.~\citenum{10.1063/5.0327478}.\\
\textsuperscript{e} The MRSF-TDDFT values use the BHHLYP functional, and the aug-cc-pVTZ basis set. 
\end{minipage}
\end{table}

\clearpage

\begin{table}[H]
\centering
\scriptsize
\setlength{\tabcolsep}{2pt}
\caption{Lowest doublet and quartet vertical transition energies for nonlinear open-shell molecules in the QUEST data set, compared with X-TDA, SA-TDDFT, and SA-TDA results. These data underlie the corresponding error statistics reported in the main text.}
\label{tab:quest_rad}
\resizebox{\textwidth}{!}{%
\begin{tabular}{llr|*{5}{r}|*{5}{r}|*{5}{r}}
\hline\hline
Molecule & State & TBE (eV)\textsuperscript{a}
& \multicolumn{5}{c|}{X-TDA (eV)}
& \multicolumn{5}{c|}{SA-TDDFT (eV)\textsuperscript{b}}
& \multicolumn{5}{c}{SA-TDA (eV)\textsuperscript{b}} \\
\cline{4-8}\cline{9-13}\cline{14-18}
 & & & SVWN5 & BLYP & BHHLYP & B3LYP & HF
 & HFLYP & HF & BHHLYP & M06-HF & M06-2X
 & HFLYP & HF & BHHLYP & M06-HF & M06-2X \\
\hline
$\mathrm{AlCH_2}$ & ${}^{2}A_1$ & 0.95 & 0.62 & 1.01 & 1.56 & 1.30 & 1.71 & 0.20 & 0.20 & 0.73 & 0.49 & 0.43 & 0.84 & 0.83 & 0.86 & 0.16 & 0.67 \\
$\mathrm{AlCH_2}$ & ${}^{4}A_2$ & 2.10 & 2.07 & 2.08 & 2.02 & 2.08 & 1.42 & 1.05 (1.52) & 1.02 & 2.07 (1.75) & -0.35 (1.47) & 1.66 (1.96) & 1.46 (1.95) & 1.42 & 2.21 (2.02) & 0.58 (1.70) & 1.87 (2.08) \\
Allyl & ${}^{2}B_1$ & 3.42 & 2.62 & 2.98 & 3.89 & 3.32 & 4.94 & 3.68 & 1.36\textsuperscript{c} & 3.21 & 2.40 & 2.86 & 3.28 & 3.29 & 3.27 & 2.50 & 2.95 \\
Allyl & ${}^{4}A_2$ & 6.01 & 5.42 & 5.31 & 5.36 & 5.38 & 4.96 & 3.61 (4.40) & 3.64 & 5.58 (4.86) & 3.70 (4.64) & 4.97 (5.30) & 4.98 (5.35) & 4.96 & 5.78 (5.36) & 3.10 (4.90) & 5.25 (5.51) \\
$\mathrm{BH_2}$ & ${}^{2}B_1$ & 1.18 & 1.09 & 1.33 & 1.30 & 1.31 & 1.42 & 1.09 & 1.13 & 1.09 & 0.03 & 0.90 & 1.24 & 1.28 & 1.14 & 0.65 & 0.98 \\
$\mathrm{BH_2}$ & ${}^{4}A_2$ & 5.37 & 4.69 & 5.05 & 5.40 & 5.19 & 5.29 & 5.11 (5.68) & 5.15 & 5.64 (5.33) & 3.31 (4.74) & 4.91 (5.04) & 5.25 (5.76) & 5.29 & 5.66 (5.40) & 3.47 (4.82) & 4.95 (5.05) \\
$\mathrm{CH_2N}$ & ${}^{2}B_2$ & 3.96 & 4.07 & 4.32 & 4.03 & 4.22 & 3.83 & 2.21 & 2.37 & 3.31 & 2.51 & 3.23 & 3.51 & 3.50 & 3.47 & 3.21 & 3.43 \\
$\mathrm{CH_2N}$ & ${}^{4}B_1$ & 5.30 & 5.51 & 5.30 & 4.80 & 5.12 & 3.98 & 0.98 (2.57) & 1.13 & 5.00 (4.02) & 3.75 (5.11) & 4.40 (5.09) & 3.96 (4.41) & 3.98 & 5.39 (4.80) & 2.73 (5.38) & 4.93 (5.34) \\
$\mathrm{CH_2O^+}$ & ${}^{2}B_1$ & 3.82 & 4.66 & 4.82 & 3.83 & 4.42 & 2.99 & 1.88 & 1.89 & 2.81 & 1.44 & 2.75 & 2.81 & 2.79 & 3.14 & 2.68 & 3.19 \\
$\mathrm{CH_2O^+}$ & ${}^{4}B_2$ & 6.93 & 6.73 & 6.58 & 6.72 & 6.60 & 6.92 & 5.70 (6.34) & 5.72 & 7.01 (6.15) & 2.09 (5.76) & 6.22 (6.12) & 6.91 (7.28) & 6.92 & 7.32 (6.72) & 4.71 (6.35) & 6.65 (6.54) \\
$\mathrm{CH_3}$ & ${}^{2}A_1^{\prime}$ & 5.88 & 5.14 & 4.82 & 5.72 & 5.29 & 6.26 & 6.46 & 6.10 & 5.28 & 6.23 & 5.53 & 6.61 & 6.22 & 5.31 & 6.40 & 5.56 \\
$\mathrm{ClO_2}$ & ${}^{2}B_2$ & 2.96 & 3.28 & 3.33 & 3.01 & 3.17 & 2.42 & 0.78\textsuperscript{c} & 4.44 & 2.01 & 2.41 & 1.68 & 2.04 & 2.02 & 2.64 & 1.64 & 2.56 \\
$\mathrm{ClO_2}$ & ${}^{4}B_2$ & 6.24 & 6.09 & 5.97 & 5.80 & 6.06 & 5.51 & 5.28 (5.49) & 5.30 & 6.18 (5.72) & 3.41 (5.38) & 5.63 (5.90) & 5.49 (5.67) & 5.51 & 6.23 (5.80) & 3.69 (5.50) & 5.69 (5.94) \\
$\mathrm{F_2BO}$ & ${}^{2}B_1$ & 0.70 & 1.37 & 1.85 & 1.17 & 1.59 & 0.68 & 2.35 & 2.34 & 0.33 & -0.52 & 2.12 & 0.46 & 0.48 & 0.43 & -0.02 & 0.26 \\
$\mathrm{F_2BS}$ & ${}^{2}B_1$ & 0.48 & 0.71 & 0.89 & 0.59 & 0.77 & 0.48 & 2.43 & 0.01 & 0.17 & -0.20 & 0.01 & 0.25 & 0.26 & 0.28 & 0.00 & 0.18 \\
$\mathrm{F_2BS}$ & ${}^{4}B_2$ & 6.40 & 5.48 & 5.56 & 6.27 & 5.86 & 6.60 & 6.27 (6.68) & 6.33 & 6.45 (6.16) & 4.47 (5.90) & 5.96 (6.03) & 6.55 (6.88) & 6.60 & 6.51 (6.27) & 4.89 (6.03) & 6.04 (6.09) \\
$\mathrm{H_2BO}$ & ${}^{2}B_1$ & 2.16 & 3.10 & 3.27 & 2.39 & 2.92 & 1.60 & 0.68 & 0.68 & 1.47 & 0.25 & 1.30 & 1.42 & 1.41 & 1.72 & 1.12 & 1.66 \\
$\mathrm{H_2BO}$ & ${}^{4}B_2$ & 6.92 & 6.27 & 6.33 & 7.50 & 6.75 & 7.56 & 7.42 (7.74) & 7.44 & 7.68 (7.28) & 5.48 (7.06) & 7.12 (6.86) & 7.54 (7.81) & 7.56 & 7.81 (7.50) & 6.24 (7.34) & 7.29 (7.08) \\
$\mathrm{H_2PO}$ & ${}^{2}A^{\prime\prime}$ & 2.85 & 2.78 & 2.92 & 2.98 & 2.92 & 3.83 & 2.03 & 1.92 & 2.24 & 0.98\textsuperscript{c} & 2.06 & 3.03 & 2.95 & 2.50 & 2.20 & 2.39 \\
$\mathrm{H_2PO}$ & ${}^{4}A^{\prime\prime}$ & 6.33 & 5.80 & 5.85 & 6.78 & 6.26 & 7.23 & 7.09 (7.43) & 7.05 & 6.92 (6.71) & 5.17 (6.32) & 6.41 (6.55) & 7.28 (7.58) & 7.23 & 6.96 (6.78) & 5.46 (6.43) & 6.46 (6.59) \\
$\mathrm{H_2PS}$ & ${}^{2}A^{\prime\prime}$ & 1.16 & 1.35 & 1.44 & 1.21 & 1.36 & 0.89 & -0.03 & 3.64 & 0.79 & 2.76 & 0.67 & 0.71 & 0.71 & 0.95 & 0.50 & 0.88 \\
$\mathrm{H_2PS}$ & ${}^{4}A^{\prime\prime}$ & 5.14 & 4.65 & 4.61 & 5.32 & 4.93 & 5.63 & 5.42 (5.83) & 5.41 & 5.45 (5.24) & 3.88 (4.96) & 5.00 (5.14) & 5.64 (5.98) & 5.63 & 5.49 (5.32) & 4.21 (5.04) & 5.05 (5.17) \\
HBCl & ${}^{2}A^{\prime\prime}$ & 1.70 & 2.02 & 2.40 & 2.19 & 2.32 & 2.06 & 1.63 & 1.67 & 1.59 & 0.92 & 1.38 & 1.84 & 1.88 & 1.67 & 1.29 & 1.49 \\
HBCl & ${}^{4}A^{\prime\prime}$ & 5.95 & 5.12 & 5.26 & 6.03 & 5.56 & 6.31 & 6.13 (6.63) & 6.15 & 6.17 (5.97) & 4.44 (5.38) & 5.62 (5.59) & 6.30 (6.74) & 6.31 & 6.20 (6.03) & 4.66 (5.48) & 5.66 (5.62) \\
HCO & ${}^{2}A^{\prime\prime}$ & 2.09 & 1.96 & 2.16 & 2.26 & 2.19 & 2.47 & 1.63 & 1.66 & 1.97 & 0.79 & 1.68 & 2.13 & 2.16 & 2.09 & 1.35 & 1.84 \\
HCO & ${}^{4}A^{\prime\prime}$ & 6.41 & 6.54 & 6.42 & 6.06 & 6.28 & 5.62 & 3.87 (4.88) & 3.88 & 6.48 (5.75) & 3.29 (5.64) & 5.77 (6.16) & 5.62 (5.97) & 5.62 & 6.54 (6.06) & 3.62 (5.80) & 5.85 (6.20) \\
HOC & ${}^{2}A^{\prime\prime}$ & 0.91 & 0.93 & 1.12 & 0.93 & 1.03 & 0.83 & 4.67 & 0.00 & 0.65 & -0.33 & 0.36 & 0.64 & 0.67 & 0.74 & 0.15 & 0.53 \\
HOC & ${}^{4}A^{\prime\prime}$ & 3.86 & 3.89 & 3.98 & 3.66 & 3.87 & 2.79 & 1.80 (2.80) & 1.91 & 4.18 (3.29) & 5.07 (3.38) & 3.24 (3.62) & 2.71 (3.36) & 2.79 & 4.34 (3.67) & 0.79 (3.48) & 3.46 (3.72) \\
$\mathrm{NH_2}$ & ${}^{2}A_1$ & 2.11 & 2.03 & 2.32 & 2.20 & 2.26 & 2.24 & 1.50 & 1.51 & 1.84 & 0.80 & 1.57 & 2.02 & 2.04 & 1.96 & 1.41 & 1.74 \\
$\mathrm{NH_2}$ & ${}^{4}B_1$ & 7.34 & 6.34 & 6.20 & 7.29 & 6.72 & 7.71 & 7.83 (8.21) & 7.58 & 7.47 (7.25) & 6.07 (7.11) & 7.04 (7.23) & 7.98 (8.31) & 7.71 & 7.49 (7.29) & 6.25 (7.18) & 7.06 (7.25) \\
$\mathrm{NO_2}$ & ${}^{2}B_1$ & 2.87 & 3.23 & 3.44 & 3.21 & 3.33 & 3.02 & 0.68\textsuperscript{c} & 1.87 & 2.41 & 0.89\textsuperscript{c} & 2.17 & 2.72 & 2.74 & 2.78 & 2.14 & 2.59 \\
$\mathrm{NO_2}$ & ${}^{4}A_2$ & 4.62 & 4.67 & 4.58 & 4.21 & 4.39 & 3.95 & 0.86 (1.87) & 1.04 & 4.37 (3.47) & 2.12 (3.19) & 3.71 (3.98) & 3.91 (4.18) & 3.95 & 4.56 (4.21) & 2.17 (3.50) & 4.09 (4.16) \\
Nitromethyl & ${}^{2}B_2$ & 2.00 & 1.12 & 1.40 & 3.08 & 2.15 & 3.78 & 1.42\textsuperscript{c} & 2.02 & 1.79 & 1.25 & 1.50 & 1.66 & 1.66 & 1.94 & 0.63 & 1.69 \\
Nitromethyl & ${}^{4}A_2$ & 4.34 & 3.95 & 3.90 & 3.67 & 4.00 & 3.36 & 1.07 (1.76) & 1.12 & 3.76 (3.03) & 3.16 (3.19) & 3.20 (3.62) & 3.35 (3.53) & 3.36 & 4.11 (3.67) & 1.88 (3.76) & 3.69 (3.97) \\
$\mathrm{PH_2}$ & ${}^{2}A_1$ & 2.76 & 2.64 & 2.81 & 2.81 & 2.81 & 2.85 & 1.84 & 1.82 & 2.56 & 0.96 & 2.34 & 2.65 & 2.64 & 2.71 & 2.17 & 2.52 \\
$\mathrm{PH_2}$ & ${}^{4}A_2$ & 6.15 & 6.19 & 6.04 & 6.06 & 6.09 & 5.51 & 5.04 (5.68) & 4.99 & 6.35 (5.84) & 3.68 (6.04) & 5.83 (6.28) & 5.56 (6.05) & 5.51 & 6.45 (6.06) & 4.36 (6.18) & 5.96 (6.34) \\
Vinyl & ${}^{2}A^{\prime\prime}$ & 3.28 & 3.38 & 3.48 & 3.30 & 3.41 & 3.56 & 2.08 & 2.28 & 2.69 & 2.49 & 2.75 & 3.11 & 3.07 & 2.81 & 2.98 & 2.90 \\
Vinyl & ${}^{4}A^{\prime}$ & 4.56 & 4.75 & 4.57 & 4.22 & 4.44 & 3.45 & 1.09 (2.57) & 1.25 & 4.38 (3.59) & 6.37 (4.47) & 3.79 (4.44) & 3.42 (3.94) & 3.45 & 4.70 (4.22) & 2.20 (4.70) & 4.24 (4.63) \\
\hline\hline
\end{tabular}%
}

\begin{flushleft}
\textsuperscript{a} TBE/AVQZ values are used when available; otherwise the QUEST TBE/AVTZ value is listed.\\
\textsuperscript{b} For SA-TDDFT and SA-TDA, doublet entries use the lowest equal-sector root with spin-unpolarized pure parts. Quartet entries are reported as spin-unpolarized values with noncollinear functional values in parentheses; HF has no spin-unpolarized variant.\\
\textsuperscript{c} The displayed SA-TDDFT root has $|\operatorname{Im}\omega| > 0.05~\mathrm{Ha}$ before taking the real part.
\end{flushleft}
\end{table}

\clearpage

\begin{table}[H]
\caption{\label{tab:cr2}
Calculated near-equilibrium relative energies of Cr$_2$ used for the
potential-energy curve in the main text. Missing entries indicate problematic SCF calculations.}
\footnotesize
\setlength{\tabcolsep}{5pt}
\renewcommand{\arraystretch}{0.95}
\begin{ruledtabular}
\begin{tabular}{ccc}
$R$ (\AA) & BHHLYP & M06-2X \\
 & $\Delta E$ (eV) & $\Delta E$ (eV) \\
1.51 & -- & 1.2332 \\
1.52 & 1.1524 & 1.0840 \\
1.53 & 1.0101 & 0.9474 \\
1.54 & -- & 0.8228 \\
1.56 & -- & 0.6066 \\
1.57 & 0.5559 & 0.5138 \\
1.58 & 0.4680 & 0.4302 \\
1.59 & 0.3893 & 0.3556 \\
1.60 & 0.3192 & -- \\
1.61 & -- & 0.2312 \\
1.62 & 0.2030 & 0.1804 \\
1.63 & 0.1561 & 0.1368 \\
1.64 & 0.1160 & 0.0998 \\
1.65 & 0.0824 & -- \\
1.66 & -- & 0.0444 \\
1.67 & 0.0333 & 0.0253 \\
1.68 & 0.0171 & 0.0116 \\
1.69 & 0.0061 & 0.0030 \\
1.70 & 0.0000 & -- \\
1.71 & -- & 0.0000 \\
1.72 & 0.0015 & 0.0051 \\
1.73 & 0.0086 & -- \\
1.74 & 0.0196 & 0.0268 \\
1.75 & 0.0343 & -- \\
1.76 & -- & 0.0634 \\
1.77 & 0.0742 & 0.0868 \\
1.78 & 0.0989 & 0.1133 \\
1.79 & 0.1267 & 0.1429 \\
1.80 & 0.1573 & -- \\
1.81 & -- & 0.2103 \\
1.82 & 0.2264 & 0.2479 \\
1.83 & 0.2646 & 0.2880 \\
1.84 & 0.3052 & 0.3304 \\
1.85 & 0.3479 & -- \\
1.86 & -- & 0.4217 \\
1.87 & 0.4394 & 0.4704 \\
1.88 & 0.4879 & 0.5209 \\
1.89 & 0.5382 & 0.5732 \\
1.90 & 0.5902 & -- \\
1.91 & -- & 0.6828 \\
1.92 & 0.6987 & 0.7399 \\
1.93 & 0.7551 & 0.7984 \\
1.94 & 0.8128 & 0.8583 \\
1.95 & 0.8717 & -- \\
1.96 & -- & 0.9818 \\
1.97 & 0.9930 & 1.0452 \\
1.98 & 1.0552 & -- \\
1.99 & 1.1183 & -- \\
2.00 & 1.1824 & -- \\
\end{tabular}
\end{ruledtabular}
\end{table}

\clearpage

\begin{table}[H]
\centering
\scriptsize
\setlength{\tabcolsep}{3.6pt}
\renewcommand{\arraystretch}{1.03}
\caption{
Relative energies $\Delta E$ of the three lowest singlet states of phenol along the
O-H dissociation coordinate. For each scan point, singlet total energies were
obtained by SA-TDDFT with BHHLYP. The reported values are shifted so that the
minimum over all listed $R_{\mathrm{OH}}$ values and $S_0$-$S_2$ roots is zero.
This zero occurs at $R_{\mathrm{OH}}=1.00$~\AA{} for the $S_0$ root. All energies
are in eV. These data are plotted in the phenol O--H scan in the main text.
}
\label{tab:phenol}
\begin{ruledtabular}
\begin{tabular}{c ccc@{\hspace{1.4em}}c ccc}
\multicolumn{4}{c}{BHHLYP} & \multicolumn{4}{c}{BHHLYP} \\
$R_{\mathrm{OH}}$ (\AA) & $S_0$ & $S_1$ & $S_2$ &
$R_{\mathrm{OH}}$ (\AA) & $S_0$ & $S_1$ & $S_2$ \\
0.50 & 21.225906 & 26.737981 & 27.024258 & 2.35 & 4.764753 & 4.832176 & 6.995958 \\
0.60 & 9.082653 & 14.573488 & 14.854520 & 2.36 & 4.773845 & 4.830829 & 6.999221 \\
0.70 & 3.392980 & 8.912211 & 9.081021 & 2.37 & 4.782679 & 4.829521 & 7.002427 \\
0.80 & 0.899760 & 6.346502 & 6.628732 & 2.38 & 4.791260 & 4.828251 & 7.005576 \\
0.90 & 0.033966 & 5.458904 & 5.743281 & 2.39 & 4.799597 & 4.827018 & 7.008663 \\
1.00 & 0.000000 & 5.399176 & 5.697055 & 2.40 & 4.807693 & 4.825822 & 7.011687 \\
1.01 & 0.024161 & 5.431800 & 5.706518 & 2.41 & 4.815557 & 4.824661 & 7.014641 \\
1.02 & 0.051952 & 5.449850 & 5.741952 & 2.42 & 4.823193 & 4.823534 & 7.017542 \\
1.03 & 0.083419 & 5.477687 & 5.773767 & 2.43 & 4.822442 & 4.830609 & 7.020337 \\
1.04 & 0.117804 & 5.505042 & 5.813411 & 2.44 & 4.821382 & 4.837808 & 7.023034 \\
1.05 & 0.155817 & 5.549528 & 5.838790 & 2.45 & 4.820355 & 4.844799 & 7.025615 \\
1.06 & 0.195520 & 5.579475 & 5.857996 & 2.46 & 4.819359 & 4.851585 & 7.028053 \\
1.07 & 0.238194 & 5.612330 & 5.871674 & 2.47 & 4.818394 & 4.858172 & 7.030312 \\
1.08 & 0.282901 & 5.659496 & 5.884529 & 2.48 & 4.817458 & 4.864567 & 7.032351 \\
1.09 & 0.330978 & 5.707681 & 5.896683 & 2.49 & 4.816552 & 4.870773 & 7.034138 \\
1.10 & 0.380964 & 5.754270 & 5.910607 & 2.50 & 4.815674 & 4.876797 & 7.035665 \\
1.11 & 0.432421 & 5.803898 & 5.924002 & 2.60 & 4.808293 & 4.928046 & 7.043844 \\
1.12 & 0.485354 & 5.854834 & 5.936927 & 2.70 & 4.803011 & 4.965909 & 7.048964 \\
1.13 & 0.539792 & 5.907152 & 5.949390 & 2.80 & 4.799290 & 4.993830 & 7.053189 \\
1.14 & 0.595660 & 5.960912 & 5.961308 & 2.90 & 4.796732 & 5.014220 & 7.057297 \\
1.15 & 0.652855 & 5.972586 & 6.015992 & 3.00 & 4.795009 & 5.029290 & 7.060484 \\
1.16 & 0.711248 & 5.983112 & 6.072241 & 3.10 & 4.793910 & 5.040372 & 7.063253 \\
1.17 & 0.770704 & 5.992787 & 6.129562 & 3.20 & 4.793287 & 5.048548 & 7.065717 \\
1.18 & 0.831098 & 6.001542 & 6.187921 & 3.30 & 4.793001 & 5.054555 & 7.067938 \\
1.19 & 0.892360 & 6.009381 & 6.247284 & 3.40 & 4.792954 & 5.059001 & 7.069926 \\
1.20 & 0.954408 & 6.016286 & 6.307640 & 3.50 & 4.793063 & 5.062295 & 7.071684 \\
1.30 & 1.306942 & 5.575989 & 6.697940 & 3.60 & 4.793254 & 5.064715 & 7.073211 \\
1.40 & 1.855265 & 5.434673 & 6.776218 & 3.70 & 4.793495 & 5.066529 & 7.074517 \\
1.50 & 2.381052 & 5.295462 & 6.782084 & 3.80 & 4.793744 & 5.068407 & 7.074602 \\
1.60 & 2.864264 & 5.177135 & 6.771765 & 3.90 & 4.793984 & 5.069404 & 7.075508 \\
1.70 & 3.293308 & 5.083103 & 6.771410 & 4.00 & 4.794199 & 5.070133 & 7.076268 \\
1.80 & 3.663371 & 5.010634 & 6.787718 & 4.10 & 4.794374 & 5.070652 & 7.076886 \\
1.90 & 3.974333 & 4.955484 & 6.818233 & 4.20 & 4.794507 & 5.071000 & 7.077383 \\
2.00 & 4.229207 & 4.913654 & 6.857428 & 4.30 & 4.794612 & 5.071235 & 7.077784 \\
2.10 & 4.433150 & 4.881906 & 6.899745 & 4.40 & 4.794682 & 5.071392 & 7.078092 \\
2.20 & 4.592731 & 4.857795 & 6.941056 & 4.50 & 4.794714 & 5.071462 & 7.078325 \\
2.30 & 4.715185 & 4.839534 & 6.978848 & 4.60 & 4.794721 & 5.071482 & 7.078496 \\
2.31 & 4.725669 & 4.837976 & 6.982372 & 4.70 & 4.794708 & 5.071478 & 7.078611 \\
2.32 & 4.735861 & 4.836462 & 6.985847 & 4.80 & 4.794679 & 5.071448 & 7.078687 \\
2.33 & 4.745768 & 4.834992 & 6.989270 & 4.90 & 4.794637 & 5.071393 & 7.078736 \\
2.34 & 4.755396 & 4.833563 & 6.992641 & 5.00 & 4.794587 & 5.071335 & 7.078758 \\
\end{tabular}
\end{ruledtabular}
\end{table}

\clearpage

\begin{table}[H]
\centering
\scriptsize
\setlength{\tabcolsep}{3pt}
\renewcommand{\arraystretch}{1.03}
\caption{
Total energies and effective numbers of unpaired electrons for the lowest
singlet state of H$_2$ along the H--H bond-stretching coordinate. Distances are in \AA{} and total energies are in hartree. These data are plotted in the H$_2$ bond-stretching figure in the main text.
}
\label{tab:h2_natural_orbitals}
\begin{ruledtabular}
\begin{tabular}{c rcc rcc rcc}
& \multicolumn{3}{c}{EOM--SF--CCSD}
& \multicolumn{3}{c}{SF--TDDFT/B5050LYP}
& \multicolumn{3}{c}{SA--TDA/BHHLYP} \\
$R_{\mathrm{H-H}}$ & $E_{S_0}$ & $n_u$ & $n_{u,\mathrm{nl}}$
& $E_{S_0}$ & $n_u$ & $n_{u,\mathrm{nl}}$
& $E_{S_0}$ & $n_u$ & $n_{u,\mathrm{nl}}$ \\
0.74 & -1.1723 & 0.071 & 0.007 & -0.9844 & 0.046 & 0.003 & -1.192045587 & 0.018385 & 0.000487 \\
0.84 & -1.1672 & 0.085 & 0.010 & -0.9813 & 0.050 & 0.004 & -1.181441135 & 0.022863 & 0.000793 \\
0.94 & -1.1549 & 0.103 & 0.015 & -0.9716 & 0.057 & 0.005 & -1.165175835 & 0.029446 & 0.001397 \\
1.04 & -1.1392 & 0.126 & 0.024 & -0.9589 & 0.066 & 0.006 & -1.146619883 & 0.038612 & 0.002522 \\
1.14 & -1.1222 & 0.156 & 0.039 & -0.9450 & 0.077 & 0.009 & -1.127545340 & 0.050992 & 0.004551 \\
1.24 & -1.1054 & 0.195 & 0.061 & -0.9309 & 0.092 & 0.012 & -1.108905556 & 0.067418 & 0.008125 \\
1.34 & -1.0892 & 0.243 & 0.096 & -0.9171 & 0.111 & 0.018 & -1.091225845 & 0.088929 & 0.014284 \\
1.44 & -1.0744 & 0.303 & 0.148 & -0.9041 & 0.136 & 0.028 & -1.074797162 & 0.116765 & 0.024658 \\
1.54 & -1.0610 & 0.375 & 0.221 & -0.8920 & 0.167 & 0.043 & -1.059773915 & 0.152317 & 0.041667 \\
1.64 & -1.0493 & 0.459 & 0.322 & -0.8809 & 0.205 & 0.065 & -1.046226756 & 0.197033 & 0.068685 \\
1.74 & -1.0393 & 0.556 & 0.450 & -0.8709 & 0.252 & 0.099 & -1.034172594 & 0.252251 & 0.110008 \\
1.84 & -1.0309 & 0.661 & 0.602 & -0.8620 & 0.308 & 0.147 & -1.023590089 & 0.318969 & 0.170457 \\
1.99 & -1.0210 & 0.829 & 0.859 & -0.8507 & 0.410 & 0.252 & -1.010358524 & 0.441213 & 0.306241 \\
2.24 & -1.0104 & 1.111 & 1.285 & -0.8367 & 0.625 & 0.532 & -0.994606114 & 0.694902 & 0.657792 \\
2.49 & -1.0048 & 1.356 & 1.606 & -0.8278 & 0.871 & 0.904 & -0.985094626 & 0.975444 & 1.087118 \\
2.74 & -1.0020 & 1.547 & 1.800 & -0.8225 & 1.114 & 1.270 & -0.979785082 & 1.235946 & 1.458425 \\
2.99 & -1.0008 & 1.686 & 1.903 & -0.8195 & 1.328 & 1.556 & -0.976982734 & 1.449147 & 1.707925 \\
3.39 & -1.0000 & 1.830 & 1.971 & -0.8173 & 1.587 & 1.821 & -0.975072581 & 1.685729 & 1.902429 \\
3.79 & -0.9997 & 1.910 & 1.992 & -0.8165 & 1.756 & 1.934 & -0.974461196 & 1.824876 & 1.969446 \\
4.19 & -0.9997 & 1.953 & 1.998 & -0.8162 & 1.862 & 1.977 & -0.974279091 & 1.903892 & 1.990774 \\
4.59 & -0.9996 & 1.987 & 2.000 & -0.8161 & 1.927 & 1.993 & -0.974232239 & 1.948067 & 1.997304 \\
4.99 & -0.9996 & 1.993 & 2.000 & -0.8160 & 1.966 & 1.998 & -0.974224236 & 1.972481 & 1.999243 \\
\hline
\multicolumn{4}{l}{ME}   & 0.1775 & -0.206 & -0.215 & 0.0078 & -0.184 & -0.177 \\
\multicolumn{4}{l}{MAE}  & 0.1775 &  0.206 &  0.215 & 0.0136 &  0.184 &  0.177 \\
\multicolumn{4}{l}{RMSE} & 0.1776 &  0.257 &  0.324 & 0.0164 &  0.222 &  0.266 \\
\end{tabular}
\end{ruledtabular}
\end{table}

\clearpage

\section{Raw Computational Files}

This Supporting Information contains the complete set of raw computational files underlying this work, organized as follows:
\begin{enumerate}
    \item \textbf{IQC:} Input files and executable scripts are provided. Calculations can be reproduced by running the scripts according to \texttt{README.md}. The corresponding script outputs constitute the raw data reported in the Appendix.
    \item \textbf{OpenQP:} Input files and the associated output files are provided.
    \item \textbf{XTDDFT:} Input files and the associated output files are provided.
\end{enumerate}

\bibliography{main}